\documentclass{article}
\pdfoutput=1
\usepackage{jcappub}
\usepackage{amsmath}
\usepackage{amssymb}
\usepackage{xcolor}
\usepackage{hyperref}
\usepackage{slashed}
\usepackage{braket}
\usepackage{physics}
\usepackage{comment}
\usepackage{cleveref}
\usepackage{wrapfig}
\usepackage[normalem]{ulem}
\usepackage{slashed}
\usepackage{mathtools}
\usepackage{adjustbox}
\usepackage{graphicx}
\usepackage{subcaption}
\usepackage{gensymb}
\usepackage{tabularx}
\usepackage{eso-pic}% http://ctan.org/pkg/eso-pic

\def\vec#1{\mathbf{#1}}
\newcommand{\unit}[1]{\vec{\hat{#1}}}
\newcommand{\rpc}{r_{\rm pc}}
\newcommand{\rns}{R_{\rm NS}}
\newcommand{\mns}{M_{\rm NS}}
\newcommand{\vesc}{v_{\rm esc}}
\newcommand{\gagg}{g_{a\gamma\gamma}}
\newcommand{\edb}{{\bf E} \cdot {\bf B}}
\newcommand{\epar}{E_\parallel}

\colorlet{blue}{black}

\begin{document}

\AddToShipoutPictureBG*{%
  \AtPageUpperLeft{%
    \hspace{0.855\paperwidth}%
    \raisebox{-4\baselineskip}{%
      \makebox[0pt][r]{CALT-TH/2025-016}
}}}%
\AddToShipoutPictureBG*{%
  \AtPageUpperLeft{%
    \hspace{0.856\paperwidth}%
    \raisebox{-5\baselineskip}{%
      \makebox[0pt][r]{N3AS-24-041}
}}}%

\title{Physics beyond the Standard Model  \\
with the DSA-2000}

\author[a]{Kim~V.~Berghaus,}
\emailAdd{berghaus@caltech.edu}
\author[a]{Yufeng~Du,}
\emailAdd{yfdu@caltech.edu}
\author[a,b,c]{Vincent~S.~H.~Lee,}
\emailAdd{vincentszehimlee@berkeley.edu}
\author[d]{Anirudh~Prabhu,}
\emailAdd{prabhu@princeton.edu}
\author[e]{Robert Reischke,}
\emailAdd{rreischke@astro.uni-bonn.de}
\author[f]{Liam~Connor,}
\emailAdd{liam.connor@cfa.harvard.edu}
\author[a]{and Kathryn~M.~Zurek}
\emailAdd{kzurek@caltech.edu}

\affiliation[a]{Walter Burke Institute for Theoretical Physics, 
California Institute of Technology, 
1200 E. California Boulevard, 
Pasadena CA 91125, USA}
\affiliation[b]{Department of Physics, University of California Berkeley, Berkeley, CA 94720, USA}
\affiliation[c]{Department of Physics, University of California, San Diego, La Jolla, CA 92093-0319, USA}
\affiliation[d]{Princeton Center for Theoretical Science, Princeton University, 400 Jadwin Hall, Princeton, NJ 08544, USA}
\affiliation[e]{Argelander-Institut für Astronomie, Universität Bonn, Auf dem Hügel 71, D-53121 Bonn, Germany}
\affiliation[f]{Center for Astrophysics | Harvard $\&$ Smithsonian, Observatory Building E, 60 Garden St, Cambridge, MA 02138, USA}

\abstract{
The upcoming Deep Synoptic Array 2000 (DSA-2000) will map the radio sky at $0.7-2$ GHz ($2.9 - 8.3 \, \mu$eV) with unprecedented sensitivity.  This will enable searches for dark matter and other physics beyond the Standard Model, of which we study four cases: axions, dark photons, dark matter subhalos and neutrino masses. 
We forecast DSA-2000's potential to detect axions through two mechanisms in neutron star magnetospheres: photon conversion of axion dark matter and radio emission from axion clouds, developing the first analytical treatment of the latter.
We also forecast DSA-2000's sensitivity to discover kinetically mixed dark photons from black hole superradiance, constrain dark matter substructure and fifth forces through pulsar timing, and improve cosmological neutrino mass inference through fast radio burst dispersion measurements. Our analysis indicates that in its planned five year run the DSA-2000 could reach sensitivity to QCD axion parameters, improve current limits on compact dark matter by an order of magnitude, and enhance cosmological weak lensing neutrino mass constraints by a factor of three. }

\maketitle

\section{Introduction }
\label{sec:intro}

The hunt for physics beyond the Standard Model (BSM) requires diverse observational approaches spanning different scales and wavelengths. Despite extensive searches, fundamental questions remain about the nature of dark matter (DM), neutrino masses, and potential new forces and particles. Radio astronomy offers unique windows into these questions, with the upcoming Deep Synoptic Array 2000 (DSA-2000) representing a considerable advancement in our observational capabilities \citep{2019BAAS...51g.255H}.

The DSA-2000 will be an interferometric array of 2000 × 5 m antennas to be built in a radio-quiet valley in Nevada, USA. Operating at 0.7-2.0 GHz ($2.9 - 8.3 \, \mu$eV), its notable feature is the combination of small dishes that provide a large field of view together with substantial collecting area, resulting in substantial survey speed advantages for specific observing scenarios. 
For example, the DSA-2000 can image a 10\,deg$^2$ field to 1 $\mu$Jy in one hour, while the proposed mid-frequency Square Kilometer Array (SKA1-Mid) would require six hours, and the Karl G. Jansky Very Large Array (VLA) would need approximately 50 days of continuous observing. 

The sensitivity, survey speed, and multi-mode observing enable several distinct science applications at the intersection of astrophysics and physics. In multi-messenger astronomy, the DSA-2000 will contribute to follow-up observations of ground-based gravitational wave detections to locate and study their electromagnetic counterparts. It could find $\sim$\,20,000 new Galactic pulsars, $10^5$ localized cosmological fast radio bursts (FRBs), and millions of slower radio transients from stellar explosions. The continuum and spectroscopic imaging surveys will provide a window into the cosmic history of the Universe. 
Additionally, about 25$\%$ of observing time will be allocated to timing millisecond pulsars with bi-weekly cadence, contributing to pulsar timing array efforts to detect nanoHertz gravitational waves from supermassive black hole binaries. The parameters of the instruments and its key surveys are summarized in Table~\ref{tab:dsa}.

Planned for operation between 2028-2033, the DSA-2000 presents an upcoming opportunity to probe BSM physics. 
\begin{table}[t]
\centering
\begin{tabular}{|l|c|}
\hline
\textbf{Parameter} & \textbf{Value} \\
\hline
Antennas & $2000\times5$\,m\\
\hline
Frequency coverage & 0.7--2.0\,GHz\\
\hline
Frequency resolution (beamforming) & $\sim$\,10\,kHz \\
\hline
Frequency resolution (all sky survey) & $\sim$\,2.6\,MHz \\
\hline
Field of view (FoV) & 10\,deg$^2$ at 1.4\,GHz\\
\hline
Angular resolution & 3.3'' at 1.4\,GHz\\
\hline
Continuum sensitivity (1 hour) & 1\,$\mu$Jy\\
\hline
Pulsar sensitivity (15 min, 10 ms MSP) & 3\,$\mu$Jy\\
\hline
System-equivalent flux density (SEFD) & 2.5\,Jy\\
\hline
\end{tabular}
\caption{Specifications of the DSA-2000.}
\label{tab:dsa}
\end{table}
In this paper, we examine the DSA-2000's discovery potential for detecting such BSM signatures through two approaches: First, by searching for radio signals produced by BSM particles such as axions and dark photons, and second, by using the DSA-2000's planned pulsar timing and FRB discovery capabilities to probe BSM physics with  gravitational imprints, such as DM substructure, fifth forces, and neutrino masses.

We present forecasts for multiple BSM scenarios, distinguishing between searches that can be conducted within the planned DSA-2000 survey program and those requiring additional dedicated observations. 
The all-sky survey, which is part of the core observing strategy, will enable searches for dark photon superradiance signals from black holes \cite{Siemonsen:2022ivj}, and improve neutrino mass constraints through FRB dispersion measurements \cite{reischke_cosmological_2023}. We forecast that DSA-2000 will improve cosmological neutrino mass constraints from weak lensing measurements by a factor of three through breaking degeneracies with baryonic feedback. 
For axion searches, we consider two distinct scenarios:~(1)~radio signals from axion DM conversion in neutron star (NS) magnetospheres~\cite{Pshirkov:2007st,Huang:2018lxq,Hook:2018iia,Safdi:2018oeu,Battye:2019aco,Leroy:2019ghm,Foster:2020pgt,Prabhu:2020yif,Buckley:2020fmh,Witte:2021arp,Battye:2021xvt,battye2021robust,Nurmi:2021xds,Foster:2022fxn,Witte:2022cjj,Battye:2023oac,McDonald:2023shx,Xue:2023ejt,Tjemsland:2023vvc,Witte:2024akb,Kouvaris:2022guf,Maseizik:2024qly,Song:2024rru, Roy:2025mqw}, for which we advocate dedicated 10-hour high-resolution beamforming observations of select magnetars. Beamforming refers to phasing up the array in a particular direction, in this case preserving a single spatial pixel at the position of a known NS, but with very high time and frequency resolution. And (2),~radio signals from axion clouds that form around NSs~\cite{Noordhuis:2023wid, Caputo:2023cpv}, which we can search for using both the planned all-sky survey data and targeted beamforming observations, potentially reaching sensitivity to the QCD axion band. Here, we derive the first analytical description of radio signals from axion clouds around NSs as a function of period and magnetic field—a signal previously quantified only through simulations~\cite{Noordhuis:2023wid}. This analytical approach captures key features of the full simulation while providing deeper physical insight into the underlying processes, enabling more efficient exploration of the parameter space relevant to DSA-2000 observations.

Lastly, the planned discovery of $\sim 130$ new millisecond pulsars suitable for pulsar timing through the DSA-2000 pulsar survey~\cite{2019BAAS...51g.255H} will improve constraints on DM substructure and fifth forces by approximately an order of magnitude compared to current limits~\cite{NANOGrav:2023hvm}. Our analysis shows that while DSA-2000's planned 5-year operation will yield significant advances, extending millisecond pulsar timing observations over a 25-year period would dramatically enhance sensitivity to DM substructure, potentially reaching levels corresponding to a significant fraction of the local DM density—a threshold that would make this approach particularly powerful for constraining theories of DM.

The remainder of this paper is organized as follows: In Section~\ref{sec:axions}, we investigate radio signals from axions, beginning with the formalism and forecast for axion DM conversion in NS magnetospheres (Section~\ref{subsec:axion_A}). We then discuss non-DM axion probes (Section~\ref{subsec:axions_DM}), and examine the axion production mechanism from vacuum gaps around NSs (Section~\ref{subsec:vacuum_gaps}), which populates axion clouds (Section~\ref{subsec:axion_clouds}). Section~\ref{sec:darkphoton} explores radio signatures of dark photons from kinetically mixed dark photon superradiance around black holes. In Section~\ref{sec:pta}, we shift focus to pulsar timing arrays (PTAs), examining how DSA-2000's pulsar discoveries can constrain both DM substructure through gravitational effects (Section~\ref{subsec:pta_grav}) and potential fifth forces between DM and baryons (Section~\ref{subsec:pta_fifth}). Section~\ref{sec:neutrinos} details how DSA-2000's FRB detections can contribute to precise neutrino mass measurements by breaking degeneracies with baryonic feedback in weak lensing observations. Finally, in Section~\ref{sec:conc}, we discuss the implications of our findings, compare the sensitivities of the presented approaches, and outline promising directions for future work.

\section{Radio signals from axions }
\label{sec:axions}

Axions are hypothesized particles arising from well-motivated extensions of the Standard Model (SM) and are leading DM candidates. With mass $m_a \lesssim 1$ eV and high occupation numbers, axion DM is well-described by a classical wave oscillating at frequency $m_a/(2\pi)$. The mass range $m_a \lesssim$ meV is particularly compelling, as the observed DM density can be produced through non-thermal mechanisms like vacuum misalignment
\cite{PhysRevLett.40.223, DINE1983137, PRESKILL1983127}
and decay of topological defects
\cite{Battye:1993jv,
PhysRevLett.73.2954, DAVIS1989167, Marsh_2016, 
Gorghetto:2018myk, Gorghetto:2020qws,Buschmann:2019icd, Buschmann:2021sdq}.
The quantum chromodynamics (QCD) axion is especially well-motivated, as it not only offers a natural explanation for DM but also elegantly resolves the strong CP problem in QCD~\cite{Peccei:1977hh, Peccei:1977ur, 
PhysRevLett.40.223, Wilczek:1977pj}. The strong CP problem concerns the seemingly unnatural absence of charge-parity (CP) violation in QCD, as evidenced by the stringent experimental limits on the neutron’s electric dipole moment~\cite{Abel:2020pzs}.
Introducing the axion field $a(x)$ with a discrete shift symmetry, coupling to QCD via
\begin{equation}
L \supset \frac{\alpha_s}{8\pi f_a}\, a(x)\, G^{\mu\nu,a}\tilde{G}_a^{\mu\nu}. \tag{2.2}
\end{equation}
where $f_a$ is the axion decay constant, $\alpha_s$ is the strong coupling constant of QCD, $G^{\mu \nu,a}$ is the QCD field strength, and $\tilde{G}^{a}_{\mu \nu} = \frac{1}{2}\epsilon_{\mu \nu \rho \sigma} G^{\rho \sigma, a}$ is its dual, induces a potential for the axion. This potential is minimized at the CP conserving value, dynamically solving the strong CP problem. 
The predicted QCD axion mass-coupling relationship is
\begin{equation}
m_a = \sqrt{\frac{m_u m_d}{(m_u + m_d)^2}}\frac{m_\pi^2 f_\pi^2}{f_a^2} \simeq 0.0057\left(\frac{10^9 \text{ GeV}}{f_a}\right)\text{eV}, \tag{2.3}
\end{equation}
where $m_{u, d}$ are the up and down quark masses, respectively, and $m_\pi$ and $f_\pi$ are the pion mass and decay constant, respectively. The axion couples to electromagnetism through the operator
\begin{equation}
\label{eqn:axion_photon_coupling}
\mathcal{L}_{a\gamma\gamma} = -\frac{g_{a\gamma\gamma}}{4}aF^{\mu\nu}\tilde{F}_{\mu\nu} = g_{a\gamma\gamma}a\mathbf{E} \cdot \mathbf{B}, \tag{2.4}
\end{equation}
where for the QCD axion
\begin{equation}
g_{a\gamma\gamma} = \frac{\alpha}{2\pi f_a}\left(\frac{E}{N} - 1.92\right), \tag{2.5}
\end{equation}
with $E/N = 8/3$ (DFSZ \cite{DINE1981199, Zhitnitsky:1980tq}) or $E/N = 0$ (KSVZ \cite{PhysRevLett.43.103,SHIFMAN1980493}) in benchmark models.
Beyond QCD axions, general axions arise ubiquitously in UV completions and string theory \cite{Svrcek:2006yi, Green_Schwarz_Witten_2012, PhysRevD.81.123530, Demirtas:2021gsq}. Both DM and non-DM axions are theoretically motivated, and we explore DSA-2000's discovery potential for radio signals from axions in both contexts.

%%%%%%%%%%%%%%%%%%%%%%%%%%%%%%%%%%%%%%%%%%%%%%%%%%

\subsection{Axion dark matter conversion in neutron star magnetospheres}

In this section, we discuss the implications of the axion-photon mixing (through the term (\ref{eqn:axion_photon_coupling})) around NSs. In particular, we discuss the plasma state around NSs, briefly review the formalism of resonant axion-photon mixing in plasma, and discuss the detectability of radio signals associated with axion DM both in individual NSs and a population of galactic pulsars. 

\subsection*{Formalism}

\begin{wrapfigure}{r}{0.4\textwidth}
  \begin{center}
    \includegraphics[width=0.4\textwidth]{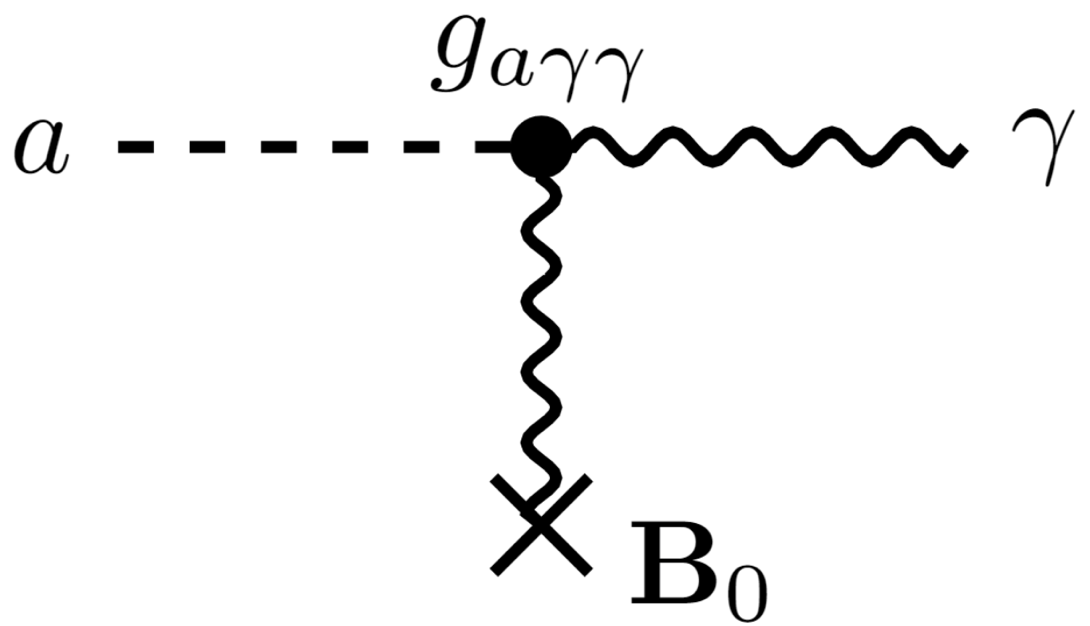}
  \end{center}
  \caption{An axion converts into a photon with frequency $f = m_a/2\pi$ in the presence of a static magnetic field $B_0$.}
  \label{fig:gagg}  
\end{wrapfigure}

NSs, the collapsed cores of massive stars, possess some of the largest magnetic fields in the observed universe, with surface field strengths as high as $10^{15}$~G.
Additionally, rotating NSs---pulsars---are surrounded by a dense plasma magnetosphere, consisting of electrons and positrons co-rotating with the NS\footnote{Importantly, co-rotation cannot be sustained everywhere in the magnetosphere. 
The implications of breakdown of co-rotation are expounded in Sec.~\ref{subsec:axion_clouds}.}\cite{1969ApJ...157..869G}, which can further enhance the mixing between axions and photons. 
A promising avenue towards indirect detection of axion DM is to search for bright, narrow radio lines associated with axion-to-photon conversion in NS magnetospheres~\cite{Pshirkov:2007st,Huang:2018lxq,Hook:2018iia,Leroy:2019ghm,Safdi:2018oeu, Battye:2019aco,Foster:2020pgt, 
Witte:2021arp,millar2021axionphotonUPDATED, battye2021robust,Foster:2022fxn, Witte:2022cjj,Battye:2023oac}, via the process depicted in Fig.~\ref{fig:gagg}.

The frequency of the radio signal is related to the axion mass through 
\begin{equation}\label{eqn:photon_frequency}
    f = \frac{m_a}{2\pi} \approx 0.2 \,\mathrm{GHz}\left(\frac{m_a}{10^{-6}\,\mathrm{eV}}\right) \, .
\end{equation}

\label{subsec:axion_A}

With a frequency coverage of $0.7\,\mathrm{GHz}\leq f \leq 2\,\mathrm{GHz}$, the DSA-2000 is sensitive to axions within the mass range
\begin{equation}
2.9\times 10^{-6}\,\mathrm{eV}\leq m_a\leq 8.3\times 10^{-6}\,\mathrm{eV}. 
\end{equation}

The expected flux of photons from axion conversion within an NS magnetosphere has been extensively derived in the literature~\cite{Pshirkov:2007st,Huang:2018lxq,Hook:2018iia,Safdi:2018oeu,Battye:2019aco,Leroy:2019ghm,Foster:2020pgt,Prabhu:2020yif,Buckley:2020fmh,Witte:2021arp,Battye:2021xvt,battye2021robust,Nurmi:2021xds,Foster:2022fxn,Witte:2022cjj,Battye:2023oac,Prabhu:2023cgb,McDonald:2023shx,noordhuis2023novel,Xue:2023ejt,Tjemsland:2023vvc,Noordhuis:2023wid,Caputo:2023cpv, Khelashvili:2024sup, Witte:2024akb,Kouvaris:2022guf,Maseizik:2024qly,Song:2024rru}.
Here we briefly summarize the assumptions and derivations.

\begin{figure}
	\includegraphics[width=1\textwidth]{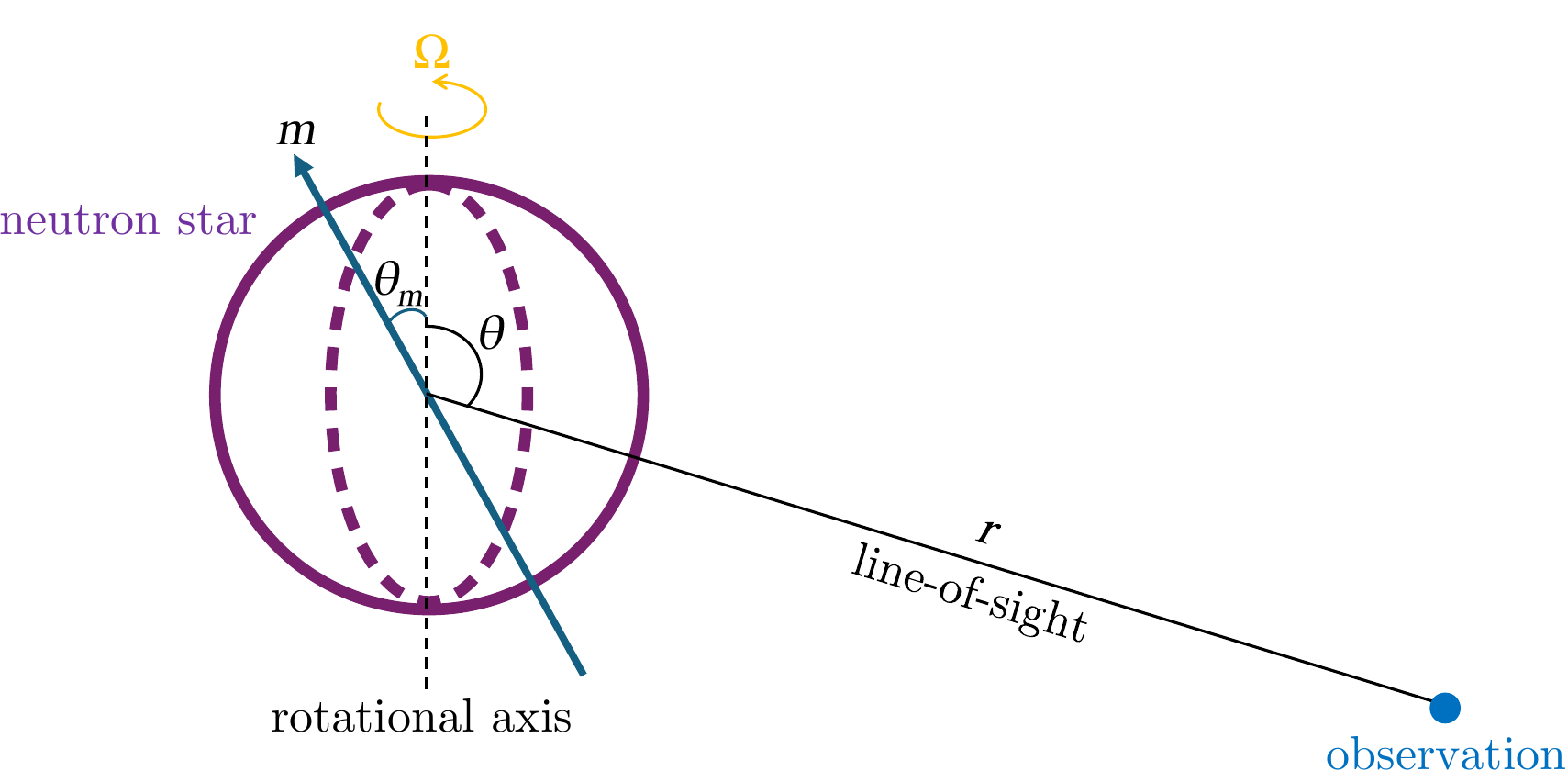} 
	\caption{Schematic of the NS geometry. The NS rotational axis ($\unit{z}$), magnetic axis ($\unit{m}$), and the line-of-sight ($\vec{r}$) are denoted by a dashed line, a blue arrow, and a solid black line, respectively. The misalignment angle between the rotational and magnetic axes, and the angle of observation with respect to the rotational axis, are defined to be $\theta_m$ and $\theta$, respectively, as described in the main text. }\label{fig:NS_schematic}
\end{figure}

The electron plasma distribution in the NS magnetosphere is well-described by the Goldreich-Julian (GJ) model~\cite{1969ApJ...157..869G}. 
The GJ model provides the minimum electron/positron number density required to screen any accelerating electric fields, which is derived by solving Maxwell's equations for particles on the magnetic field lines rotating around the star:
\begin{equation}\label{eqn:electron_density}
    n_e(\vec{x})=\frac{2\boldsymbol{\Omega} \cdot \vec{B}(\vec{x})/e}{1-\Omega^2r^2\sin^2\theta}\approx \frac{2\Omega B_z}{e} \, ,
\end{equation}
where $e$ is the positron charge, $r$ and $\theta$ are the radial distance (measured from the center of the NS) and polar angle in spherical coordinates, and $\boldsymbol{\Omega}$ (taken to be parallel to $\hat{z}$) and $\vec{B}(\vec{x})$ are the NS rotation frequency vector and magnetic field, respectively. We use the approximation $(\Omega r)^2=4\times 10^{-8}(\Omega/(2\pi\,\mathrm{Hz}))^2(r/(10\,\mathrm{km}))^2\ll 1$ for the second equality, which is valid for the range of NS rotational frequencies and radial distances of interest. The argument of Goldreich and Julian provides a condition on the net charge density, $\rho = - 2 \boldsymbol{\Omega} \cdot \vec{B}(\vec{x})$, required to screen the rotationally-generated electric field, with Eq.~\eqref{eqn:electron_density} indicating the minimum number density of charges required to provide this charge density.
It may be possible that the same charge density is provided by a much larger number density of particles, $n  = \kappa n_e$, where $\kappa$ is known as the ``multiplicity'' of the magnetosphere (which may itself be position dependent). 
High multiplicity is associated with regions of active pair creation. For example, the region supporting open field lines is expected to host a very high multiplicity, possibly reaching $10^5$~\cite{timokhinharding2015, timokhinharding2018}. 
By contrast, the closed magnetosphere is expected to be relatively dormant, which supports the assertion that the pair multiplicity should be relatively low therein. 
This is supported by 3D global simulations, although determining the exact closed-zone multiplicity is an active area of research.
For a dipole magnetic field, the $z$-component is

\begin{equation}\label{eqn:dipole_field}
    B_z=\frac{B_0}{2}\left(\frac{\rns}{r}\right)^3[3\cos\theta \unit{m}\cdot\unit{r}-\cos\theta_m] \, ,
\end{equation}
where $B_0$ is the magnetic field strength at the magnetic pole, $\rns$ is the NS radius, taken to be approximately 10 km, $\unit{m}$ is the unit vector in the direction of the magnetic axis, and $\theta_m$ is the polar angle between the rotation axis and the magnetic axis. 
In Fig.~\ref{fig:NS_schematic} we schematically show the geometry. 

The surface magnetic field of a generic NS can be computed by equating the radiation power of a rotating dipole with the change in NS rotational kinetic energy, giving~\cite{2012hpa..book.....L,Philippov:2013aha}
\begin{equation}\label{eqn:magnetic_field_P_Pdot_relation}
    B_0\approx 3.2\times 10^{19}\,\mathrm{G}\sqrt{\frac{P\dot{P}}{\mathrm{s}}} \, .
\end{equation}
Associated with the GJ density in (\ref{eqn:electron_density}) is a plasma frequency, $\omega_p(\vec{x}) = \sqrt{e^2 n_e(\vec{x})/m_e}$ (where $m_e$ is the electron mass), which can be understood as an effective photon mass.
When the plasma frequency is comparable to the axion mass, i.e. $m_a\sim \omega_p$, axions can resonantly convert into photons. In vacuum, axion-photon mixing suffers from a mismatch in the dispersion relations of the massive axion and the massless photon. In a magnetized plasma, the axion mixes with a particle eigenmode, called the O-mode, whose dispersion relation is (see, e.g.,~\cite{Witte2021})

\begin{align} \label{eqn:o_mode}
    \omega^2 = {1\over 2} \left(k^2 + \omega_p^2 + \sqrt{k^4 + \omega_p^4 + 2 k^2 \omega_p^2 (1 - 2 \cos^2\theta_B)} \right),
\end{align}
where $\theta_B$ is the angle between the mode wave vector and the magnetic field. In the limit of perpendicular propagation, the dispersion relation simplifies to $\omega^2 = k^2 + \omega_p^2$, exactly the dispersion relation of a massive particle with mass set by the plasma frequency. When the plasma frequency equals the axion mass, the dispersion relations coincide, making axion-photon interconversion kinematically favorable. This takes place at a distance of $r_c$ from the NS center, commonly referred to as the conversion radius, given by
\begin{align}\label{eqn:conversion_radius}
    r_c &= \rns \left(\frac{ e\Omega B_0}{m_a^2m_e}[3\cos\theta \unit{m}\cdot\unit{r}-\cos\theta_m]\right)^{1/3} \nonumber \\ 
    &\approx 14\ {\rm km} \left(\frac{R_{\mathrm{NS}}}{10\,\mathrm{km}}\right)
    \left(\frac{B_0}{10^{12}\,\mathrm{G}}\right)^{1/3}\left(\frac{\mathrm{sec}}{P}\right)^{1/3} \left(\frac{\mathrm{GHz}}{m_a/2\pi}\right)^{2/3}[3\cos\theta \unit{m}\cdot\unit{r}-\cos\theta_m]^{1/3} \, .
\end{align}

The radiated power of photons depends on the DM density near $r_c$, which is enhanced by the NS's gravitational field.
Assuming that DM velocity satisfies the Maxwell-Boltzmann distribution with virial velocity of $v_0$, by conservation of energy, the DM mass density near $r_c$ is related to its asymptotic DM density at infinity, $\rho_{\mathrm{DM}}$, by~\cite{Hook:2018iia}
\begin{align}\label{eqn:rho_rc}
    \rho^{r_c}_{\mathrm{DM}} = \rho_{\mathrm{DM}}\frac{2}{\sqrt{\pi}}\frac{1}{v_0}\sqrt{\frac{2GM_{\mathrm{NS}}}{r_c}} \, ,
\end{align}
where $M_{\mathrm{NS}}$ is the NS mass. 
The asymptotic DM mass density $\rho_{\mathrm{DM}}$ is dependent on its position within the galaxy.
In this work, for simplicity, we model the DM density profile with the Navarro-Frenk-White (NFW) profile, given by~\cite{Navarro:1995iw, Navarro:1996gj}
\begin{align}\label{eqn:NFW}
    \rho_{\mathrm{DM}}(R_{\mathrm{GC}}) = \frac{\rho_0}{(R_{\mathrm{GC}}/r_s)(1+R_{\mathrm{GC}}/r_s)^2}\, ,
\end{align}
where $R_{\mathrm{GC}}$ is the distance from the galactic center, $r_s\approx 8$ kpc is the scale radius of the profile~\cite{2019MNRAS.487.5679L}, and $\rho_0$ is a normalization constant set to produce $\rho_{\mathrm{DM}}(R_{\odot})=0.4$~GeV/cm$^3$ at the solar radius, $R_{\odot}=8$~kpc.

During most of the axion's trajectory through the magnetosphere, the dispersion relation of the axion $\omega^2 = k^2 + m_a^2$ differs from that of the O-mode~\eqref{eqn:o_mode}, with the momentum mismatch $\delta k \equiv |\vec{k}_a - \vec{k}_\gamma| \gg L^{-1}$, where $L$ is a typical length scale associated with the NS. Therefore, axion-photon mixing is highly suppressed. On resonance, when $\delta k \ll L^{-1}$, the conversion probability can be derived in the stationary-phase approximation (see, e.g.,~\cite{Hook:2018iia, Witte2021} for a derivation). In the non-relativistic limit, the conversion probability is~\cite{McDonald:2023ohd}

\begin{align} \label{eqn:conversion_probability}
    P_{a\to\gamma} &= {\pi \over 2} {\gagg^2 B(r_c)^2 \sin^2\theta_c \over |\vec{v}_c \cdot \boldsymbol{\nabla} \omega_p(r_c)|} \approx \frac{\pi}{3}g_{a\gamma\gamma}^2B(r_c)^2\frac{r_c}{m_av_c}
\end{align}
where $v_c\approx \sqrt{2GM_{\mathrm{NS}}/r_c}$ is the DM velocity at $r_c$, and $\theta_c$ is the angle between the magnetic field and the photon momentum at the conversion point. For simplification, we compute the conversion probability at the equator, where the angle between radially infalling axions and the magnetic field is $\theta_c = \pi/2$, and adopt the GJ model, wherein $\omega_p(r) \propto r^{-{3/2}}$. Since the conversion occurs resonantly in a small region around $r_c$, the radiated power of converted photons from the NS is obtained by considering a flux of DM through a surface with solid angle $d\Omega_{\mathrm{ang}}$ at $r=r_c$, i.e. $d\mathcal{P}\approx 2P_{a\to\gamma}\rho^{r_c}_{\mathrm{DM}}v_cr_c^2d\Omega_{\mathrm{ang}}$, and is hence given by (cf. Eqs.~\eqref{eqn:dipole_field}-\eqref{eqn:rho_rc} and Eq.~\eqref{eqn:conversion_probability})~\cite{Zhou:2022yxp}
\begin{align}\label{eqn:radiated_power}
    \frac{d\mathcal{P}}{d\Omega_{\mathrm{ang}}} = 1.3\times 10^9\,\mathrm{W}&\left(\frac{g_{a\gamma\gamma}}{10^{-12}\,\mathrm{GeV}^{-1}}\right)^2 \left(\frac{R_{\mathrm{NS}}}{10\,\mathrm{km}}\right)^{5/2} \left(\frac{m_a/2\pi}{\mathrm{GHz}}\right)^{4/3}\left(\frac{B_0}{10^{12}\,\mathrm{G}}\right)^{5/6} \left(\frac{P}{\mathrm{sec}}\right)^{7/6} \nonumber \\
    &\left(\frac{\rho_{\mathrm{DM}}}{0.4\,\mathrm{GeV\,cm}^{-3}}\right) \left(\frac{M_{\mathrm{NS}}}{M_{\odot}}\right)^{1/2}\left(\frac{200\,\mathrm{km/s}}{v_0}\right)|3\cos\theta\unit{m}\cdot\unit{r}-\cos\theta_m|^{5/6} \, .
\end{align}
The time-dependence of the power enters in the $\unit{m}\cdot\unit{r}$ term, which is periodic with a period that equals the pulsar period $P$.
For simplicity in our forecast, we consider the time-averaged signal. 
However, we note that since the timing resolution of DSA-2000 exceeds the frequency of a typical NS with a large magnetic field (and $P \gtrsim 1$~s). We expect an improved sensitivity by weighting the signal across the rotation phase, though this requires a more detailed analysis of the signal’s time dependence, since Eq.~\eqref{eqn:radiated_power} provides only an order-of-magnitude estimate on the signal amplitude. We leave such an analysis to future work.

The photon spectral flux density as measured on Earth, in units of Janskys (Jy), is given by
\begin{align}\label{eqn:flux}
    F = &\frac{1}{d^2\mathcal{B}}\frac{d\mathcal{P}}{d\Omega_{\mathrm{ang}}} \nonumber \\
    = & 1.4\times 10^{-5}\,\mathrm{mJy}\left(\frac{g_{a\gamma\gamma}}{10^{-12}\,\mathrm{GeV}^{-1}}\right)^2 \left(\frac{R_{\mathrm{NS}}}{10\,\mathrm{km}}\right)^{5/2} \left(\frac{m_a/2\pi}{\mathrm{GHz}}\right)^{1/3}\left(\frac{B_0}{10^{12}\,\mathrm{G}}\right)^{5/6} \left(\frac{P}{\mathrm{sec}}\right)^{7/6} \nonumber \\
    &\left(\frac{\rho_{\mathrm{DM}}}{0.4\,\mathrm{GeV\,cm}^{-3}}\right) \left(\frac{M_{\mathrm{NS}}}{M_{\odot}}\right)^{1/2}\left(\frac{200\,\mathrm{km/s}}{v_0}\right)\left(\frac{1\,\mathrm{kpc}}{d}\right)^2|3\cos\theta\unit{m}\cdot\unit{r}-\cos\theta_m|^{5/6}\, ,
\end{align}
where $d$ is the Earth-pulsar distance and orange$\mathcal{B}$ is the signal bandwidth, which we estimate as
\begin{equation}
\label{eq:axionDMbandwidth}
\mathcal{B}\sim 10^{-5}\left(\frac{m_a}{2\pi}\right) = 10 \, \text{kHz} \left(\frac{m_a/2\pi}{\mathrm{GHz}}\right)\,  .
\end{equation}
Here we have taken into account the interactions of axion‑sourced photons with magnetospheric plasma, which leads to a broadening of the signal bandwidth~\cite{Battye:2021xvt, Witte:2021arp, McDonald:2023shx}.
The signal-to-noise ratio (SNR) is given by 
\begin{equation}\label{eqn:SNR}
    \mathrm{SNR}=\frac{\langle F\rangle_{\varphi}}{\sigma(\Delta t_{\mathrm{obs}},\mathcal{B})} \, ,
\end{equation}
where $\sigma(\Delta t_{\mathrm{obs}},\mathcal{B})$ is the radio flux sensitivity limit as defined in Eq.~\eqref{eqn:SEFD}, $\varphi\equiv\Omega t$ is the NS phase, related to Eqs.~\eqref{eqn:radiated_power}--\eqref{eqn:flux} by $\unit{m}\cdot\unit{r}=\cos\theta_m\cos\theta+\sin\theta\sin \theta_m\cos\varphi$, and $\langle \cdots\rangle_{\varphi}$ denotes the phase average. Although the equations above hold for arbitrary polar angles, $\theta$, we are particularly interested in the value corresponding to the line of sight to Earth.

\subsection*{Target identification}

\begin{figure}[h]
	\includegraphics[width=1\textwidth]{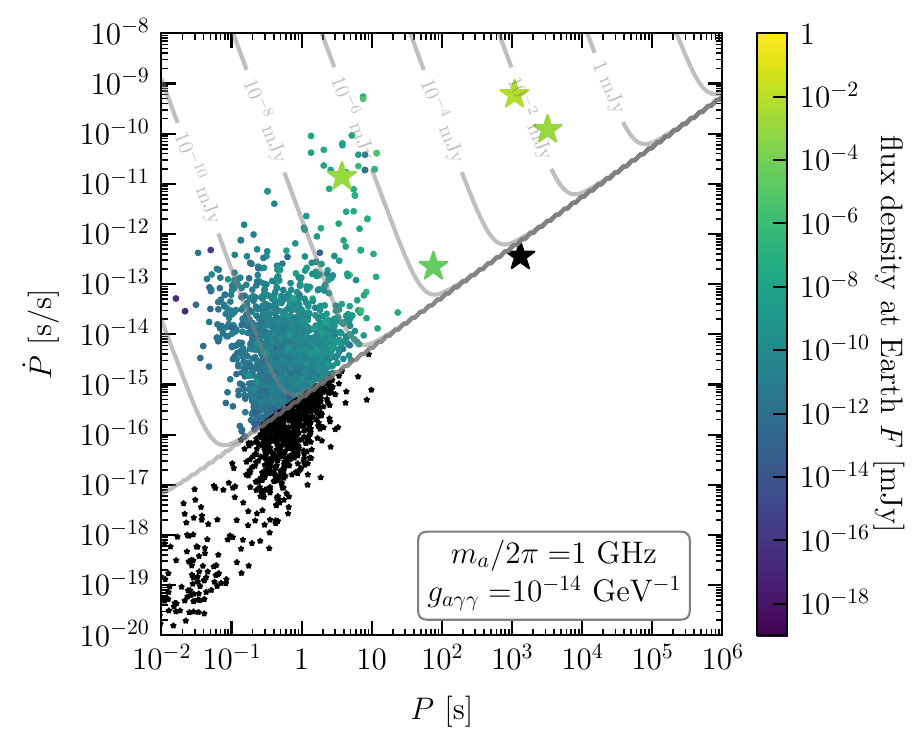} 
	\caption{Expected radio spectral flux density from axion-photon conversions in NSs, assuming $m_a/(2\pi)=1$~GHz and $g_{a\gamma\gamma}=10^{-14}$~GeV$^{-1}$, 
       for each pulsar in the Australia Telescope National Facility pulsar catalog~\cite{Manchester:2004bp}, magnetars in the McGill Magnetar Catalog~\cite{Olausen:2013bpa}, the Galactic Center magnetar SGR J1745-2900 with $P=3.76$~s~\cite{Kennea:2013dfa, Mori:2013yda, Shannon:2013hla}, and four recently identified long-period magnetars, including PSR J0901-4046 with $P=75.89$~s~\cite{Caleb:2022xyo}, GLEAM-X J162759.5-523504.3 with $P=18.2$~min~\cite{Hurley-Walker:2022}, GPM J1839-10 with $P=22.0$~min~\cite{Hurley-Walker:2023}, and ASKAP J1935+2148, with $P=53.8$~min~\cite{Caleb:2024nrq}.
       Pulsars and magnetars in the catalogs are denoted by dots, while long-period magnetars are denoted by stars. The background contours in grey are derived assuming $M_{\mathrm{NS}}=1\,M_{\odot}$, $R_{\mathrm{NS}}=10$ km, $\theta=120^{\circ}$, $\theta_m=10^{\circ}$, $\rho_{\mathrm{DM}}=0.4$~GeV/cm$^3$, $d=1$~kpc, and $B_0$ satisfying Eq.~\eqref{eqn:magnetic_field_P_Pdot_relation}. The NSs that are not expected to produce any observable flux with these parameters are colored in black. Here we note that PSR J0901-4046 and GLEAM-X J162759.5-523504.3 are not within DSA-2000's field of view.}\label{fig:flux_heatmap}
\end{figure}

To detect a photon signal from axion-photon conversion, one could either target specific NSs expected to exhibit a strong axion-induced flux or measure the total flux from a population of NSs (see, \textit{e.g.}, Ref.~\cite{Safdi:2018oeu}). However, in the case of DSA-2000, achieving the required frequency resolution to detect the DM axion signal (\textit{cf.} Eq.~\eqref{eq:axionDMbandwidth}) necessitates beamforming with high temporal and spectral resolution on specific sky positions. 
This limitation makes a collective observation of an NS population unfavorable. Therefore, for the remainder of this Section, we focus on identifying individual targets that offer the most promising sensitivity to DM axion signals.

It is clear from the expression in Eq.~\eqref{eqn:radiated_power} that an ideal target will have a long period $P$ and a strong magnetic field strength $B_0$, close to the galactic center (for enhanced DM density), and proximity to Earth. We identify potential NS targets using the pulsar catalog in the Australia Telescope National Facility~\cite{Manchester:2004bp} and the McGill Magnetar Catalog~\cite{Olausen:2013bpa}\footnote{Online version of the McGill Magnetar Catalog: \url{https://www.physics.mcgill.ca/~pulsar/magnetar/main.html}.}.  
Since DSA-2000 is located in Nevada, we consider only NSs with declination above $-30^{\circ}$.

In Fig.~\ref{fig:flux_heatmap}, we show the expected radio flux density, computed from Eqs.~\eqref{eqn:conversion_probability}-\eqref{eqn:flux}, considering targets from the ATNF pulsar catalog~\cite{Manchester:2004bp} and the McGill Magnetar Catalog~\cite{Olausen:2013bpa}.
We take the representative axion benchmark parameters to be $m_a/(2\pi)=1$~GHz and $g_{a\gamma\gamma}=10^{-14}$~GeV$^{-1}$.
We also show the expected flux density from NSs with other values of $P$ and $\dot{P}$ with grey background contours, with $B_0$ computed using Eq.~\eqref{eqn:magnetic_field_P_Pdot_relation}, and fiducial values of $\rho_{\mathrm{DM}}=0.4$~GeV/cm$^3$ and $d=1$~kpc. Note that while increasing $P$ leads to a larger radio flux in general, it also shrinks the conversion radius $r_c$ in Eq.~\eqref{eqn:conversion_radius}, which can completely block the axion-photon conversion if $r_c$ falls below the NS radius $R_{\mathrm{NS}}$. These NSs, which are colored in black in Fig.~\ref{fig:flux_heatmap}, will not produce any observable flux.

Additionally, we note that DSA 2000 has the potential of discovering slow magnetars. 
Indeed, several slow magnetars have been identified in the past few years, including PSR J0901-4046 with $P=75.89$~s~\cite{Caleb:2022xyo}, GLEAM-X J162759.5-523504.3 with $P=18.2$~min~\cite{Hurley-Walker:2022}, GPM J1839-10 with $P=22.0$~min~\cite{Hurley-Walker:2023}, and ASKAP J1935+2148 with $P=53.8$~min~\cite{Caleb:2024nrq}.
Among these four magnetars, GPM J1839-10 and ASKAP J1935+2148 are within DSA-2000's field of view.
These slow magnetars also generically have strong surface magnetic fields ($B_0\sim10^{14}-10^{16}$~G), and are thus promising targets for axion search thanks to the $P$ and $B_0$ scaling in Eq.~\eqref{eqn:radiated_power}.
We have denoted these slow magnetars as well as the galactic center magnetar with star symbols in the scatter plot in Fig.~\ref{fig:flux_heatmap}. Due to their long periods and strong magnetic fields, the expected radio flux density from axion-photon conversion can be quite high, ranging from $F\sim 0.01-1$~mJy.
We expect DSA-2000 to be able to discover additional slow magnetars with similar physical parameters. 

\subsection*{Forecast}
\begin{figure}[h]
	\includegraphics[width=1\textwidth]{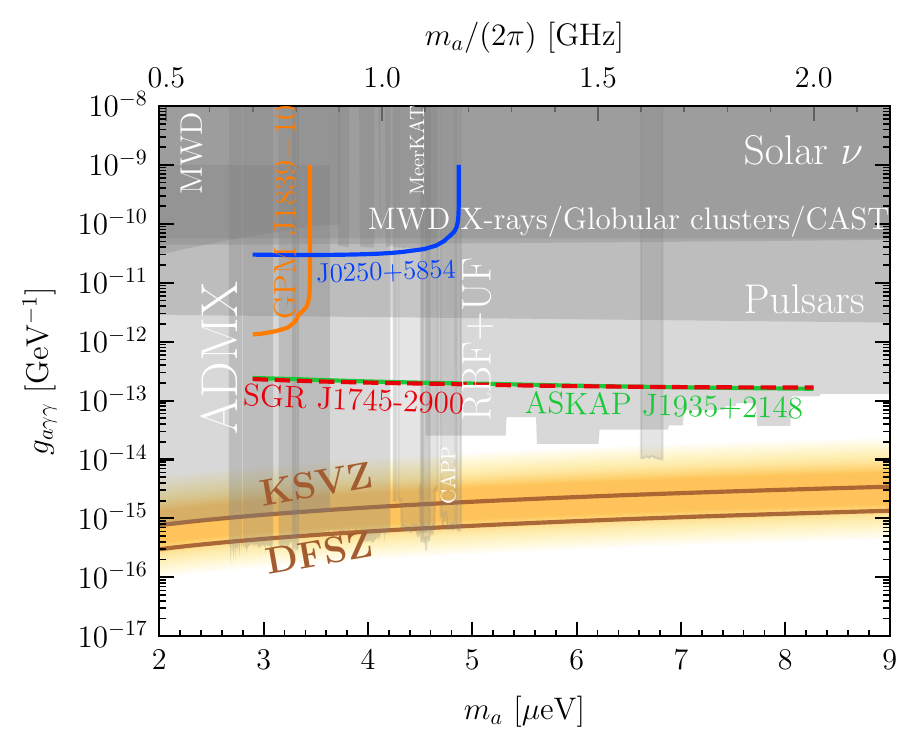} 
    \caption{Projected $5\sigma$ upper limits on $g_{a\gamma\gamma}$ are shown 
    assuming 10 hours of observation time ($\Delta t_{\mathrm{obs}}=10$ hrs) on the Galactic Center magnetar SGR J1745-2900~\cite{Kennea:2013dfa, Mori:2013yda, Shannon:2013hla} (red dashed), two long-period magnetars: GPM J1839-10 with $P=22.0$ min~\cite{Hurley-Walker:2023} (orange solid) and ASKAP J1935+2148 with $P=53.8$ min~\cite{Caleb:2024nrq} (green solid), and an ordinary pulsar J0250+5854 with $P=23.5$~s~\cite{Tan:2018rhg} (blue solid), observed with DSA 2000. The properties of each target are listed in Table~\ref{tab:axion_DM_targets}. Here, we assume $M_{\mathrm{NS}}=1 M_{\odot}$, $R_{\mathrm{NS}}=10$ km, $\theta=120^{\circ}$, $\theta_m=10^{\circ}$, and $\mathrm{SEFD} = 2.5$ Jy. The DM density near the NS is estimated using an NFW profile with a scale radius of $r_s=8$ kpc, with the Galactic Center magnetar's distance from the Galactic Center taken to be $R_{\mathrm{GC}}=0.1$ pc. Existing constraints from ADMX~\cite{Asztalos_2010, ADMX:2018gho, ADMX:2018ogs, ADMX:2019uok, ADMX:2021mio, ADMX:2021nhd, ADMX:2024xbv}, RBF~\cite{PhysRevLett.59.839,Wuensch:1989sa}, UF~\cite{PhysRevD.42.1297, Hagmann:1996qd}, CAPP~\cite{Lee:2020cfj, Jeong:2020cwz, CAPP:2020utb, Lee:2022mnc, Yoon:2022gzp, Kim:2022hmg, Yi:2022fmn, Yang:2023yry, Kim:2023vpo, CAPP:2024dtx}, CAST~\cite{CAST:2007jps, CAST:2017uph, CAST:2024eil}, globular clusters~\cite{Ayala:2014pea, Dolan:2022kul}, pulsar polar cap~\cite{noordhuis2023novel}, solar neutrinos~\cite{Vinyoles_2015}, MWD~\cite{Dessert:2021bkv, Dessert:2022yqq, Benabou:2025jcv}, and MeerKAT~\cite{Battye:2023oac} are shaded in grey (see Ref.~\cite{AxionLimits}). The QCD axion parameter space is shaded in yellow~\cite{Saikawa:2024bta}. 
}\label{fig:reach}
\end{figure}

In Fig.~\ref{fig:reach}, we report the projected $5\sigma$ upper limits on $g_{a\gamma\gamma}$ after 10 hours of observation, each on Galactic Center magnetar SGR J1745-2900~\cite{Kennea:2013dfa, Mori:2013yda, Shannon:2013hla}, and two long-period magnetars, GPM J1839-10 with $P=22.0$~min~\cite{Hurley-Walker:2023} and ASKAP J1935+2148 with $P=53.8$~min~\cite{Caleb:2024nrq} with DSA-2000. For comparison, we also show upper limits from an ordinary (\textit{i.e.} non-magnetar) pulsar J0250+5854, with $P = 23.5$~s~\cite{Tan:2018rhg}, which is the second-slowest-spinning radio pulsar known to date.\footnote{The slowest-spinning radio pulsar known to date is PSR J0901-4046 with $P=75.9$~s~\cite{Caleb:2022xyo}, which is one of the long-period magnetars mentioned in the previous paragraph. It is not within DSA-2000's field of view.} The upper limits are derived using Eqs.~\eqref{eqn:conversion_probability}-\eqref{eqn:flux}, and setting SNR$=5$ in Eq.~\eqref{eqn:SNR}. The parameters for each target are listed in Table~\ref{tab:axion_DM_targets}.

\begin{table}[h]
\centering
\small
\begin{tabular}{>{\centering\arraybackslash}p{3.6cm}|>{\centering\arraybackslash}p{0.6cm}|>{\centering\arraybackslash}p{1.35cm}|>{\centering\arraybackslash}p{.6cm}|>{\centering\arraybackslash}p{1.3cm}|>{\centering\arraybackslash}p{1.5cm}|>{\centering\arraybackslash}p{1.2cm}|>{\centering\arraybackslash}p{1.8cm}}  
\hline
\hline
\textbf{Target} & \textbf{$\mathbf{P}$} & {$\mathbf{B_{0}}$} & \textbf{$\mathbf{d}$} & \textbf{RA} & \textbf{DEC} & \textbf{$\mathbf{R_{\mathrm{GC}}}$} & \textbf{$\mathbf{\rho_{\mathrm{DM}}}$} \\
 & \textbf{[s]} & \textbf{[G]} & \textbf{[kpc]} & & & \textbf{[pc]} & \textbf{[GeV/cm$^3$]} \\
\hline
\hline
SGR J1745-2900~\cite{Kennea:2013dfa, Mori:2013yda, Shannon:2013hla} & $3.764$ & $2.3\times 10^{14}$ & $8.0$ & $17^{\text{h}}45^{\text{m}}40^{\text{s}}$ & $-29^\circ00'30''$ & $0.1$ & $1.6\times 10^5$ \\
\hline
GPM J1839-10~\cite{Hurley-Walker:2023} & $1318$ & $7.0\times 10^{14}$ & $5.7$ & $18^{\text{h}}39^{\text{m}}2^{\text{s}}$ & $-10^\circ31'49''$ & $3.7\times 10^3$ & $2.0$ \\
\hline
ASKAP J1935+2148~\cite{Caleb:2024nrq} & $3225$ & $2.0\times 10^{16}$ & $5.0$ & $19^{\text{h}}35^{\text{m}}5^{\text{s}}$ & $21^\circ48'42''$ & $7.0\times 10^3$ & $0.7$ \\
\hline
J0250+5854~\cite{Tan:2018rhg} & $23.54$ & $2.6\times 10^{13}$ & $1.6$ & $02^{\text{h}}50^{\text{m}}18^{\text{s}}$ & $58^\circ54'01''$ &$9.7\times 10^3$ & $0.3$ \\
\hline
\hline
\end{tabular}
\caption{Proposed pulsar and magnetar targets for a DM axion search with DSA-2000. Here, RA and DEC denote the right ascension and declination of the star, respectively. The values of $R_{\mathrm{GC}}$ for GPM J1839-10 and ASKAP J1935+2148 are inferred using $d$, RA, and DEC, while the value of $R_{\mathrm{GC}}$ for SGR J1745-2900 is taken from Ref.~\cite{Shannon:2013hla}. The value of $\rho_{\mathrm{DM}}$ for each star is computed assuming an NFW halo profile (\textit{cf}. Eq.~\eqref{eqn:NFW}).}
\label{tab:axion_DM_targets}
\end{table}

We assume the polar angles of the magnetic axis and line-of-sight to the Earth are $\theta_m=10^{\circ}$ and $\theta = 120^{\circ}$, respectively, for illustrative purposes.
In general, we find that magnetars are better targets than pulsars thanks to their strong magnetic field. Here we note that the converted photon power depends linearly on the local DM density around the NS, which in turn depends on the density profile of the DM halo (taken to be an NFW profile in this work as described in Eq.~\eqref{eqn:NFW}). 

We find that DSA-2000 will be sensitive to axions with masses in the range $2.9\,\mathrm{\mu eV} \lesssim m_a \lesssim 8.3\,\mathrm{\mu eV}$ and coupling strengths $g_{a\gamma\gamma} \gtrsim 2\times 10^{-13}\,\mathrm{GeV}^{-1}$, assuming axions constitute all of DM. However, the axion mass range within DSA-2000's sensitivity has already been ruled out by a combination of direct detection and astrophysical constraints.
As expected, magnetars can generally place tighter constraints on the axion coupling strength thanks to their strong magnetic field, although ordinary pulsars are subjected to less modeling uncertainties.
It is also interesting to note that both SGR J1745-2900 and ASKAP J1935+2148 are projected to place almost identical upper limits on \( g_{a\gamma\gamma} \) in a log-log plot. This is because the photon flux density measured on Earth scales parametrically as \( \propto B_0^{5/6} P^{7/6} \rho_{\mathrm{DM}}^{\infty} d^{-2} \) (\textit{cf}. Eqs.~\eqref{eqn:radiated_power}-\eqref{eqn:flux}), for which these two stars yield numerical values within \(\sim 20\%\) of each other based on their parameters listed in Table~\ref{tab:axion_DM_targets}. 

Our analysis is based on a number of simplifying assumptions.
We model axion-photon mixing using a 1D approximation in which photon trajectories are radial such as in~\cite{Hook:2018iia}. More detailed computations that account for, e.g., 3D effects, non-radial axion and photon trajectories, mode mixing, small scale plasma fluctuations, have been performed in~\cite{Witte2021, millar2021axionphotonUPDATED, McDonald:2023ohd, McDonald:2023shx, Gines:2024ekm}.
The numerical analysis of axion-photon mixing performed in~\cite{Gines:2024ekm} agrees well with the results of~\cite{McDonald:2023ohd}, on which our calculations are based. We also neglect Euler–Heisenberg corrections to the photon dispersion relation, which are expected to become relevant for relativistic axions~\cite{Long:2024qvd}. This omission is justified in our case, as the axions we consider are at most semi-relativistic.
An additional complication in modeling magnetar magnetospheres is the presence of magnetic ``twist'' -- that is, regions where $\boldsymbol{\nabla} \times \mathbf{B} \ne 0$. Such twist generates strong currents in the magnetosphere, supported by plasma with densities far exceeding the GJ value. The implications of twisted magnetospheres were first explored in~\cite{McDonald:2023shx}, with a more detailed analysis presented in~\cite{Roy:2025mqw}. The latter study found that in certain regimes—particularly as the magnetar enters quiescence—large regions of the magnetosphere may resemble the untwisted, GJ-like configuration. This behavior supports the estimates used in our modeling, but more observations of the considered sources are required to reduce systematic uncertainties.

\subsection{The need for non-dark matter axion probes}
\label{subsec:axions_DM}

A major obstacle in the way of axion DM detection is the theoretical uncertainty in predicting the present-day axion abundance.
For instance, the cosmology of the QCD axion depends critically on when PQ symmetry is broken relative to inflationary reheating.
For pre-inflationary symmetry breaking, the present-day axion abundance is degenerate with the initial misalignment angle, providing little information on the axion mass~\cite{Marsh_2016}.
For post-inflationary symmetry breaking, there is a unique axion mass that provides the observed DM abundance, but determining that mass has been the subject of decades of intensive numerical simulations~\cite{Klaer:2017ond, Gorghetto:2018myk,
Vaquero:2018tib, Drew:2019mzc, Gorghetto:2020qws, Dine:2020pds, Buschmann:2021sdq, Drew:2022iqz, Drew:2023ptp, Kim:2024wku, Saikawa:2024bta, Kim:2024dtq, benabou2024axionmasspredictionadaptive}. Despite disagreements on the determination of the axion mass by different groups, most predict the axion mass to be $m_a \gtrsim \mu$eV. 
Moreover, it is possible that the axion mass lives outside the aforementioned range and makes up only a subcomponent of DM, reducing the sensitivity of probes that rely on axions making up all of DM.

Apart from being a compelling DM candidate, the axion holds significant theoretical interest in its own right.
As discussed in the beginning of Sec.~\ref{sec:axions}, the QCD axion provides an elegant solution to the strong CP problem in the Standard Model, releasing the fine-tuning constraint on the CP-violating parameter $\bar{\Theta}$. Additionally, in string theory, axions arise naturally in the compactifications of Calabi-Yau manifolds, and the detection of a QCD axion could facilitate meaningful statements about the string theory landscape \cite{Gendler:2024adn}.
Therefore, the discovery of an axion, especially the QCD axion, could have serious theoretical implications even if it does not constitute a significant fraction of DM.
Such opportunities motivate Light-Shining-through-Wall (LSW) experiments where axions are generated from coupling to SM fields, such as $\mathcal{L}_{a}\supset -g_{a\gamma\gamma}a\edb$.
Unlike in laboratories where the magnetic field is well controlled but weak, astrophysical systems like NSs offer unique advantages with their strong magnetic fields.
In the following subsection, we explore signals from NS-originated axion clouds converting to photons through the NS magnetosphere. 

\subsection{Vacuum gaps around neutron stars}
\label{subsec:vacuum_gaps}

\begin{figure}[h]
    \centering
    \includegraphics[width=\linewidth]{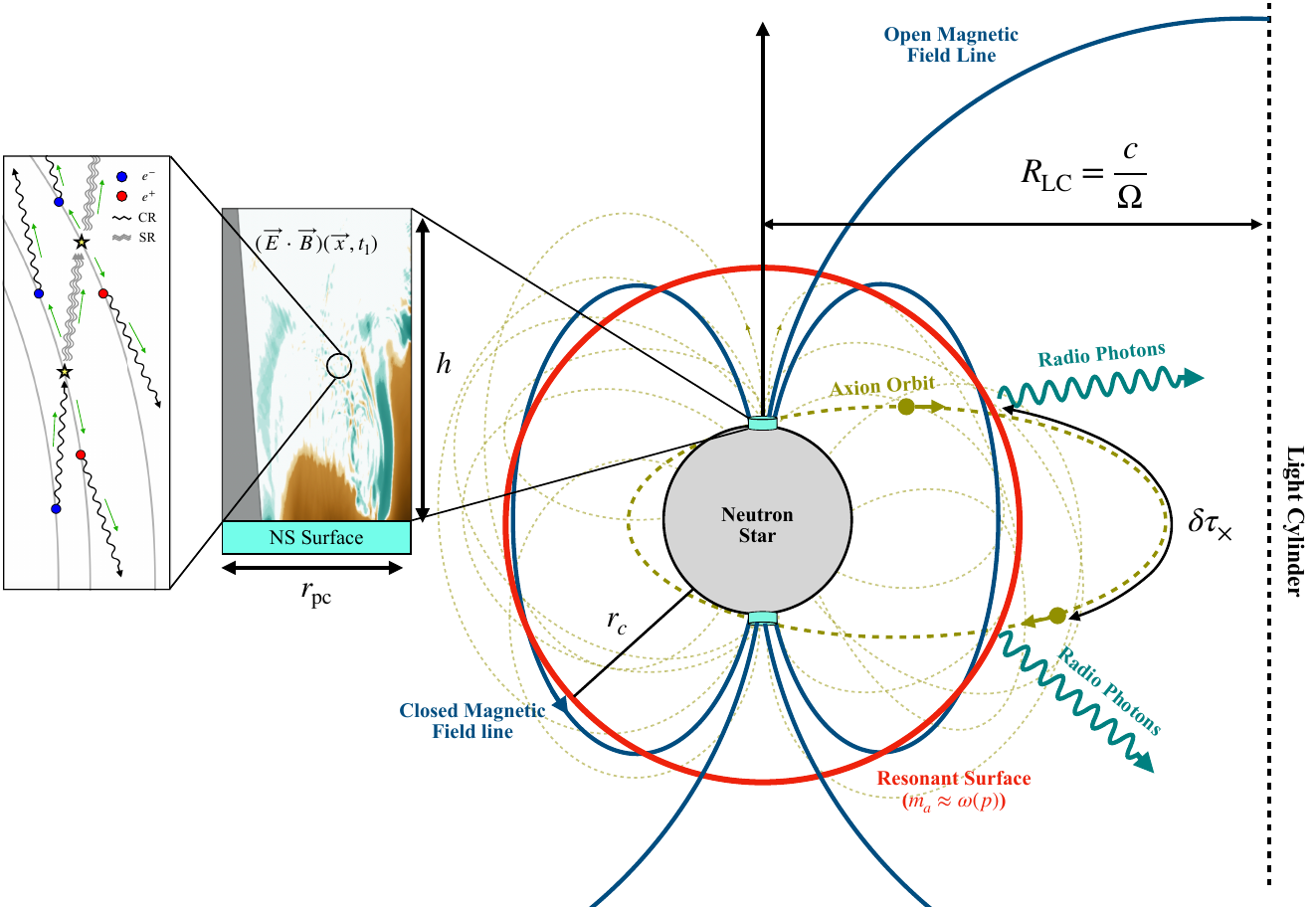}
    \caption{ 
    Schematic of axion cloud formation and associated radio emission. Vacuum gaps (cyan cylinders) are formed at the footpoints of open field lines, defined as field lines that extend beyond the light cylinder (black, dashed). These gaps host large and time-varying $\edb$. The right inset shows a snapshot of $\edb$ within the gap region, and the left inset shows a schematic of microphysical processes related to pair production (``CR'' stands for curvature radiation and ``SR'' for synchrotron radiation). A fraction of axions produced in the gap execute bound gravitational orbits around the NS (gold, dashed) and resonantly convert to radio photons when they encounter the resonant conversion surface (red). The characteristic time between encounters of the orbit with the resonant surface is represented as $\delta \tau_{\cross}$. }
    \label{fig:cartoon}
\end{figure}

The GJ model describes the minimum charge density required to screen the rotation-induced electric field component, $\epar$, parallel to the background magnetic field. 
Crucially, this model is insufficient to describe regions that do not co-rotate with the neutron star. Many important deviations from the GJ model are reviewed below. The GJ model breaks down near the light cylinder (LC), a cylinder that is coaxial with the NS rotation axis and with radius $R_{\rm LC} = c/\Omega$, where $\Omega$ is the NS rotation frequency. 
The LC defines the boundary beyond which co-rotation requires superluminal propagation. In the equatorial region near the LC, the GJ density diverges, signaling a breakdown of the model. The LC divides the region surrounding the NS into the ``closed'' and ``open'' magnetospheres illustrated in Fig.~\ref{fig:cartoon}, where the former contains all magnetic field lines whose maximum extent from the NS lies within the magnetosphere (thus allowing the field lines to close) and the latter contains field lines that exit the LC and whose closure within the magnetosphere is forbidden by causality. Application of the GJ model is appropriate only in the closed magnetosphere and well within the LC.
The footpoints of all open field lines form a region on the NS surface called the polar cap (PC), represented as cyan cylinders in Fig.~\ref{fig:cartoon}. For a dipolar magnetic field, the PC is described by the region on the NS surface with $\theta \le \theta_{\rm pc} \equiv \sqrt{\Omega \rns}$, where $\theta$ is the polar angle in spherical coordinates (with $\theta = 0$ corresponding to the magnetic pole, which here and below is assumed to be aligned with the rotation axis, such that $\theta_m = 0$).\footnote{This expression may be derived by noticing that dipolar field lines are contours of fixed $r/\sin^2(\theta)$. Each contour may be defined by the polar angle of its footpoint, $\theta_f$, such that $r/\sin^2(\theta) = \rns/\sin^2(\theta_f)$.
The maximum radius a field line extends from the NS occurs when $\theta = \pi/2$, giving $R_{\rm max} = \rns/\sin^2(\theta_f)$. 
The last closed field line, which demarcates the edge of the polar cap has maximum radius, $R_{\rm max} = R_{\rm LC}$, which gives the quoted expression for $\theta_{\rm pc}$ in the limit that $\Omega \rns \ll 1$, which is always the case for pulsars of interest.}
Particles flowing along open field lines are evacuated from the polar region, leading to the formation of a vacuum gap right above the PC. The dearth of pairs in the gap leads to a large longitudinal voltage drop along the gap, which grows until it becomes unstable to runaway $e^\pm$ pair creation, screening the gap~\cite{1971ApJ...164..529S}. The competition between pair evacuation along open field lines and screening leads to a rich spatio-temporal evolution of $\epar$ (see right panel in Fig.~\ref{fig:cartoon}), with important implications for axion production as we describe in the following subsection ~\cite{Prabhu:2021zve, Noordhuis:2022ljw}.

\subsection{Axion clouds around neutron stars}
\label{subsec:axion_clouds}

It was recently shown that NSs are efficient emitters of axions in the radio frequency range~\cite{Prabhu:2021zve}.
The mechanism for axion generation is related to the formation of vacuum gaps in NS magnetospheres. 
Vacuum gaps host large, accelerating electric fields $\epar$, parallel to the background magnetic field,  and are believed to be the sites of non-thermal electromagnetic emission. 
A large $\epar$ is unstable to runaway $e^\pm$ pair creation, which dynamically screens $\epar$, giving rise to a space and time-varying $\edb$, which acts as a classical source of axions, according to the Klein-Gordon equation \cite{Prabhu:2021zve},

\begin{equation}
\label{eq:axionEOM}
(\square+m_a^2)a=g_{a\gamma\gamma}  \Vec{E}\cdot\Vec{B} \, .
\end{equation}
Dynamical screening of $\epar$ in pulsar polar caps is believed to be the origin of pulsed radio emission observed in rotating neutron stars~\cite{Philippov2020}. Axions are produced with a broad spectrum, set by the spectrum of plasma oscillations in the gap. A fraction of produced axions resonantly convert to photons, giving rise to broadband radio emission. 
Non-observation of these radio signals from NSs has yielded some of the most stringent constraints on axions with masses between $10^{-8} \ {\rm eV} \lesssim m_a \lesssim 10^{-5}$ eV~\cite{Noordhuis:2022ljw}.
If the axion mass lies in the range $10^{-9}$ eV $\lesssim m_a \lesssim 10^{-4}$ eV, a significant fraction of produced axions remain gravitationally bound to the NS~\cite{Noordhuis:2023wid}.
The population of bound axions forms a cloud whose density can grow over astrophysical timescales, reaching densities  over twenty orders of magnitude larger than the local DM density ($\rho_{\rm DM} \approx 0.4$ GeV/cm$^3$)~\cite{Noordhuis:2023wid}.
The cloud also experiences dissipation through a number of processes, with the dominant effect being resonant axion-to-photon conversion at large distances. 
The dissipation of the cloud gives rise to bright, narrow-band radio emission that may be observable with sensitive telescopes and arrays such as the DSA-2000.
The radio lines emanating from axion clouds have a fractional bandwidth of $\mathcal{B}/f \sim \mathcal{O}(10^{-2})$, which is set by the kinematic requirement for resonant axion-to-photon conversion as described in detail below. This signal is much broader than the expected signal associated with DM axion conversion in the magnetosphere. Crucially, the line-width of radio emission from axion clouds exceeds the broadening from individual NS motions, thus allowing a simple search strategy from populations of NSs. Below we estimate the sensitivity of DSA-2000 to radio signals from axion clouds. 

In this subsection, we review the semi-analytic formalism for axion production in pulsar polar caps, with a focus on the gravitationally bound population~\cite{Noordhuis:2023wid, Caputo:2023cpv}. We then derive analytic estimates for the luminosity and bandwidth of the associated radio signal arising from axion cloud dissipation. Finally, we use these estimates to forecast the sensitivity of DSA-2000 to both individual pulsars and the broader population expected to be discovered by the array. A visual summary of the formation of axion clouds and their associated radio signals is shown in Fig.~\ref{fig:cartoon}.

\subsection*{Formalism}

The axions produced by dynamical screening of the accelerating electric field can be obtained by solving the axion equation of motion \eqref{eq:axionEOM},
giving a differential production rate of~\cite{Peskin:1995ev}:
\begin{equation} \label{eqn:dndotdk}
    \frac{d\dot{E}_a}{d^3k} = \frac{{|\widetilde{\mathcal{S}}(\Vec{k})|}^2}{2{(2\pi)}^3 T} \, ,
\end{equation}
where $\dot{E}_a$ is the axion power, and $T$ is the periodicity of the discharge process, and  $\widetilde{\mathcal{S}}(\Vec{k})$ is the Fourier transform of the source:
\begin{equation} \label{eqn:stilde}
    \widetilde{\mathcal{S}}(\Vec{k}) = g_{a\gamma\gamma} \int d^4 x e^{ik\cdot x}(\Vec{E}\cdot\Vec{B})(x),
\end{equation}
where $k\cdot x = -\omega_k t + \Vec{k} \cdot \Vec{x}$, and $\omega_k^2 = \Vec{k}^2 + m_a^2$.
The time integral in (\ref{eqn:stilde}) is performed over the limit cycle of the pair cascade, and the spatial integral is performed over the volume of the polar cap, here approximated as a cylinder of radius $\rpc = \rns \sqrt{\Omega\rns}$, where $\Omega = 2\pi/P$ is the rotation frequency of the pulsar, and height~\cite{RudermanSutherland} 

\begin{align}
    h = 55 \ {\rm m} \left({P \over {\rm sec}}\right)^{3/7} \left( {B_0 \over 10^{12} \ {\rm G}}\right)^{-4/7} \theta^{-2/7}
    \label{eq: gap_height} \,.
\end{align}

The height of the gap is set by the voltage threshold for pair production. 
The maximum electric field in the gap is $\epar \approx 2 \Omega h B_0$. As the gap height increases, so too does the potential drop, which leads to particles being accelerated to very high energies. 
Since these particles move along curved field lines, they emit high-energy curvature radiation photons, which initiate a cascade of pair production through the process $\gamma_{\rm CR} + B \to e^+ + e^-$, where $\gamma_{\rm CR}$ is a curvature radiation photon, and $B$ is the background magnetic field (see left inset of Fig.~\ref{fig:cartoon}).
The height quoted in \eqref{eq: gap_height} corresponds to the mean free path of the initially sourced curvature radiation photon. Since the field line radius of curvature increases towards the pole, 
the emission of curvature radiation is suppressed, leading to larger gap heights, as shown in (\ref{eq: gap_height}). 
The spectrum of axions produced by the gap is determined by a combination of the axion mass, $m_a$, and the spectrum of oscillations of $\edb(x)$.
For low-mass axions with $m_a h \ll 1$, the axion field is coherent over the gap, and axion production is dominated by variations of $\edb$ on the scale of the gap. 
In the high-mass limit, $m_a h \gg 1$, axions are produced most efficiently by the small-scale plasma oscillations within the gap.

%%%%%%%%%%%%%%%%%%%%%%%%%%%%%%%%%%%%%%%%%%

Equation (\ref{eqn:dndotdk}) can be rewritten as

\begin{align} \label{eqn:dedotdk2}
    {d \dot{E}_a \over d^3 k} = {1 \over 4\pi^3} \gagg^2 B_0^4 \Omega^2 \rpc^4 h^4 {t_{\rm open}^2 \over T} |\tilde{g}(\vec k )|^2 ,
\end{align}
where 

\begin{align}
    \tilde{g}(\vec{k}) = {1 \over \pi }\displaystyle\int d^4 \tilde{x} \ e^{- i \omega_k t_{\rm open} \tilde{t} + i (k_x \rpc) \tilde x + i (k_y \rpc) \tilde y + i (k_z h) \tilde z } \ {\edb(\tilde x) \over (\edb)_0},
\end{align}
where $\omega_k = \sqrt{k^2 + m_a^2}$, $\tilde x = x/\rpc$, $\tilde y = y/\rpc$, $\tilde z = z/h$, $\tilde t = t/t_{\rm open}$, $t_{\rm open}$ is the typical timescale over which the gap is open, and $(\edb)_0 = 2 \Omega h B_0^2$ is the maximum value of $\edb$ in the gap. The gap opening timescale is related to the light crossing time of the gap, $t_{\rm open} = h$. The duty cycle of the gap opening, $t_{\rm open}/T$ is taken to be $10\%$ as in~\cite{Caputo:2023cpv}. 
 Equation~\eqref{eqn:dedotdk2}  can be rewritten as 

\begin{align}
{d \dot{E}_a \over d^3 k } = {1 \over 40 \pi} \gagg^2 B_0^4 \Omega^4 \rns^6 h^5 |\tilde{g}(\vec k )|^2.
\end{align}

The function $\tilde{g}(\vec{k})$ depends on both the spectrum of spatial fluctuations in $\edb$ and the axion mass.
The former has two major components: the first coming from the geometry of the gap, and the second from small-scale plasma oscillations that arise due to non-stationary pair creation and dynamical screening in the gap~\cite{Caputo:2023cpv}.
For light axions, with wavelengths larger than the characteristic gap size, $h$, the geometric contribution to $\tilde{g}(\vec k)$ dominates. In this low-momentum limit, $|\tilde{g}(\vec k)| \to 1$. 
For heavy axions, with wavelengths much shorter than the typical gap height, axion production depends on details of the plasma oscillation spectrum, understanding which requires kinetic particle-in-cell (PIC) simulations. Semi-analytic modeling of the plasma structure during the discharge phase shows a frequency spectrum $\epar (\omega) \approx E_{\parallel, 0} (\omega/\omega_0)^{-1/2}$, where $\omega_0$ is an IR scale, taken to be $\omega_0 \sim 1/h$. Assuming spatial fluctuations have the same spectrum\footnote{See~\cite{Caputo:2023cpv} for a plausibility argument for this assumption and a statement of the requirement of simulations for definitive justification.},~\cite{Caputo:2023cpv} demonstrated that in the small-$k$ limit, relevant for gravitationally bound axions, the effect of small-scale oscillations leads to a suppression of $\sim (m_a h)^{-4}$ in the axion luminosity\footnote{The expression quoted in~\cite{Caputo:2023cpv} is $(m_a h)^{-3} (\omega_0/m_a)$, but we assume $\omega_0 \sim 1/h$.}.
Putting together the high-mass and low-mass regimes, we obtain

\begin{align} \label{eqn:dEadotdktot}
    {d \dot{E}_a \over d^3 {\bf k}} &\approx {\gagg^2 \over 40\pi} B_0^4 \Omega^4 h^5 \rns^6 \min\left(1, (m_a h)^{-4} \right) \Theta(k_{\rm esc} - k),
\end{align}
where $\Theta(x)$ is the Heaviside theta function, which selects axions with velocity less than the escape velocity of the neutron star, $v_{\rm esc} = \sqrt{2 G M_{\rm NS}/\rns} \approx 0.5 c$. A simplifying assumption made in~\eqref{eqn:dEadotdktot} is that the differential axion production rate is approximately flat at small $k$. The distribution of axion velocities near the NS surface was studied using ray-tracing simulations in~\cite{Caputo:2023cpv}, and is approximately flat at low-$k$, before dropping at about $v \approx 0.9 \vesc$. Such simulations are required to make more robust predictions about axion cloud formation and dissipation. The total luminosity of axions emitted into bound states is then

\begin{align} \label{eqn:Eadot}
    \dot{E}_a = {1 \over 30} \gagg^2 m_a^3 B_0^4 \Omega^4 h^5 \rns^6 \vesc^3 \min\left(1, (m_a h)^{-4} \right).
\end{align}
Equation~\eqref{eqn:Eadot} agrees with the result in~\cite{Caputo:2023cpv} up to an order unity factor that arises from the approximation $d^3 k \sim 4\pi \bar{k}^3/3 \sim 4\pi m_a^3/81$, where $\bar{k} \sim m_a/3$ is the average cloud axion momentum. 

As low-velocity ($v < \vesc$) axions are produced, they accumulate around the NS to form a dense axion cloud. 
Early in the lifetime of a NS, the injection rate of axions into the cloud dominates dissipative processes, leading to unimpeded growth of the cloud. 
Axions with velocity sufficiently close to the escape velocity have orbits that extend far from the NS, allowing them to cross the resonant conversion surface, where the axion mass coincides with the local plasma frequency. 
In contrast to Eq.~\eqref{eqn:conversion_radius}, we assume here a spherical conversion surface with radius 

\begin{align} \label{eqn:conversion_radius_spherical}
    r_c = \rns \left( {2 e \Omega B_0 \over m_a^2 m_e} \right)^{1/3} \approx 46 \ {\rm km} \left({B_0 \over 10^{12} \ {\rm G}} \right)^{1/3} \left({\Omega \over 2\pi \times \ {\rm Hz}} \right)^{1/3} \left({m_a \over \mu{\rm eV}} \right)^{-2/3}.
\end{align}

The choice of a spherically symmetric conversion surface is adopted to avoid the regions of artificially low plasma frequency predicted near null surfaces ($\vec \Omega \cdot \vec B = 0$) in the GJ model. Axions passing through the resonant conversion surface convert to radio photons with probability given in Eq.~\eqref{eqn:conversion_probability}.
In the case of axions produced in polar caps, the axion velocity at the conversion surface is

\begin{align} \label{eqn:vc}
    v_c = \sqrt{v_0^2 - 2 G M_{\rm NS} \left( {1 \over \rns} - {1 \over r_c} \right)},
\end{align}
where $v_0$ is the initial velocity of the axion launched from the NS surface. 
Therefore, only bound state axions with velocity $v_{\rm esc}(1 - \rns/r_c)^{1/2} \lesssim v_0 \le v_{\rm esc}$ contribute to radio emission from axion clouds. This kinematic requirement imposes a characteristic energy (frequency) spread of the resulting radio signal. For non-relativistic axions, the characteristic bandwidth is set by $\mathcal{B} = (m_a/2\pi) \Delta (v^2)/2$, where $\Delta(v^2)$ is the spread in $v^2$. Using~\eqref{eqn:vc} and recalling that the central frequency is $f \approx m_a/2\pi$, the fractional width of the produced radio signal is

\begin{align} \label{eq:band_cloud}
    {\mathcal{B} \over f} = 3 \times 10^{-2} \left({B_0 \over 10^{12} \ {\rm G}} \right)^{-1/3} \left({\Omega \over 2\pi \times \ {\rm Hz}} \right)^{-1/3} \left({m_a \over \mu{\rm eV}} \right)^{2/3}.
\end{align}

Bound state axions execute elliptical orbits 
around the neutron star and encounter the resonant conversion surface many times, yielding an ``effective'' conversion probability of $P_{a\to\gamma} N_{\cross}$, where $P_{a\to\gamma}$ is the conversion probability for a single crossing, which is given by (\ref{eqn:conversion_probability}) 
and $N_{\cross}$ is the total number of times the axion crosses the conversion surface. 
At early times, when $N_{\cross} \ll 1/P_{a\to\gamma}$, the axion production rate dominates the dissipation rate and the radio signal grows linearly with time. However, when $ N_{\cross} \sim 1/P_{a\to\gamma}$, axion production
and dissipation to radio emission achieve equilibrium and the effective conversion probability is unity, leading to a dramatic enhancement in the radio flux from the axion cloud. 
Taking into account the dynamics of the system, we find that the growth of the radio signal can be characterized by the effective conversion probability as a function of time, given the dependence on both the axion production rate and the dissipation rate at early and late times. 
A general orbit whose apoapsis is greater than $r_c$ can cross the conversion surface one, two, or four times per orbital period, depending on the eccentricity of the orbit.
For simplicity, we set the time between resonant crossings to be equal to half the average orbital period, $P_{\text{orbit}}$ , of axions that reach the conversion surface, since orbits with a single crossing per period form a negligible  
subset of phase space.
The average time between resonant crossings, $\left\langle \delta \tau_{\cross} \right\rangle$, is then 

\begin{align}
    \left\langle \delta \tau_{\cross} \right\rangle =\frac{P_{\text{orbit}}}{2} = {2\pi G \mns \sqrt{2} \over (\vesc^2 - v^2)^{3/2}}.
\end{align}

The saturation time for the axion cloud is defined as $\tau_{\rm sat} \equiv \left\langle \delta \tau_{\cross} \right\rangle/P_{a \to \gamma}$. For young pulsars with $t_{\rm pulsar} \ll \tau_{\rm sat}$ (where $t_{\rm pulsar}$ is the age of the pulsar), the cloud is still growing, and the effective axion-photon conversion probability is $\propto \gagg^2$, while for $t_{\rm pulsar} \gg \tau_{\rm sat}$ axion production and conversion are in equilibrium and the effective conversion probability is of order unity. The radio luminosity from the axion cloud for a given NS
is then given by

\begin{align} \label{eqn:Edotgamma}
    \dot{E}_\gamma &\approx \displaystyle\int d^3 {\bf v}{d \dot{E}_a \over d^3 {\bf v}} \min \left(1, {P_{a \to \gamma}(\vec v) t_{\rm pulsar} \over \left\langle \delta \tau_{\cross}(\vec{v})\right\rangle} \right) \nonumber \\
    &= \dot{E}_a \times \left[ 1 - \left(1 - {\rns \over r_c} \right)^{3/2} \right] \times \min \left(1, {\frac{g_{a\gamma\gamma}^2 B_0^2 t_\text{pulsar}}{15 \sqrt{2} \  m_a}} g\left({r_c \over \rns}\right) \right)
\end{align}
 where the integral is performed over the initial axion velocity $\vec{v}$, with $v_{\rm esc}(1 - \rns/r_c)^{1/2} \lesssim |{\bf v}| \le v_{\rm esc}$, $\dot{E}_a$ is given in Eq.~\eqref{eqn:Eadot}, and

\begin{align}
    g(x) &= \frac{\sqrt{x - 1} \left[ (8 - x (3 + 2 x)) E\left(\frac{1}{1 - x}\right) + 2 x (2 + x) K\left(\frac{1}{1 - x}\right) \right]}{\left(1 - \left(1 - \frac{1}{x}\right)^{3/2}\right) x^{15/2}},
\end{align}
and $K(x)$ and $E(x)$ are the complete elliptic integrals of the first and second kind, respectively. The second term in (\ref{eqn:Edotgamma}) is a phase space suppression associated with the kinematic requirement that gap-produced axions must reach the resonant conversion surface. 

The photon flux density from the axion cloud of an individual NS, measured on Earth, is:
\begin{equation}
    \begin{split}
        F_\mathrm{NS}=&\frac{1}{d^2 \mathcal{B}}\frac{\dot{E}_\gamma}{4\pi} \\
        \approx \, & 1.6\times 10^{-9}\,\mathrm{Jy}\\
        &\times
        \begin{dcases}
        &{\left(\frac{m_a}{5\,\mu\mathrm{eV}}\right)}^{-2} {\left(\frac{g_{a\gamma\gamma}}{10^{-15}\,\mathrm{GeV}^{-1}}\right)}^{4}{\left(\frac{d}{1.9\, \mathrm{kpc}}\right)}^{-2} {\left(\frac{B_0}{10^{12}\,\mathrm{G}}\right)}^{38/7} {\left(\frac{P}{33\,\mathrm{ms}}\right)}^{-25/7} \, , \quad  \text{Growing},\\
        & {\left(\frac{m_a}{5\,\mu\mathrm{eV}}\right)}^{-1} {\left(\frac{g_{a\gamma\gamma}}{10^{-15}\,\mathrm{GeV}^{-1}}\right)}^{2}{\left(\frac{d}{1.9\, \mathrm{kpc}}\right)}^{-2} {\left(\frac{B_0}{10^{12}\,\mathrm{G}}\right)}^{24/7} {\left(\frac{P}{33\,\mathrm{ms}}\right)}^{-25/7}\, , \quad   \text{Saturated}.
        \end{dcases}
    \end{split}
\end{equation}
The axion cloud signal appears as a broadened spectral line centered at frequency corresponding to the axion mass, with natural bandwidth set by the axion velocity distribution (see Eq.~\eqref{eq:band_cloud}). Several effects can broaden this signal, similar to the case of radio emission from axion DM, but these are generally subdominant to the natural width. For instance, asphericity of the conversion surface produces both Doppler and gravitational broadening. The Doppler effect arises from co-rotation of the magnetosphere: at different radii, the plasma’s linear velocity projected along the line of sight varies. The typical width scales as $(\Delta f / f)_{\rm Doppler} \sim \Omega,\Delta r$, where $\Delta r$ is the difference between the minimum and maximum radii of the conversion surface. Since conversion generally does not occur for $r > 10,R{\rm NS}$, the maximal Doppler contribution is $(\Delta f / f)_{\rm Doppler} \sim 2\times 10^{-3}$ for a pulsar with rotation period $P = 1$ s—well below the natural width. Gravitational redshift adds an additional broadening of order $(\Delta f / f)_{\rm grav} \sim \sqrt{1 - 2 G M_{\rm NS}/r_{\rm max}} - \sqrt{1 - 2 G M_{\rm NS}/r_{\rm min}}$. This is typically at the sub-percent level for order-unity deviations from sphericity, but can approach $\sim 10 \%$ for highly aspherical surfaces that extend very close to the neutron star surface. A more detailed treatment of these effects is left to future work; here we proceed with the estimate in~\eqref{eq:band_cloud}. For generality, the scaling equation above has assumed a constant signal bandwidth $\mathcal{B}\sim 60\,\text{MHz}$,
suitable for both a population survey and a fixed target observation; the latter case could be made more precise by incorporating the full parametric dependence as given in Eq.~\eqref{eq:band_cloud}. Note that in the DSA-2000 frequency range $ 0.7\, \mathrm{GHz}\lesssim f \lesssim 2\,\mathrm{GHz}$ (corresponding axion masses $2.9\,\mu\mathrm{eV} \lesssim m_a\lesssim 8.3\,\mu\mathrm{eV}$), the axion wavelength is almost always less than the gap height $h$ estimated by Eq.~\eqref{eq: gap_height}, such that taking $\min{(1, {(m_a h)}^{-4})} \approx {(m_a h)}^{-4}$ is justified in the estimate of $\dot{E}_a$ \footnote{In this regime, an additional radio line is sourced due to the oscillating current associated with the axion field, which may have interesting observational consequences~\cite{Caputo:2023cpv}}. The magnitude of the signal is primarily determined by the axion-photon coupling strength, the NS surface magnetic field, its distance to Earth, and its rotation period. Optimal candidates are nearby, fast-spinning pulsars with strong magnetic fields. 

\subsection*{Target identification}

The preceding section focuses  on radio signals from individual NSs. Another approach to axion detection involves searching for signals from populations of NSs~\cite{Safdi:2018oeu, Foster:2022fxn, Bhura:2024jjt}. An advantage of population signals is that the total received flux is bounded from below by the flux from a single NS. However, in the case of axion DM, the signal is Doppler broadened to $\mathcal{B}/f \sim v_{\rm gal}/c \approx 7 \times 10^{-4}$, where $v_{\rm gal}$ is the galactic velocity dispersion. Therefore, the expected signal is much broader than that of an individual NS's $\mathcal{B}/f \sim v_0 \approx 10^{-6}$\footnote{Even signals from individual NSs are expected to have bandwidth greater than $\sim 10^{-6}$ due to the NS's rotation (see, e.g.,~\cite{Witte:2021arp}), but still less than the Doppler-broadened population signal.}. Here, we consider radio signals from a population of NSs surrounded by axion clouds. The expected bandwidth for axion cloud signals in~\eqref{eq:band_cloud} dominates over Doppler broadening. Therefore, we can estimate the total flux density received from a population of NSs by simply summing the luminosity density of each individual NS:
\begin{equation}\label{eq:flux_pop}
    F_\mathrm{pop} = \sum_i F_\mathrm{NS,\, i} = \frac{1}{\mathcal{B}}\sum_i \frac{\dot{E}_{\gamma, \, i}}{4\pi {d_i}^2}
\end{equation}

We consider two observation scenarios, including a fixed target search for individual NS sources and an all-sky survey for a population of NSs, which is already part of the DSA-2000 observing strategy. 
For the single target search, we focus on the Crab pulsar, which is one of the most energetic pulsars within our current pulsar catalog. We also include a predicted target from our population model (elaborated on below), which represents new targets expected to be discovered by the DSA-2000 telescope. The parameters of the Crab pulsar and the best predicted target are shown in Table~\ref{tab:crab}. 

\begin{table}[h]
\centering
\begin{adjustbox}{max width=1.1\textwidth,center}
\begin{tabular}{c|c|c|c|c}
\hline
\hline
\textbf{Target} & \textbf{$\mathbf{P}$ [s]} & \textbf{$\mathbf{B_{0}}$ [G]} & \textbf{$\mathbf{d}$ [kpc]} &  \textbf{$\mathbf{\mathcal{B}}$ [MHz]} \\
\hline
\hline
Crab pulsar (PSR B0531+21) & $0.033$ &  $3.8\times 10^{12}$ & $1.9$ & 21.8 \\
\hline
Best predicted target
& $0.098$ &  $3.2\times 10^{13}$ & $3.0$ & 17.5 \\
\hline
\hline
\end{tabular}
\end{adjustbox}
\caption{Parameters used for axion cloud signal estimate for a fixed target search. We include the Crab pulsar (PSR B0531+21), as well as a potentially better target predicted by our population model (see text for details). The last column shows the signal bandwidth (i.e., optimal resolution for a single target observation), taking a benchmark value at $m_a=5\,\mu\text{eV}$. 
}
\label{tab:crab}
\end{table}
 
For the all-sky survey, we consider the population documented by the ATNF pulsar catalog that are within the field of view of DSA-2000 (DEC $\gtrsim -30\degree$). We neglect the pulsars whose conversion radius is smaller than the NS radius ($r_c<R_\text{NS}$) for any mass in the DSA-2000 range ($2.9\,\mu\text{eV}\lesssim m_a \lesssim 8.3\,\mu\text{eV}$). In addition to NSs in the ATNF pulsar catalog, 
the DSA-2000 is expected to discover $\approx 20,000$ new pulsars 
in its wide-field pulsar search. 
During the all-sky survey the DSA-2000 will identify new radio NSs per sky patch, and will follow up on the identified objects.
With 16 epochs planned for the cadenced all-sky survey, the total observation time amounts to roughly 4 hours per radio source. 
Based on the large number of NSs observed, we expect the data collected in the all-sky survey to be the most sensitive probe to date of the hypothesized axion-cloud radio signal, comparable to a $10$-hr fixed-target observation of the Crab pulsar. Moreover, the all-sky survey may discover more distant pulsars, potentially with stronger magnetic fields, which could serve as even better targets for detecting radio signals from axion clouds. 
Since the characteristic signal bandwidth $\mathcal{B}$ from axion clouds is broad enough to be resolvable by the all sky-survey, we forecast the sensitivity DSA-2000 by utilizing population modeling to estimate the properties of the yet to be discovered NSs. One would search for the $\mathcal{O}(50\,\rm MHz)$ after stacking frequency spectra on the positions of known pulsars.

\subsection*{Population model}

\begin{figure}[h]
    \centering
    \includegraphics[width=\textwidth]{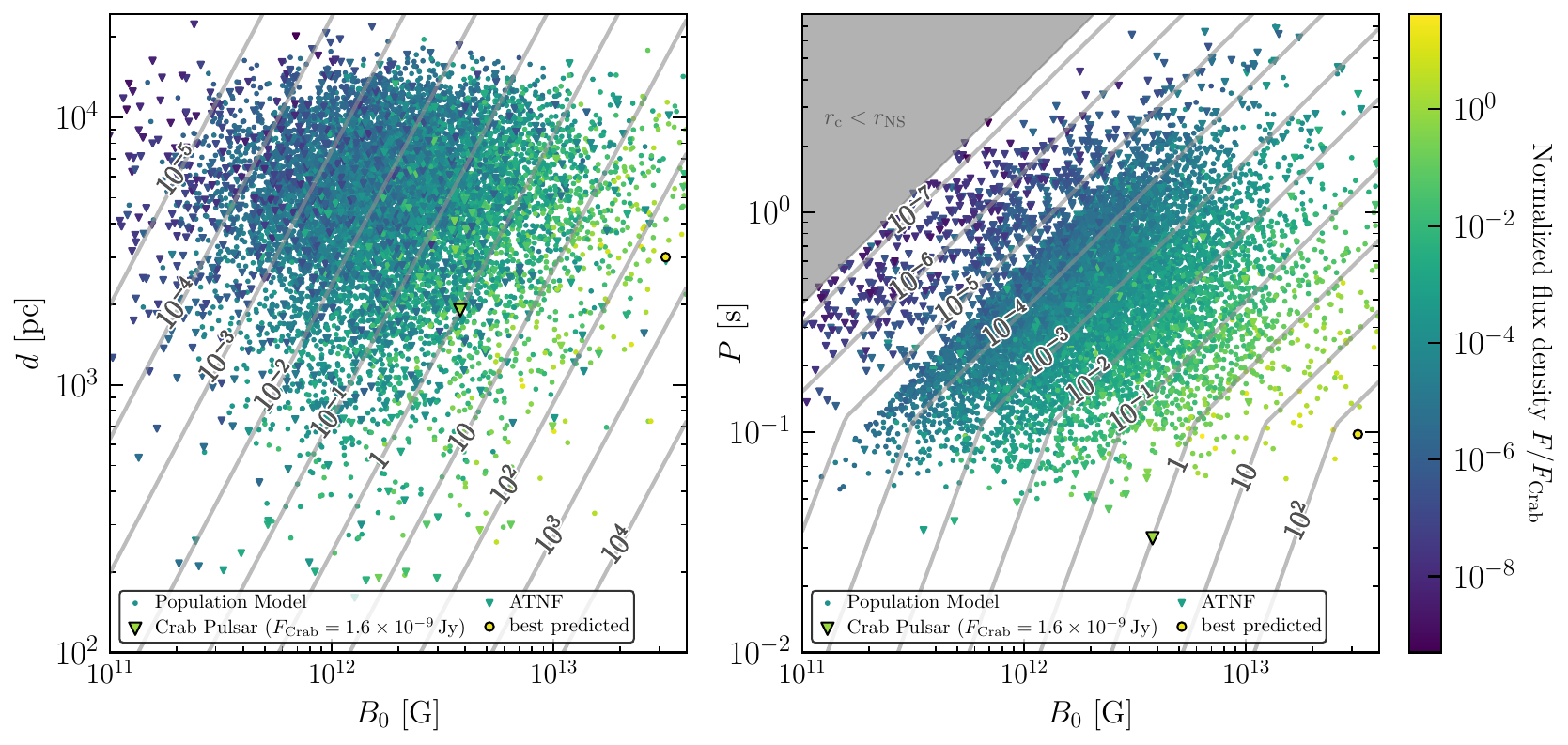}
    \caption{Distribution plots of the population model (dots), the ATNF catalog (triangles), the Crab pulsar (circled triangle), and the best predicted pulsar to be discovered by the DSA-2000 (circled dot). The color bar shows the axion cloud signal flux density from each individual source, normalized to the flux density from the Crab pulsar. The grey lines show the contours of the axion cloud signal flux density in the corresponding parameter space, normalized to the axion cloud signal flux density from the Crab pulsar. The contour lines are calculated with the varying parameters shown in the plots ($B_0, d$ for the left panel and $B_0, P$ for the right panel) and the rest of the parameters fixed to be identical to that of the Crab pulsar. We have taken benchmark values of $m_a=5\, \mu\text{eV}$ and $g_{a\gamma\gamma}=10^{-15}\, \text{GeV}^{-1}$ in our estimation of the flux, and the Crab pulsar signal flux density is $F_\text{Crab}=1.6 \times10^{-9}\, \text{Jy}$. The left panel shows the distribution in surface magnetic field ($B_0$ [G]) and distance to Earth ($d$ [pc]). The distribution plot is bounded from above by the size of the Milky Way, at $d\sim 1.5\times 10^4$ pc. The right panel shows the distribution in surface magnetic field ($B_0$ [G]) and rotation period ($P$ [s]). Equivalently, one can obtain the distribution on a $P\dot{P}$ diagram using the relation in Eq.~\eqref{eqn:magnetic_field_P_Pdot_relation}.  The gray region is where the conversion radius $r_c$ falls inside the NS. We have discarded NSs older than $10^7$ yrs in the population model due to their deviations in the magnetosphere from the GJ model. 
    This age cut keeps the population (dots) away from the region with high period $P$ and weak magnetic field $B_0$.}
    \label{fig:population}
\end{figure}

For our population modeling, we rely on a pulsar population snapshot generated from the \texttt{PrsPopPy} simulation \cite{Bates:2013uma}, which provides NS parameters, including their Period $P$ and distance $d$, and only includes objects discoverable with DSA-2000. 
We generate their present-day surface magnetic field, $B_0\,$[G], by drawing from a log-normal distribution centered at $11.98$ and with a standard deviation of $0.52$.  
These parameters are consistent with the \texttt{PsrPopPy} simulation \cite{Bates:2013uma}, as well as the ATNF catalog \cite{Manchester:2004bp}. 
Since the magnetosphere of old pulsars is not well-described by the GJ model~\cite{Electrosphere}, we discard the NSs with age  $t_{\text{NS}} \geq 10^7\, \text{yrs}$. 
We also do not include  millisecond pulsars with rotation periods $ P \lesssim 0.05\, \text{s}$. After imposing the above additional criteria on the population model for new detections, we are left with $\approx 9000$ pulsars that satisfy those criteria, which we use for the axion cloud forecast. In Fig.~\ref{fig:population}, we show the distributions of the expected detection population (dot) along with the ATNF catalog (triangle). The axion cloud signal flux density from these sources, normalized to the Crab pulsar (circled triangle), is shown in the color bar.
The left panel shows the distribution in surface magnetic field, $B_0$ and distance $d$, where we have cut off NSs outside of the Milky Way ($d\gtrsim1.5\times 10^4\,$pc). The right panel shows the distribution in surface magnetic field and pulse period $P$\footnote{Equivalently, one can obtain the distribution on a $P\dot{P}$ diagram using the relation in Eq.~\eqref{eqn:magnetic_field_P_Pdot_relation}. Here we show a $PB_0$ diagram instead to illustrate the effect of surface magnetic field strength, a dominant factor affecting the signal flux density.}, where we cut off the region where the conversion surface falls inside the NS surface ($r_c<R_\mathrm{NS}$), as well as NSs older than $10^7\,$yrs. 
We also plot contours of the signal flux density $F_\mathrm{NS}$ in the corresponding parameter space with benchmark parameters $m_a=5\,\mu\mathrm{eV}$ and $g_{a\gamma\gamma}=10^{-15}\,\mathrm{GeV}^{-1}$ (while fixing the other parameters to be identical to the Crab), normalized against the Crab pulsar ($F_\mathrm{Crab}\approx 1.6 \times 10^{-9}\, \mathrm{Jy}$). We have chosen the above described cuts on distance, period, and magnetic field for the NSs included in our forecast to be conservative. However, once DSA-2000 discovers new pulsars, the assumptions on the new population made here can be updated and validated before using the collected data to set a constraint.

We also show the best predicted target from the population model (circled dot, parameters shown in Table~\ref{tab:crab}), which represents a prospective candidate that could be discovered by the DSA-2000, whose axion cloud signal flux density is higher than that of the Crab pulsar but has not been observed before. The best predicted target from the population model has a stronger magnetic field, which enables a stronger axion cloud signal than the Crab pulsar. However, due to its larger distance from the Earth, the new candidate produces a smaller non-axion radio flux density than the Crab pulsar \cite{Tolman_2022}. We expect data from the DSA-2000 to discover targets farther from the Earth that could be potential better candidates for observing axion signals.  We note that the parameters shown in Table~\ref{tab:crab} for the best predicted target, informed by our population modeling, are speculative. The data to be taken with the DSA-2000 will ultimately unveil the best discovered target, whose measured parameters will
determine the sensitivity to the QCD axion band that can be reached with targeted follow-up observations.

\begin{figure}
	\includegraphics[width=1\textwidth]{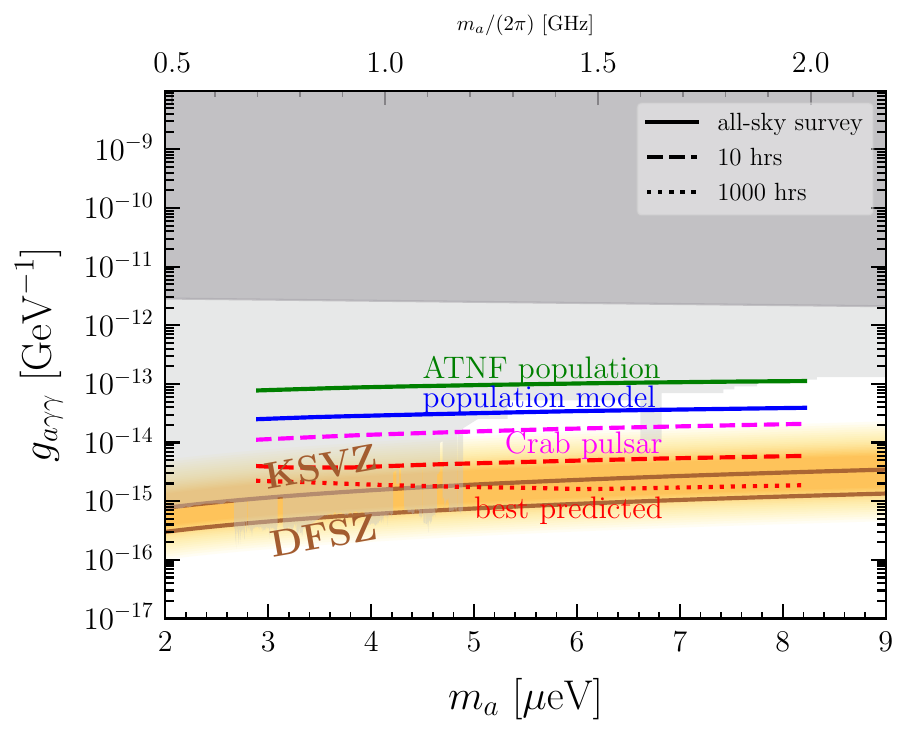} 
	\caption{Projected QCD axion discovery reach (SNR $\geq 5$) with the DSA-2000 from radio emission of axion clouds around NSs. We make projections of the Crab pulsar (dashed magenta, $\Delta t_{\text{obs}} = 10$ hours), the ATNF population (solid green), and the population model forecasting the DSA-2000 detection (solid blue), based on the planned all sky survey which will observe each NS for 4 hours. We also include projections of a predicted candidate that could potentially be discovered by the DSA-2000 (see Sec.~\ref{subsec:axion_clouds} for more discussions), with a $\Delta t_{\text{obs}} = 10$ hours (dashed red) and a $\Delta t_{\text{obs}} = 1000$ hours (dotted red) observation. Other relevant astrophysical constrains are plotted in darker gray (dominated by unbounded polar cap axions from NS~\cite{Noordhuis:2022ljw}), and relevant haloscope constraints (ADMX~\cite{Asztalos_2010, ADMX:2018gho, ADMX:2018ogs, ADMX:2019uok, ADMX:2021mio, ADMX:2021nhd, ADMX:2024xbv}, CAPP~\cite{Lee:2020cfj, Jeong:2020cwz, CAPP:2020utb, Lee:2022mnc, Yoon:2022gzp, Kim:2022hmg, Yi:2022fmn, Yang:2023yry, Kim:2023vpo, CAPP:2024dtx}, RBF+UF~\cite{PhysRevLett.59.839,Wuensch:1989sa, PhysRevD.42.1297, Hagmann:1996qd}) on axion DM are plotted in lighter gray. The QCD axion parameter space is shaded in yellow~\cite{Saikawa:2024bta}. }\label{fig:reach_bsa}
\end{figure}

\subsection*{Forecast}

The SNR for a fixed target search focusing on an identified NS is:
\begin{equation}
    \mathrm{SNR}_\text{NS}=\frac{F_\mathrm{NS}}{\sigma(\Delta t_{\mathrm{obs}},\mathcal{B}(m_a, B_0, P))} \, ,
\end{equation}
where the radio flux sensitivity limit $\sigma$ can be obtained from Eq.~\eqref{eqn:SEFD}. The bandwidth $\mathcal{B}(m_a, B_0, P)$ for a fixed target search is calculated from Eq.~\eqref{eq:band_cloud} with the corresponding NS parameters. The signal bandwidths for our two sample targets are shown in Table~\ref{tab:axion_DM_targets}. For a single target search, we take $\Delta t_\text{obs}=10\,\text{hrs}$, and an optimistic $\Delta t_\text{obs}=1000\,\text{hrs}$ for any particularly promising target.

The SNR for a multi-target observation from the all-sky survey is:
\begin{equation}
    \mathrm{SNR}_\text{pop}=\frac{F_\mathrm{pop}}{\sqrt{N}\sigma(\Delta t_{\mathrm{obs}},\mathcal{B}(m_a))} \, ,
\end{equation}
where the radio flux sensitivity limit $\sigma$ can be obtained from Eq.~\eqref{eqn:SEFD}, and $N$ is the number of targets.
For simplicity, we take a fixed bandwidth $\mathcal{B}(m_a)$ for observation at each frequency to be $\mathcal{B}(m_a)=0.05\, (m_a/2\pi)$, which approximates Eq.~\eqref{eq:band_cloud} and represents a larger signal bandwidth over the range of pulsar parameters. In the DSA-2000 range, we expect the signal bandwidth to be $\mathcal{B}\sim 35-100\,$MHz.
For the population search, we take $16$ of $15\,\mathrm{min}$ observations, amounting to a total of $\Delta t_\text{obs}=4\,\text{hrs}$ for each target.

Fig.~\ref{fig:reach_bsa} shows the projected reach for DSA-2000 observation of the Crab pulsar (dashed magenta), ATNF catalog (solid green), and the new population model (solid blue), requiring an $\text{SNR}$ of $5$. We also include the projected reach for a $10$-hr observation of the best target predicted by our population model to be discovered by the DSA-2000 (dashed red), as well as an optimistic $1000$-hr observation of the same target (dotted red). 
Existing constraints in this mass range are shown gray. 
Leading astrophysical constraints on non-DM axion (darker gray) include the radio signal from polar cap axion conversion to photon exiting the NS~\cite{Noordhuis:2022ljw}. Leading terrestrial constraints on axion DM (lighter gray) includes ADMX~\cite{Asztalos_2010, ADMX:2018gho, ADMX:2018ogs, ADMX:2019uok, ADMX:2021mio, ADMX:2021nhd, ADMX:2024xbv}, CAPP~\cite{Lee:2020cfj, Jeong:2020cwz, CAPP:2020utb, Lee:2022mnc, Yoon:2022gzp, Kim:2022hmg, Yi:2022fmn, Yang:2023yry, Kim:2023vpo, CAPP:2024dtx}, and RBF+UF~\cite{PhysRevLett.59.839,Wuensch:1989sa, PhysRevD.42.1297, Hagmann:1996qd}. Our projections indicate that DSA-2000 will reach a sensitivity of order $g_{a\gamma\gamma} \sim 10^{-14}\, \text{GeV}^{-1}$, which will set the most competitive non-DM axion constraints in the mass range $2.9\,\mu\mathrm{eV}\lesssim m_a\lesssim 8.3\,\mu\mathrm{eV}$, including the uncovered parameter space in the mass range $5\,\mu\mathrm{eV}\lesssim m_a\lesssim 8\,\mu\mathrm{eV}$ in the axion DM landscape. This can be achieved with either a $10$-hr fixed-target search pointing at the Crab pulsar, or data from the planned all-sky survey assuming discovery of the expected new pulsars. Independent of the discovery of new pulsars, data from the planned all-sky survey will allow us to set the most competitive non-DM axion bound at $\sim 10^{-13}\, \text{GeV}^{-1}$ in the mass range $3\,\mu\mathrm{eV}\lesssim m_a\lesssim 8\,\mu\mathrm{eV}$, using the ATNF catalog. If DSA-2000 is able to discover a pulsar with parameters comparable to the best predicted target (see Table~\ref{tab:crab}), a $10$-hr dedicated observation would allow us to reach $g_{a\gamma\gamma} \approx 5 \times 10^{-15}\, \text{GeV}^{-1}$, and a $1000$-hr dedicated observation would probe the KSVZ axion hypothesis.

\section{Radio signals from dark photons}
\label{sec:darkphoton}
Many of the observational signatures of axions are shared by another well-motivated DM candidate, the dark photon (sometimes referred to as the ``hidden photon'')~\cite{Jaeckel_2010, jaeckel2013forcestandardmodel, Fabbrichesi_2021, Caputo_2021}.
Indeed, resonance conversion into radio photons near NSs is also possible for dark photon DM
 \cite{Hardy:2022ufh}, in close analogy with the axion DM case discussed in Sec.~\ref{sec:axions}. 
 However, the limits one could obtain from such  radio signals in the DSA-2000 frequency band are not competitive.
 Here we focus on using the DSA-2000 to probe radio signals generated through dark photon superradiance \cite{Siemonsen:2022ivj, Mirasola:2025car}, with sensitivity to dark photon masses that range from $5 \times 10^{-14} - 10^{-12} \, \text{eV}$.

The dark photon, $\gamma'$, is a hypothetical vector  particle with mass $m_{\gamma'}$ that interacts with the Standard Model via kinetic mixing with ordinary photons. 
This is a well-motivated extension to the Standard Model that could, for example, arise from a new $U'(1)$ gauge group.
The dark photon itself is a viable DM candidate, but could also constitute a vector portal to a dark BSM sector which contain the stable DM.
It is extensively searched for using laboratory probes, for example, fifth force \cite{PhysRevLett.61.2285, Kroff:2020zhp} and light shining through wall experiments \cite{Ehret:2010mh, Tang:2023oid}, astrophysical probes \cite{Schwarz:2015lqa, Linden:2024uph}, as well as cosmological probes such as cosmic microwave background measurements (CMB)\cite{Mirizzi:2009iz, McDermott:2019lch, Caputo:2020bdy}. 
Its interaction with the SM is described by the Lagrangian
\begin{equation}
\mathcal{L_\gamma'} = - \frac{1}{4} F_{\mu \nu}F^{\mu \nu} + A_{\mu}J^{\mu} 
 - \frac{1}{4} F'_{\mu \nu} F'^{\mu \nu} +\frac{1}{2} m^2_{\gamma'} A'_{\mu}A'^{\mu}  - \frac{\epsilon}{2} F_{\mu \nu} F'^{\mu \nu} + A'_{\mu}J'^{\mu} , 
 \end{equation}
where $A^{\mu}$ denotes the SM photon, $F^{\mu \nu}$ is the electromagnetic field strength, $A'^{\mu}$ is the dark photon, $F'^{\mu \nu}$ is the dark photon field strength, $\epsilon$ is the kinetic mixing parameter, and $J^{\mu}$ ($J'^{\mu}$) is the electromagnetic (dark) current.
While a priori the mixing parameter $\epsilon$ could be order one, experimental probes constrain it to be much smaller.
Dark photons can oscillate during propagation in vacuum or matter from their sterile state to visible photons. 
This oscillation can be resonantly enhanced if the dark photon mass, $m_{\gamma'}$, coincides with the plasma mass of the visible photon, which scales with the electron number density. This process, analogous to resonant axion-photon conversion in plasma, sets the most stringent constraints on non-DM dark photons, $\epsilon \lesssim 5 \times 10^{-7} - 5 \times 10^{-8}$ on dark photons in the relevant mass range from $10^{-14} - 10^{-11} \, \text{eV}$ from measurements of spectral distortions \cite{Fixsen:1996nj, Caputo:2020bdy} and spectral anisotropies \cite{
Planck:2018nkj, unwise, McCarthy:2024ozh}
in the CMB.\footnote{We disregard the constraints from \cite{Cardoso:2018tly} who argue that black hole superradiance rules out dark massive vector particles in the  mass range from $10^{-13} - 3 \times 10^{-12}\, \text{eV}$ based on the observed stability of the black holes' inner disk. 
Non-gravitational interactions of the cloud perturb 
the disk dynamics as explored in \cite{Siemonsen:2022ivj} which invalidates an extrapolation of the result of \cite{Cardoso:2018tly} to the finite values of $\epsilon $ we consider.}

\subsection{Kinetically mixed dark photon  superradiance}

\subsubsection*{Formalism}
The phenomenon of Black Hole (BH) superradiance describes the formation of a bosonic cloud  around a spinning black hole \cite{Zeldovich:1971abc, PhysRevLett.28.994, Starobinsky:1973aij, PhysRevD.22.2323,Bekenstein:1998nt,Brito:2015oca}. The bosonic cloud is a coherent gravitationally-bound state of ultralight bosons, which forms if an ultralight boson exists whose Compton wavelength is of the order of the BH horizon size \cite{PhysRevD.81.123530,Arvanitaki:2010sy,Brito:2015oca}.
This phenomenon is independent of any initial abundance such that superradiance  occurs even if the boson does not comprise any sub-component of DM. 

The dark photon superradiance signatures described below rely on theoretical modeling that involves significant uncertainties, particularly in the plasma physics and energy dissipation mechanisms. The theoretical framework we utilize, developed in \cite{Siemonsen:2022ivj}, makes several assumptions about: cascade pair production, resistive magnetohydrodynamic plasma modeling adapted from pulsar studies, and turbulent energy dissipation processes, which require further validation in the extreme electromagnetic environments around black holes. Given these theoretical uncertainties, we focus on discovery-oriented searches rather than setting exclusion limits, as the absence of predicted signals could result from the breakdown of theoretical assumptions rather than the non-existence of dark photons.

For a dark photon, 
the superradiance mechanism sources a dark photon superradiance cloud around a spinning BH
\cite{Baryakhtar:2017ngi}, which if it possesses a large enough kinetic mixing with the SM photon, induces a rotating visible electromagnetic field, accompanied by a plasma of charged particles \cite{Siemonsen:2022ivj}. 
The number of dark photons in the cloud grows exponentially, following the birth of a BH with initial dimensionless spin $a_*$, mass $M$, and angular 
velocity\footnote{The angular velocity of a BH indicates how fast its event horizon is rotating.}\cite{Arvanitaki:2014wva, Siemonsen:2022ivj}  
\begin{equation}
\label{eq:angular_velocity}
\Omega_{\text{BH}}(a_*) =\frac{1}{2 MG} \left(\frac{a_*}{1+\sqrt{1-a^2_*}} \right) \,,
\end{equation}
until the superradiance condition $\Omega_{\rm BH} > \omega$ is saturated, i.e., 
\begin{equation}
\label{eq:sat}
\Omega_{\text{BH}}(a_{*,\text{fin}}) = \omega \, .
\end{equation}
Here $G$ denotes Newton's gravitational constant, and 
\cite{Siemonsen:2022ivj}
\begin{equation}
\label{eq:sr_condition}
\omega \simeq  m_{\gamma'} \left(1 - \frac{m^2_{\gamma'} M^2 G^2 }{2} \right) \, ,
\end{equation}
denotes the dark photon's energy at leading order in the gravitational coupling. Note that the dark photon's energy in the superradiance cloud is lowerered by being in a bound state with the BH, in close analogy to an electron cloud around a nucleus,
which is why superradiance is often referred to as a ``gravitational atom".

The conducting plasma within the superradiance cloud is seeded by initial synchrotron radiation from environmental electrons accelerated in the EM field of the superradiance cloud \cite{Siemonsen:2022ivj}.
They induce photon-assisted Schwinger pair production, efficiently populating an EM plasma for kinetic mixing parameters that exceed about \cite{Siemonsen:2022ivj} 
\begin{equation}
\epsilon \gg 10^{-10} \left(\frac{0.1}{m_{\gamma'} MG } \right)^\frac{5}{2} \left(\frac{0.1}{a_* - a_{*,\text{fin}}}\right)^{\frac{1}{2}} \left( \frac{10^{-12} \,  \text{eV}}{m_{\gamma'}}  \right)^{\frac{1}{2}}  \, .
\end{equation}

This astrophysical environment resembles a pulsar magnetosphere
with distinguishing ``smoking gun" features.
In particular, the electromagnetic luminosity can be as high as $\sim 10^{42} 
\, \text{erg/s}$ \cite{Siemonsen:2022ivj}, and is expected to have a continuous flux component in the radio frequency band of $\mathcal{O}(10^{-4})$ of the total luminosity \cite{Kaspi:2017fwg, Manchester:2004bp, Siemonsen:2022ivj}, resulting in a broadband radio flux ($\mathcal{B}\sim \text{GHz}$) that could be as large as
\begin{equation}
F_{\text{obs}} \lesssim 300 \times \left( \frac{500 \, \text{Mpc}}{d} \right)^2 \mu \text{Jy} \, ,
\end{equation}
where $d$ denotes the distance from the Earth to the superradiant BH. 
Additionally, the radio flux has a periodically enhanced component pulsing with a period, $P$, that at leading order is determined by the dark photon mass $P = 2\pi/\omega$. 
Due to gravitational wave emission that depletes the superradiance cloud upon formation, the signal is expected to be strongest when the superradiant cloud saturates, which occurs when the dark photon energy matches  that of the rotating black hole (i.e. Eq.~\eqref{eq:sat}), after which the luminosity decays.

Unlike normal pulsar signals, the time derivative of the pulsing frequency is positive, indicating a spin-up,
\cite{Siemonsen:2022ivj} 
\begin{equation}
\dot{\omega} \simeq \frac{5}{4} M G^2 m^3_{\gamma'} P_{GW},
\end{equation}
where $P_{GW}$ is the power emitted in gravitational waves. It is the decrease in the BH mass, $M$, over time due to gravitational wave emission that decreases the gravitational correction to the dark photon energy in Eq.~\eqref{eq:sat}, and results in the spinning up of the ``pulsar" signal.

Since the pulsing frequency at leading order is determined by the dark photon mass, identifying several objects with these characteristics clustered around the same frequency would be a smoking gun for detecting a kinetically mixed dark photon superradiance cloud, particularly if the signals can be spatially correlated with prior BH merger events. 
The characteristic dark photon masses that would lead to this observable superradiance phenomenon are around $m_{\gamma'} \sim 10^{-12} \, \text{eV}$ for stellar BHs \cite{Siemonsen:2022ivj}, which corresponds to dark photons whose Compton wavelengths are of the order
of the observed BH horizon sizes.

DSA--2000 can probe this dark photon signal through different search strategies. In particular, electromagnetic follow-ups of compact binary mergers, with characteristics that allow for a timely formation ($t_{\text{growth}} \lesssim \mathcal{O} \, (\text{years})$) of the superradiance cloud, are promising targets.  The growth timescale which indicates how long it takes after a BH merger for EM emission to start is of order \cite{Siemonsen:2022ivj}
\begin{equation}
\label{eq:tgrowth}
t_{\text{growth}} \sim 10^4 \, \text{s}\left(\frac{M}{10 \, M_{\odot}} \right) \left(\frac{0.7} {a_{*}} \right) \left( \frac{0.1}{G M m_{\gamma'}} \right)^7 \, ,
\end{equation}
corresponding to $\mathcal{O}(100)$ superradiance times, the timescale over which superradiant instabilities grow exponentially around a rotating BH.
Simulations have inferred the expected electromagnetic luminosity in terms of the BH parameters and the properties of the dark photon to scale as  \cite{Siemonsen:2022ivj}
\begin{equation}
\label{eq:LEM}
L_{\text{EM}}(t_{\text{growth}}) \simeq 4 \times 10^{41} \,\text{erg}/\text{s} \left(\frac{\epsilon}{10^{-7}} \right)^2 \left(\frac{G M m_{\gamma'}}{0.1} \right)^2  \left(\frac{a_{*}-a_{*,\text{fin}}}{0.1}\right)  , 
\end{equation}
where
\begin{equation}
a_{*,\text{fin}} \approx  \frac{4 G M m_{\gamma'}}{1+4 G^2 M^2 m^2_{\gamma'}} \, ,
\end{equation}
is the final spin that  saturates the BH superradiance condition in Eq.~\eqref{eq:sr_condition}.
The dark photon mass that can be probed via superradiance is limited from above as $a_*$ approaches $a_{*,\text{fin}}$ until eventually the BH's spin is too small, such that its angular velocity falls below the dark photon mass, and superradiance does not occur \cite{Siemonsen:2022ivj}. The dark photon mass that can be probed by following up after a BH merger event is limited from below by the growth timescale -- if the signal only appears after decades, the planned five year run-time of DSA-2000 is too short to capture it. In our forecasts featuring BH mergers from the last few years, we take this cutoff to be $t_{\text{growth}} < 8$ years, using the estimate in Eq.~\eqref{eq:tgrowth}.

After the growth of the superradiance cloud the electromagnetic signal persists, while the cloud depletes through gravitational wave and electromagnetic emission.
When more energy in released in gravitiational waves the decay of the luminosity proceeds via a power-law in time, while in the opposite scenario the decay is exponential, such that \cite{Siemonsen:2022ivj}
\begin{equation}
\label{eq:Ldecay}
\frac{L_{\text{EM}}(t_{\text{growth}}+ t)}{L_{\text{\text{EM}}}(t_{\text{growth}})} = 
\begin{cases}
\left(1 + t/\tau_{GW} \right)^{-1} &  \, \tau_{\text{GW}} \ll \tau_{EM} \, ,\\
e^{(-t/\tau_{\text{EM}}) \ln 2} &   \tau_{GW}  \gg \tau_{\text{EM}} \, ,
\end{cases}
\end{equation}
where the respective lifetimes scale as \cite{Siemonsen:2022ivj}
\begin{equation}
\tau_{\text{GW}} \approx 10^6 \, \text{s} \left(\frac{M}{10 M_{\odot}}\right) \left(\frac{0.1}{ G M m_{\gamma'}} \right)^{11} \left( \frac{0.1}{a_{*} -a_{*,\text{fin}}}  \right) \, ,
 \end{equation}
\begin{equation}
\tau_{\text{EM}} \approx 10^{11} \, \text{s} \left(\frac{M}{10 M_{\odot}}\right) \left(\frac{10^{-7}}{\epsilon} \right)^{2} \left( \frac{10^{-2}}{(0.13 (GM m_{\gamma'}) -0.19 (G M m_{\gamma'})^2 )}  \right) .
 \end{equation} 

The expected radio flux $F_{\text{obs}}$ is related to this total  electromagnetic luminosity by 
\begin{equation}
\label{eq:radioeff}
F_{\text{obs}} = \eta_{\text{radio}} \frac{L_{\text{EM}}}{4 \pi d^2 \mathcal{B} } \, .
\end{equation}
The decay constants exceed the observational timescale of 15 minute intervals we 
consider. Thus, we estimate the observable radio flux using Eq.~\eqref{eq:LEM} for the electromagnetic luminosity.

\subsection*{Target identification and forecast}

In Table~\ref{tab:superradiance} we have identified several promising BH-BH merger events and NS-BH merger events with good spatial localization within the DSA-2000 field of view
that could have lead to dark photon superradiance in the final BH \cite{LIGOScientific:2018mvr, KAGRA:2021vkt}. Many of these events have good spatial localization due to multi-messenger signals detected shortly after the merger event, which are particularly expected for NS-BH mergers
\cite{2019ApJ...887L..13D, 2021ApJ...923...66A, Dobie:2021khu}. The predicted signal from superradiance is distinct from these multi-messenger detections due to the predicted time delay from the initial merger that is needed for the superradiance cloud to grow and produce electromagnetic signals. We expect the O5 Observing Runs from the LIGO/VIRGO/KAGRA collaboration starting in 2028 \cite{KAGRA:2013rdx} to identify several more promising targets relevant for a search with the DSA-2000 beyond the already discovered ones listed in Table~\ref{tab:superradiance}. The spatial localization will also improve in these future runs, which is particularly promising for spatially correlating the described radio signal from superradiance with BH merger events. 

\begin{figure}[t]
	\includegraphics[width=0.8\textwidth]{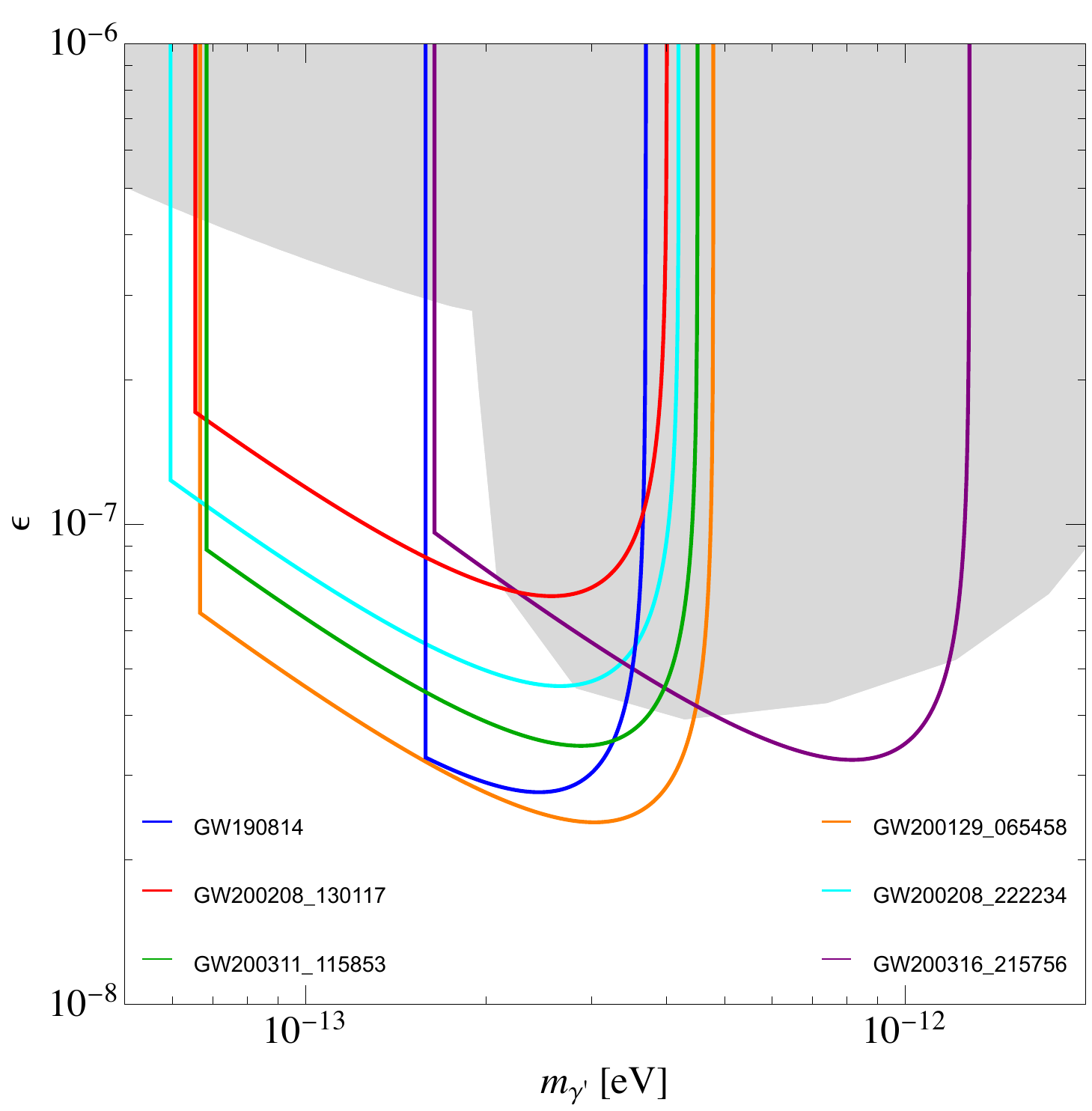} 
\caption{Projected discovery potential (SNR = 5) of the continuous radio signal from a dark photon BH superradiance cloud with mixing parameter $\epsilon$ and dark photon mass $m_{\gamma'}$
for 15 min intervals of observations as part of the all-sky survey with DSA-2000.
The lower mass range is bounded by the requirement that the superradiance cloud forms on a timescale shorter than 8 years.  
The gray shaded region is excluded by measurements of spectral distortions in the CMB performed by COBE/FIRAS \cite{Fixsen:1996nj,Caputo:2020bdy}, and Planck CMB and unWise galaxy survey data cross-correlation \cite{Planck:2018nkj, unwise, McCarthy:2024ozh}. Limits on dark photon DM extend to $\epsilon \gtrsim  10^{-14}$ and are not shown \cite{AxionLimits}.
}
\label{fig:darkphoton}
\end{figure}

\begin{table}[h]
\centering
\small
\begin{tabular}{>{\centering\arraybackslash}p{1.8cm}|>{\centering\arraybackslash}p{2.8cm}|>{\centering\arraybackslash}p{1.5cm}|>{\centering\arraybackslash}p{1.2cm}|>{\centering\arraybackslash}p{1.3cm}|>{\centering\arraybackslash}p{1.5cm}}  
\hline
\hline
\textbf{Event ID} & \textbf{Name} & \textbf{Mass} & \textbf{Spin} & \textbf{Distance} & \textbf{$\Delta \Omega$} \\
 & & \textbf{[$M_{\odot}$]} & \textbf{$a_{*}$} & \textbf{[Mpc]} & \textbf{[deg$^2$]} \\
\hline
\hline
S190814bv & GW190814 & $25.7$ & $0.28$ & $230$ & $19$ \\
\hline
S200208q & GW200208\_130117 & $62.5$ & $0.66$ & $2230$ & $48$ \\
\hline
S2003116g & GW200311\_115853 & $59.0$ & $0.69$ & $1170$ & $35$ \\
\hline
S20019m & GW200129\_065458 & $60.2$ & $0.73$ & $890$ & $54$ \\
\hline
S200208q & GW200208\_222234 & $68.7$ & $0.73$ & $1710$ & $51$ \\
\hline
S200316bj & GW200316\_215756 & $21.2$ & $0.7$ & $1120$ & $190$ \\
\hline
\hline
\end{tabular}
\caption{Proposed targets for a dark photon superradiance search with DSA-2000.}
\label{tab:superradiance}
\end{table}

In Figure \ref{fig:darkphoton} we show the discovery reach 
($\text{SNR} \equiv F_{\text{obs}}/\sigma = 5 $) in kinetic mixing $\epsilon$ that DSA-2000 will achieve through correlating detected broadband radio sources in its all sky survey with the spatial locations of the compact object mergers. For the shown forecast, we evaluate $L_{\text{EM}}$ at $t = t_{\text{growth}}$ in Eq.~{\eqref{eq:LEM}} and Eq.~\eqref{eq:Ldecay}, and take $\eta_{\text{radio}} = 10^{-4}$ in Eq.~\eqref{eq:radioeff}. This choice for $\eta_{\text{radio}}$ is motivated by
the emission spectrum of standard pulsars \cite{Kaspi:2017fwg, Manchester:2004bp, Siemonsen:2022ivj}. We have also assumed that, as part of DSA-2000's all sky survey, a sensitivity of $\sigma = 2\mu$Jy is reached for every sky patch within the field of view in the 15 minute intervals of observation.  
The detection of positive correlations would motivate targeted follow-ups with the DSA-2000 with increased frequency resolution that can resolve the spinning up pulsing frequency of the superradiance cloud radio signal. Figure \ref{fig:darkphoton} shows that BH superradiance radio signals can exceed the current dark photon limits by about one order of magnitude in sensitivity up to couplings as small as $\epsilon \gtrsim 3\times 10^{-8}$ for dark photon masses between $ 6\times 10^{-14} \, \text{eV} - 5 \times 10^{-13} \, \text{eV}$.
However, we note that Figure~\ref{fig:darkphoton} shows the potential of a discovery search, rather than a projection for a robust constraint. Due to the uncertainty in the spectral composition and total fraction of the luminosity emitted in radio frequencies, quantified by $\eta_{\text{radio}}$,
the absence of a signal does not rule out a dark photon in the shown mass range.   
However, its distinguishing features, bright radio sources spatially correlated with compact merger events turning on a timescale of months to years after the merger, in combination with a spinning up pulsing component with a pulsing frequency clustered around the dark photon mass, make it an exciting signal for a discovery search.\footnote{A discovery search for 
a continuous gravitational wave signal from dark photon superradiance, a complementary signal to the radio signal highlighted here, has recently been performed in  \cite{Mirasola:2025car} and did not find evidence for dark photon superradiance in the parameter space relevant for the radio signal.}

\section{Pulsar timing arrays}
\label{sec:pta}

In the preceding Sections, we have explored how the DSA-2000 can directly detect radio signals from BSM particles—specifically axions and dark photons—through their interactions with astrophysical environments. These approaches rely on the direct conversion of BSM particles into observable electromagnetic radiation. We now shift our focus to a complementary set of approaches that probe BSM physics through its gravitational imprints, beginning with pulsar timing. 

Millisecond pulsars (MSPs) have long been used as detectors of nanoHertz gravitational waves (GWs)~\cite{Taylor:2021yjx}. By measuring the timing of radio pulses from an array of MSPs accurately, one can measure ripples of spacetime manifesting as deviations in pulsar timings. In June 2023, The North American Nanohertz Observatory for Gravitational Waves (NANOGrav) announced positive evidence for a $\sim 3\sigma$ evidence of the existence of a stochastic gravitational wave background, possibly due to a population of merging supermassive BHs~\cite{NANOGrav:2023gor}.
The DSA-2000 is projected to detect approximately 3,000 millisecond pulsars~\cite{2024AAS...24326104S}, and to increase the total number of high-quality pulsars being timed to 200, with a monthly cadence~\cite{2019BAAS...51g.255H}, significantly improving NANOGrav’s sensitivity to gravitational waves.

In addition to gravitational waves, PTAs have also been proposed as probes of DM substructure and primordial black holes (PBHs)~\cite{Siegel:2007fz, Seto:2007kj, Baghram:2011is,Kashiyama_2012,Clark:2015sha,Schutz:2016khr, Dror:2019twh, Kashiyama:2018gsh, Ramani:2020hdo, Gresham:2022biw, NANOGrav:2023hvm}. DM substructure are self-gravitating DM objects formed on small ($\lesssim$~kpc) scales. While the large-scale matter power spectrum is accurately measured from the cosmic microwave background (CMB), DM substructure on small scales is much more difficult to detect. More importantly, different DM theories can predict different small-scale structures~\cite{Hogan:1988mp, Kolb:1993zz, Erickcek:2011us, Fan:2014zua, Graham:2015rva}, making the detection of these objects an important step in understanding the nature of DM. In this section, we review the idea of detecting DM substructure using PTAs, and explain how to calculate the projected sensitivity for a future PTA survey, including timing data acquired by DSA-2000. We find that the detection of DM substructure may be possible with sufficient observation time.

\begin{figure}
	\includegraphics[width=1\textwidth]{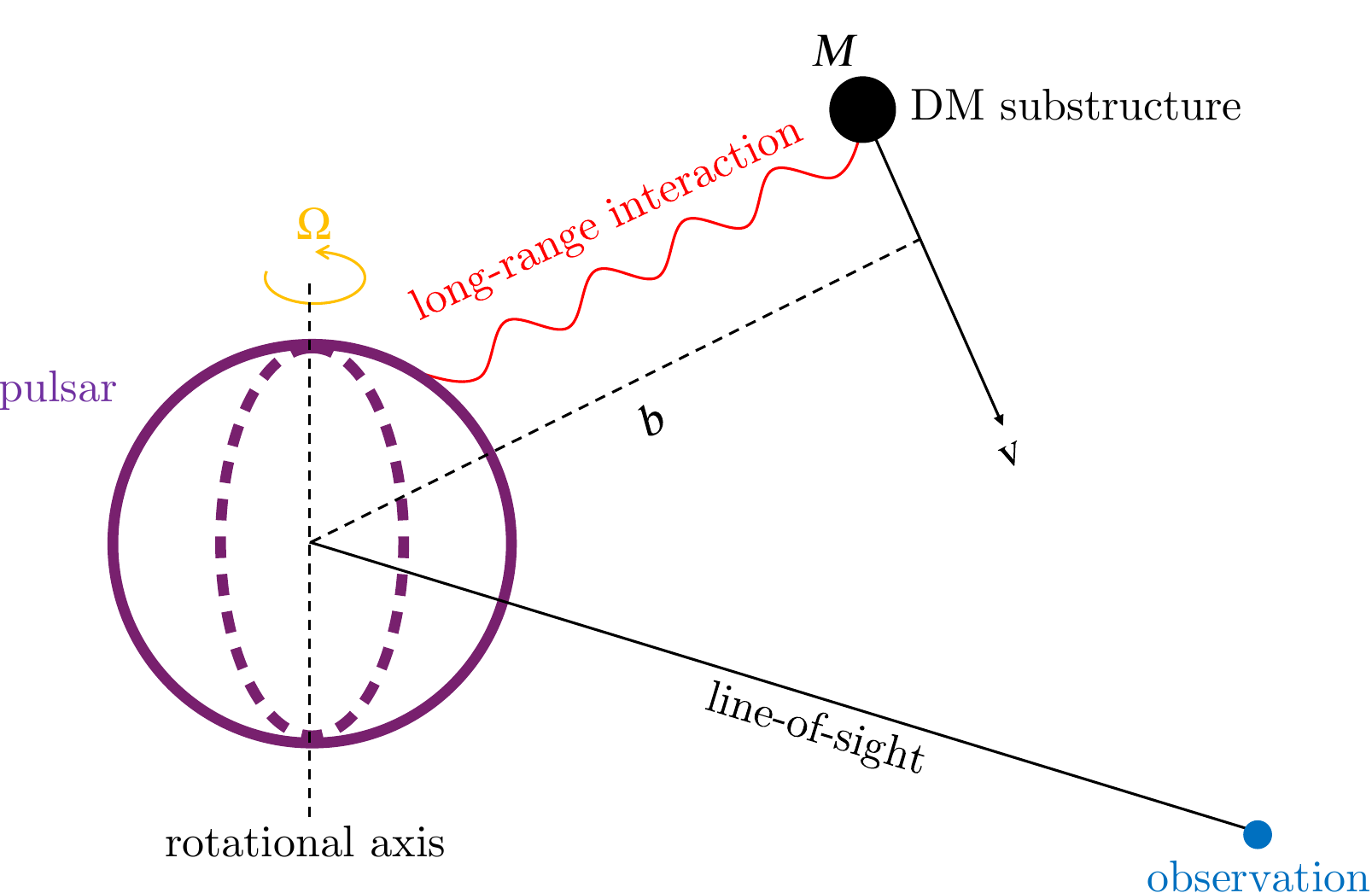} 
	\caption{Schematic of pulsar timing derivations induced by a transiting DM substructure. The DM impact parameter ($\vec{b}$) and velocity ($\vec{v}$) are represented by a dashed black line and a solid black line, respectively. The long-range interaction between DM and the pulsar, which can be either gravitational or involving an additional interaction between nucleons and DM, is depicted by a wavy red line.}\label{fig:PTA_DM_Schematic}
\end{figure}

The schematic diagram of DM substructure detection with PTA is shown in Fig.~\ref{fig:PTA_DM_Schematic}. The basic idea is that a generic long-range interaction between a pulsar and a transiting DM substructure—which can be either gravitational or an additional BSM interaction—will accelerate the pulsar.
This acceleration leads to Doppler shifts in the observed time-of-arrival of the pulses. This allows us to place an upper limit on the DM substructure abundance from purely gravitational interactions, as well as constrain the strength of the coupling constant in the case of a BSM long-range force.

\subsection{Gravitational effects}
\label{subsec:pta_grav}
Consider a transiting DM substructure with mass $M$ passing close to a pulsar with an impact parameter $\vec{b}$ and velocity $\vec{v}$, moving close to a single pulsar in the array. The trajectory of the DM in the vicinity of the pulsar can be parameterized as $\vec{r}(t) = \vec{b} + \vec{v}(t - t_{0})$, where $t_{0}$ denotes the time at which the DM is closest to the pulsar. The gravitational field, $\Phi_{\text{grav}}$, exerted on the pulsar by the transiting DM substructure is described by the Newtonian potential:
\begin{equation}\label{eqn:Newtonian_potential}
    \Phi_{\mathrm{grav}}(t) = -\frac{GM}{r(t)}  \, .
\end{equation}
The shifts in pulse arrival times, as measured by a pulsar timing experiment, $\delta t(t)$, can be obtained by integrating the pulsar frequency shifts, $\delta \Omega$:
\begin{equation}\label{eqn:timeshifts_def}
    \delta t(t) = \int_0^t dt'\, \frac{\delta \Omega(t')}{\Omega}  \, ,
\end{equation}
where the \textit{timing residuals}, $h(t)$, are defined to be the measured pulsar time shifts subtracting the best-fitted pulsar timing evolution consistent with the timing model of a naturally evolving pulsar~\cite{Taylor:2021yjx},
\begin{equation}\label{eqn:timing_residual_def}
    h(t) \equiv \delta t(t) - \delta t_{\mathrm{fitted}}(t)  \, .
\end{equation}
In the case of a transiting DM substructure, the gravitational attraction between DM and the pulsar induces an acceleration in the pulsar. This acceleration, in turn, induces a frequency shift in the apparent rotation frequency of the pulsar as measured on Earth. The frequency shift is given by the usual Doppler effect formula,
\begin{equation}\label{eqn:Doppler_effect_formula}
    \frac{\delta \Omega(t)}{\Omega} = \int_0^t dt'\, \unit{d}\cdot\nabla\Phi_{\mathrm{grav}}(t')  \, ,
\end{equation}
where $\vec{d}$ is the vector along the line-of-sight between the pulsar and Earth. Combining Eqs.~\eqref{eqn:Newtonian_potential}-\eqref{eqn:Doppler_effect_formula}, one obtains the timing residual due to the Doppler effect from a transiting DM substructure~\cite{Dror:2019twh, Lee:2020wfn},
\begin{equation}\label{eqn:DM_residual}
    h(t) = \frac{GM}{v^2}\unit{d} \cdot \left\{\unit{b}\sqrt{1+\left[\frac{v(t-t_{0})}{b}\right]^2}-\unit{v}\sinh^{-1}\left[\frac{v(t-t_{0})}{b}\right]\right\} - \delta t_{\mathrm{fitted}}(t) \, .
\end{equation}
Here we have absorbed all terms up to $\mathcal{O}(t^2)$ into $\delta t_{\mathrm{fitted}}(t)$, since these terms are typically included in a fit within the timing model. Practically, when searching for physics signals within the PTA dataset, the fitting part is handled by software packages such as \texttt{ENTERPRISE}~\cite{2019ascl.soft12015E}, \texttt{ENTERPRISE\_EXTENSIONS}~\cite{enterprise}, and a recently developed wrapper, \texttt{PTArcade}~\cite{Mitridate:2023oar}. See Ref.~\cite{Lee:2021zqw} for the Bayesian inference framework used to search for DM substructure with PTA data. Hence, the $\delta t_{\mathrm{fitted}}(t)$ term in Eq.~\eqref{eqn:DM_residual} can generally be neglected when the signal shape is interfaced into these programs for searching the actual dataset. However, for the purpose of making sensitivity projections for future experiments, the $\delta t_{\mathrm{fitted}}(t)$ term can lead to a reduction in sensitivity if a portion of the signal is degenerate with the timing model and gets absorbed into the fit~\cite{Ramani:2020hdo}.

Given a total observation time of $T$ on a particular pulsar, in the limit where $b \gg vT$ (which is applicable for DM masses sufficiently large), Eq.~\eqref{eqn:DM_residual} can be expanded in a power series in $t$.\footnote{One can also consider signals in the opposite limit (\textit{i.e.}, the dynamic limit) and effects such as the Shapiro delay. However, their upper limits are generally weaker than that of the Doppler static effect when analyzing real data~\cite{NANOGrav:2023hvm}, so we do not include them in this work.} The terms up to quadratic order in $t$ are degenerate with the intrinsic pulsar timing model and will be absorbed into the fit. Therefore, we parameterize the timing residual starting from the cubic order in $t$~\cite{Lee:2020wfn, Lee:2021zqw},
\begin{equation}\label{eqn:DM_residual_cubic}
    \textcolor{blue}{h(t) = \frac{A}{\mathrm{yr}^2} t^3 - \delta t_{\mathrm{fitted}}(t)} \, ,
\end{equation}
where $A_i$ is a dimensionless amplitude parameter, given by expanding Eq.~\eqref{eqn:DM_residual}
\begin{equation}\label{eqn:A_expansion}
    A = \mathrm{yr}^2\frac{GMv\unit{v}\cdot\unit{d}}{6b^3} \, .
\end{equation}
Here we have taken $t_0=0$ for simplicity. \textcolor{blue}{In addition, part of the signal in the cubic term of Eq.~\eqref{eqn:A_expansion} will also be absorbed by the fit due to the finite observation time, $T$. This can be estimated by considering the general form of the quadratic timing model, $\delta t_{\mathrm{fit}}(t) = a_0 + a_1 t + \tfrac{1}{2} a_2 t^2$, and solving for the coefficients $a_0$, $a_1$, and $a_2$ that minimize the difference, $\int_0^Tdt\,\left[\frac{A}{\mathrm{yr}^2} t^3 - \delta t_{\mathrm{fitted}}(t)\right]^2$. The resulting timing residual from Eq.~\eqref{eqn:DM_residual_cubic} is given by~\cite{Ramani:2020hdo}}
\begin{equation}\label{eqn:ht_best_fit}
    \textcolor{blue}{h(t) = \frac{A}{\mathrm{yr}^2} \left(t^3-\frac{3T}{2}t^2+\frac{3T^2}{5}t-\frac{T^3}{20}\right)} \, .
\end{equation}
\color{blue} If the observed data contain a signal from DM, the measured timing residual, $d(t)$, can be written as the sum of contributions from noise and background, $n(t)$, and the DM signal
\begin{equation}\label{eqn:d}
    d(t) = n(t) + h(t) \, .
\end{equation}
The term $n(t)$ can generically be written as a sum of three distinct components
\begin{itemize}
    \item pulsar intrinsic white noise (frequency-independent),
    \item pulsar intrinsic red noise (frequency-dependent)~\cite{2010ApJ...725.1607S}, 
    \item a common spectrum red noise process (frequency-dependent) across all pulsars, possibly due to a stochastic gravitational wave background~\cite{1983ApJ...265L..39H}.
\end{itemize}
For the majority of pulsars in NANOGrav's 15-year data release, the common-spectrum process is the dominant contribution at low frequencies, while pulsar-intrinsic white noise dominates at high frequencies once the common-spectrum process power falls off sufficiently~\cite{NANOGrav:2023ctt}. Defining the one-sided noise power spectral density $S_n(f)$ as $\langle \tilde{n}(f)\tilde{n}^*(f')\rangle = \tfrac{1}{2} S_n(f),\delta(f-f')$~\cite{Moore:2014lga}\footnote{\textcolor{blue}{We follow the Fourier transform convention defined in Ref.~\cite{Moore:2014lga}, namely $\tilde{x}(f)=\int_0^T dt\,x(t)\exp(-2\pi ift)$.}}, we can parameterize $S_n(f)$ as
\begin{equation}\label{eqn:S_nf}
    S_n(f) = \left[N_n + \frac{A_n^2}{12\pi^2}\left(\frac{f}{\mathrm{yr}^{-1}}\right)^{-\gamma_n}\right]\mathrm{yr}^{3} \, ,
\end{equation}
where $N_n$, $A_n$, and $\gamma_n$ are dimensionless constants. Here we neglect spatial correlations of the common-spectrum process, since the DM pulsar-term signal under consideration is uncorrelated across pulsars. The theoretical prediction for the noise spectral index due to a stochastic gravitational wave background produced by merging super massive black hole binaries is $\gamma_n = 13/3 \approx 4.3$~\cite{Phinney:2001di}. However, we note that the median and the $5$--$95$\% quantiles of the posterior distribution reported in the NANOGrav 15-year analysis are $\gamma_n = 3.2 \pm 0.6$, with the best-fit value of $\gamma_n \approx 3.2$, assuming the Hellings–Downs spatial correlation~\cite{NANOGrav:2023gor}.

The squared SNR can be estimated using the usual expression~\cite{Moore:2014lga}
\begin{equation}\label{eqn:SNR_formula}
    \mathrm{SNR}^2 = 4\int_{f_{\text{min}}}^{f_{\text{max}}}df\,\frac{|\tilde{h}(f)|^2}{S_n(f)} \, .
\end{equation}
For a finite observation time $T$ and cadence $\Delta t$ for a given pulsar, the time integral in the Fourier transform of the signal is cut off at $t = T$, while the frequency integral runs from $f_{\min} = 1/T$ to the Nyquist frequency $f_{\max} = 1/(2\Delta t)$. The (squared) Fourier transform of the DM signal can be readily computed from Eq.~\eqref{eqn:ht_best_fit}
\begin{align}\label{eqn:hf2}
    |\tilde{h}(f)|^2 &= \frac{A^2}{400\pi^8\mathrm{yr}^4}\frac{\left[\pi f T(\pi^2f^2T^2-15)\cos(\pi f T)-3(2\pi^2f^2T^2-5)\sin(\pi fT)\right]^2}{f^8} \nonumber \\
    &\sim \frac{A^2}{400\pi^2\mathrm{yr}^4}\frac{T^6\cos^2(\pi fT)}{f^2} \, ,
\end{align}
where we have taken $fT\gtrsim 1$ in the second line. Computing the SNR in Eq.~\eqref{eqn:SNR_formula} using the DM signal expression in Eq.~\eqref{eqn:hf2} and the noise power spectral density in Eq.~\eqref{eqn:S_nf}, we find\footnote{\textcolor{blue}{We thank Abhiram Cherukupalli for pointing out the importance of including the red noise when deriving the scaling of the SNR with powers of $T$.}}
\begin{align}\label{eqn:SNR_single_estimate}
    \mathrm{SNR} &\approx \frac{GMT^3v|\unit{v}\cdot\unit{d}|}{120\sqrt{2}\pi b^3\,\mathrm{yr}}\sqrt{\frac{1}{N_n}\left(\frac{\mathrm{yr}^{-1}}{f_*}\right)+\frac{12\pi^2}{(\gamma_n-1)A_n^2}\left(\frac{f_*}{\mathrm{yr}^{-1}}\right)^{\gamma_n-1}} \nonumber \\
    &\equiv \frac{GMT^3v|\unit{v}\cdot\unit{d}|}{6b^3}\frac{1}{K_n} \, ,
\end{align}
where we assume $\gamma_n > 1$, define $f_*$ as the frequency where the white-noise and red-noise contributions are equal, and define the quantity $K_n$ (with dimensions of [time]) as the effective noise level that enters the SNR. 
Here we note that the SNR computed in Eq.~\eqref{eqn:SNR_single_estimate} is dominated by the contribution at the noise turnover frequency, $f_*$. This can be easily understood by noting that $|\tilde{h}(f)|^2\sim f^{-2}$ for $f\gtrsim 1/(2\pi T)$ as in Eq.~\eqref{eqn:hf2}, and that the noise scales as $S_n(f)\propto f^{-\gamma_n}$ for $f\lesssim f_*$, and $S_n(f)\sim \mathrm{constant}$ for $f\gtrsim f_*$. Therefore, the integrand in Eq.~\eqref{eqn:SNR_formula} scales as $f^{\gamma_n-2}$ and $f^{-2}$ at low and high frequencies, respectively, and thus as long as $\gamma_n>1$, the SNR integral will be dominated by frequency near $f=f_*$.
\color{black}

The above analysis considers the SNR from timing measurements of a single pulsar. We now generalize this to the case of PTAs, where timing measurements from multiple pulsars are available. Since the DM phase space is a statistical distribution near each pulsar, the expected SNR from each pulsar also follows a statistical distribution. The variance in DM velocity can generally be neglected in an order-of-magnitude estimate due to its linear scaling in the SNR. Statistically, the signal is dominated by the DM with the smallest impact parameter. For a given pulsar, this can be derived by considering putting random points in a spherical volume, in which the cumulative distribution function (CDF) of the impact parameter of each DM is given by $F_B(b)=(b/R)^3$, where $R$ is the radius of a large (fictitious) spherical volume, and thus the CDF of the \textit{minimal} impact parameter is~\cite{Dror:2019twh} 
\begin{align}\label{eqn:minimal_parameter}
    F_{B_{\min}}(b_{\min}) 
    &= 1-\left[1-F_B(b_{\min})\right]^N \nonumber \\
    &= 1-\left[1-\left(\frac{b_{\min}}{R}\right)^3\right]^N \nonumber \\
    &\to 1-\exp\left(-\frac{4\pi}{3}n_{\mathrm{DM}}b_{\min}^3\right) \, ,
\end{align}
where $n_{\mathrm{DM}}$ is the local DM number density, and $N$ is the total DM number in the volume, hence $N=(4\pi/3)n_{\mathrm{DM}}R^3$. We take the large $N$ limit in the second line of Eq.~\eqref{eqn:minimal_parameter} and use the product limit of the exponential function. We see that both $R$ and $N$ drop out of the CDF. The CDF of the SNR from this pulsar induced by the DM with impact parameter $b=b_{\min}$ in Eq.~\eqref{eqn:SNR_single_estimate} can be found by inverting Eq.~\eqref{eqn:SNR_single_estimate} to solve for the corresponding SNR, and using Eq.~\eqref{eqn:minimal_parameter}
\begin{align}\label{eqn:SNR_CDF_single}
    F_{\mathrm{SNR}}(\mathrm{SNR}) &= 1 - F_{B_{\min}}(b_{\min}(\mathrm{SNR})) \nonumber \\
    &=\textcolor{blue}{\exp\left(-\frac{4\pi}{3}\frac{G\rho_{\mathrm{DM}}f_{\mathrm{DM}}}{6}\frac{1}{\mathrm{SNR}}\frac{T^3v|\unit{v}\cdot\unit{d}|}{K_n}\right)} \, .
\end{align}
Here we have defined
\begin{align}\label{eqn:f_DM_def}
    f_{\mathrm{DM}} \equiv \frac{n_{\mathrm{DM}}M}{\rho_{\mathrm{DM}}} \, ,
\end{align}
to be the fraction of dark matter in the form of compact objects, which are the components that can contribute to PTA signals. The theoretically compelling parameter space is $f_{\mathrm{DM}} \leq 1$, although this assumes the robustness of the canonical local dark matter density value, $\rho_{\mathrm{DM}} = 0.4~\mathrm{GeV/cm}^3$. Current experimental constraints, however, allow values as large as $f_{\mathrm{DM}} \lesssim 2 \times 10^3$, based on limits from solar system ephemerides~\cite{Pitjev_2013}, since there is currently no direct local observational measurement of $\rho_{\mathrm{DM}}$.

If we are now given timing data from $N_P$ pulsars, each labeled with subscript $i$, the optimal sensitivity to compact DM is achieved by considering the statistics of the pulsar with the highest SNR~\cite{Lee:2021zqw}, denoted as $\mathrm{SNR}_{\max}$, whose CDF can be readily derived from Eq.~\eqref{eqn:SNR_CDF_single}, and the fact that the SNR$_i$ from each pulsar is independent:
\begin{align}\label{eqn:max_SNR_statistics}
    F_{\mathrm{SNR}_{\max}}(\mathrm{SNR}_{\max}) &= \prod_{1\leq i\leq N_P}F_{\mathrm{SNR}_i}(\mathrm{SNR}_{\max}) \nonumber \\
    &= \textcolor{blue}{\exp\left(-\frac{4\pi}{3}\frac{G\rho_{\mathrm{DM}}f_{\mathrm{DM}}}{6}\frac{1}{\mathrm{SNR}_{\max}}\sum_{i=1}^{N_P}\frac{T_i^3v_i|\unit{v}_i\cdot\unit{d}_i|}{K_{n,i}}\right)} \, ,
\end{align}
\textcolor{blue}{where $K_{n,i}$ is the effective noise level as defined in Eq.~\eqref{eqn:SNR_single_estimate} for each pulsar.} Using Eq.~\eqref{eqn:max_SNR_statistics}, for any $0\leq p\leq 1$, we can derive 
the $100p^{\mathrm{th}}$-percentile $\mathrm{SNR}_{\max}$, denoted as $\mathrm{SNR}_{\max,p}$, by setting $F_{\mathrm{SNR}_{\max}}(\mathrm{SNR}_{\max,p})=p$,
\begin{align}\label{eqn:max_SNR_percentile}
    \mathrm{SNR}_{\max,p} &= \textcolor{blue}{\frac{4\pi}{3}\frac{G\rho_{\mathrm{DM}}f_{\mathrm{DM}}}{6}\frac{1}{\log p^{-1}}\sum_{i=1}^{N_P}\frac{T_i^3v_i|\unit{v}_i\cdot\unit{d}_i|}{K_{n,i}}} \nonumber \\
    &\propto f_{\mathrm{DM}} \sum_{i=1}^{N_P}T_i^{\textcolor{blue}{3}} \, ,
\end{align}
where $N_P$ is the number of pulsars used in the analysis. 
We have assumed that the \textcolor{blue}{effective} noise level for timing data from each pulsar is \textcolor{blue}{comparable}, and that the noise level for new pulsars to be discovered and timed by DSA-2000 is the same as that of existing pulsars timed by NANOGrav. Additionally, in the second line, we have neglected the variance in DM velocity and the variance $\mathcal{O}(1)$ angular factors 
%\kb{do we mean here $\mathcal{O}$(1)  angular factors or variance in angular factor. The former should be already explicit by using $\propto$ }\vl{here I meant the variance}.
Note in particular that the pulsar observing time $T_i$ has a particularly strong impact on the SNR, since a longer observation time leads to a smaller expectation value of $b_{\min}$, and thus a stronger signal.

\begin{figure}
	\includegraphics[width=\textwidth]{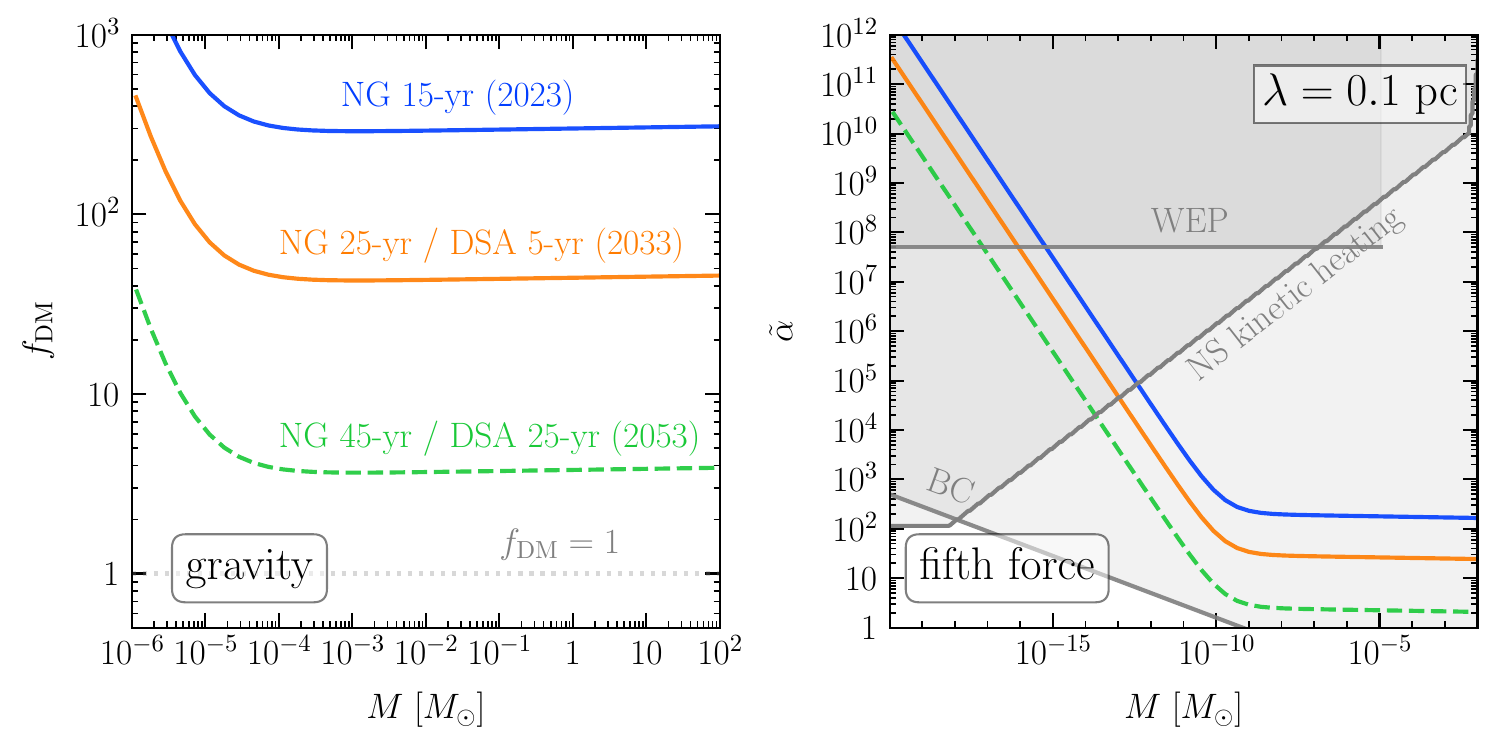} 
    \caption{
     Upper limits at the 95\% credible level on the local PBH abundance (left panel) and fifth force coupling strength (right panel). The blue line is the existing constraint published by the NANOGrav collaboration~\cite{NANOGrav:2023hvm}, while the orange and green lines are the projected sensitivity by 2033 and 2053, corresponding to 5 and 25 years after DSA-2000 begins taking timing data, obtained by rescaling the blue line using the projected parameters for a combined NANOGrav 25-yr / DSA 5-yr and NANOGrav 45-yr / DSA 25-yr observation dataset, and Eq.~\eqref{eqn:stacking} and Eq.~\eqref{eqn:stacking_2} (see main text). Following Ref.~\cite{NANOGrav:2023hvm}, existing constraints on $\tilde{\alpha}$ are also shown in the right panel, including bounds from the weak equivalence principle (WEP)~\cite{Wagner:2012ui, Shao:2018klg, Sun:2019ico}, NS kinetic heating (assuming an additional short-range DM–baryon interaction) due to DM capture~\cite{Gresham:2022biw}, and bounds derived from a combination of fifth-force constraints on baryon–baryon interactions~\cite{Berge:2017ovy, PhysRevD.99.055043} and Bullet Cluster (BC) limits on DM self-interactions~\cite{Spergel:1999mh, Kahlhoefer:2013dca} (see also Refs.~\cite{Coskuner:2018are, Gresham:2022biw}). We note that while the BC bound is very competitive, it is an indirect constraint on $\tilde{\alpha}$ and does not apply if only a subcomponent of DM interacts through the fifth force, where, in contrast, the other constraints degrade only linearly with the DM subcomponent fraction.}\label{fig:PTA_reach}
\end{figure}

\begin{figure}
	\includegraphics[width=1\textwidth]{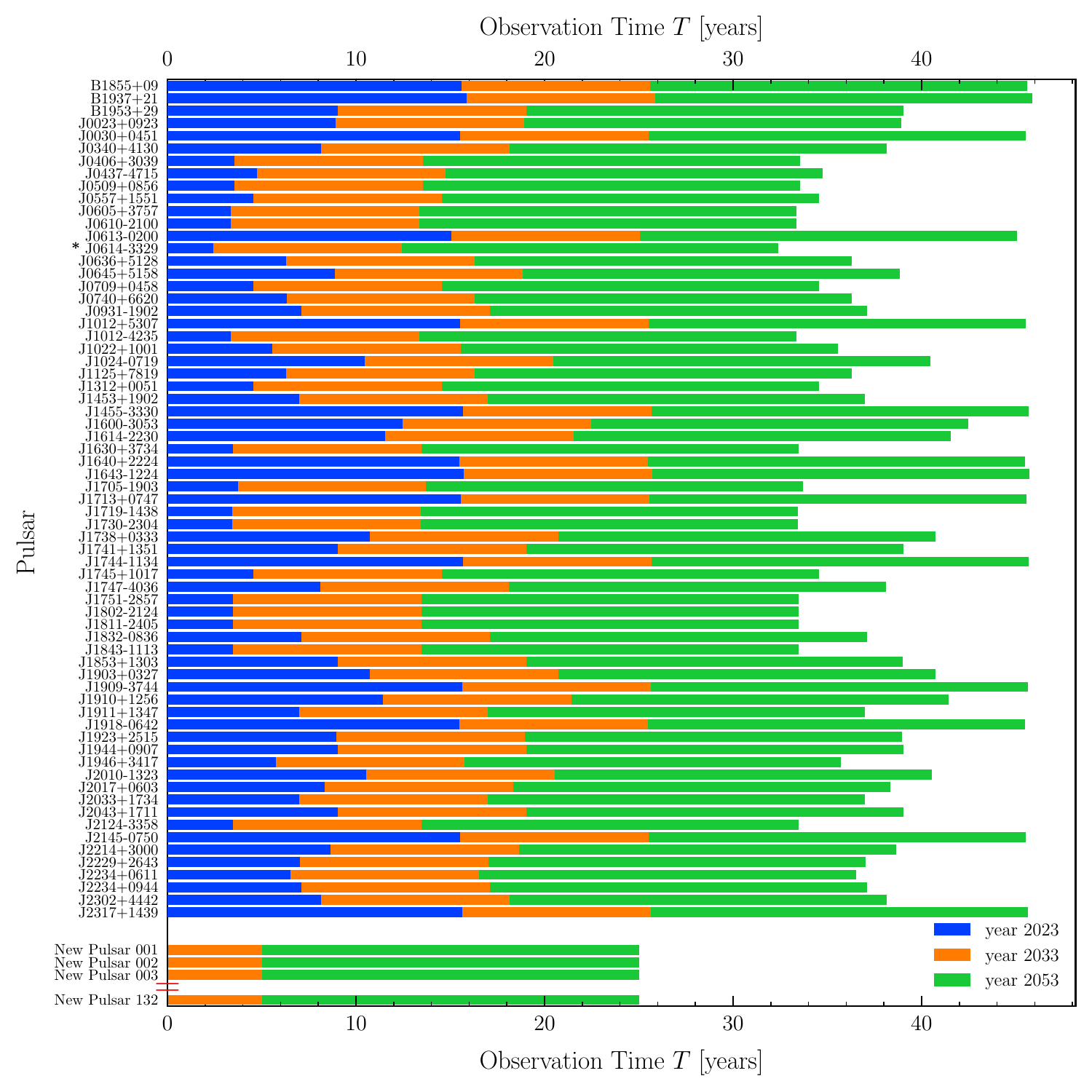} 
    \caption{Observation times for each pulsar considered in the analysis are shown for three benchmark years: 2023 (blue), 2033 (orange), and 2053 (green). The 68 existing pulsars being timed by NANOGrav in their 15-year data release (2023) are shown at the top~\cite{NANOGrav:2023hde}, although J0614-3329 (marked by an asterisk) is excluded from their DM substructure analysis since its observation time has been less than three years~\cite{NANOGrav:2023hvm}. The new pulsars that DSA-2000 is expected to discover are shown at the bottom, labeled with placeholder pulsar names starting with ``New Pulsar".}\label{fig:pulsar_times}
\end{figure}

The actual data analysis process performed by NANOGrav is a Bayesian inference process~\cite{Lee:2021zqw, NANOGrav:2023hvm}.
A Monte-Carlo Markov Chain (MCMC) analysis is first performed on the dataset to obtain a posterior distribution of $A_i$. 
It is then compared against a theoretical prediction of $A_i$ obtained by a Monte Carlo simulation assuming various values of $f_{\mathrm{DM}}$. This translates to a posterior distribution of $f_{\mathrm{DM}}$, and thus, a 95-th$\%$ confidence upper limit on $f_{\mathrm{DM}}$.
Since the actual data analysis involves complicated numerical simulations and noise marginalization over a high-dimensional parameter space, we utilize here Eq.~\eqref{eqn:max_SNR_percentile} to estimate the projected sensitivity of DSA-2000 using its expected experimental parameters, assuming the absence of a positive signal.
In practice, we take the upper limits derived by NANOGrav in their 15-yr dataset~\cite{NANOGrav:2023hvm}, denoted as $f_{\mathrm{DM}}^{(\mathrm{NG-15})}$, and the timing observation duration of each pulsar, $T^{(\mathrm{NG-15})}_j$, as listed in NANOGrav's data release~\cite{NANOGrav:2023hde}, to obtain a projected upper limit with when the DSA-2000 data are stacked on the existing NANOGrav data, by solving for $f_{\mathrm{DM}}$ in Eq.~\eqref{eqn:max_SNR_percentile}
\begin{equation}\label{eqn:stacking}
    \frac{f_{\mathrm{DM}}}{f_{\mathrm{DM}}^{(\mathrm{NG-15})}} = \frac{\sum_j^{N_P^{\mathrm{NG-15}}} (T^{(\mathrm{NG-15})}_j)^{\textcolor{blue}{3}}}{\sum_i^{N_P}T_i^{\textcolor{blue}{3}}}\, ,
\end{equation}
where $N_P^{(\mathrm{NG-15})}=67$ is the number of pulsars used in the NANOGrav 15-yr new physics analysis (one pulsar, J0614-3329, was excluded from the analysis since it was observed for less than three years)~\cite{NANOGrav:2023hvm}, released in 2023. 

% Note that the derivation of Eq.~\eqref{eqn:max_SNR_percentile} assumes that only uncorrelated white noise \vl{revise this} is present in the timing data. In the presence of colored and correlated noise, the SNR from potential DM sources will be diminished, as shown in Ref.~\cite{Lee:2021zqw}.
% However, we expect this drop in sensitivity will be common in both the NANOGrav 15-yr dataset analysis, and any future analyses combining the NANOGrav and DSA-2000 datasets. 
% Hence, its effect should be minimal in the projection obtained by a rescaling of the upper limit. 

The above analysis shows that the sensitivity to DM substructure can be improved by (i) obtaining timing data from more pulsars, and (ii) a longer cumulative observation time for each pulsar. Both of these conditions can be achieved by utilizing DSA-2000's power in discovering new pulsars. DSA-2000 is expected to bring the total number of pulsars being timed within the NANOGrav Collaboration to 200. In the left panel of Fig.~\ref{fig:PTA_reach}, we show the projected sensitivity from the outlined procedure, assuming that the new pulsars start being timed by the collaboration in 2028. We show the projected sensitivity by 2033 and 2053, corresponding to 5 and 25 years after DSA-2000 begins taking timing data. We find that with the combined data from the NANOGrav 25-yr and DSA-2000 5-year datasets, which is the currently planned timeframe for DSA-2000, the upper limit that one can place on the local DM density in the form of compact objects can improve by roughly an order of magnitude, to $f_{\mathrm{DM}}\lesssim \textcolor{blue}{50}$ for $M\gtrsim 10^{-5}$~$M_{\odot}$. If we consider the combined data from the NANOGrav 45-year and DSA-2000 25-year datasets instead, the upper limit on the local DM density in the form of compact objects can improve by more than two orders of magnitude compared to the published NANOGrav 15-yr dataset, reaching $f_{\mathrm{DM}}\lesssim \textcolor{blue}{4}$ for $M\gtrsim 10^{-5}$~$M_{\odot}$. We summarize the total observation time of each pulsar considered in the above analysis in Fig.~\ref{fig:pulsar_times}.

We note that while we have assumed a total of $N_P = 200$ pulsars in the projected timing dataset, the upper limit on $f_{\mathrm{DM}}$ scales only linearly on $N_P$, and in practice the scaling is even weaker, since newly added pulsars typically have shorter observation time spans. For instance, if the total number of pulsars in the NANOGrav 45-year / DSA-2000 25-year dataset (green line in Fig.~\ref{fig:PTA_reach}) is 175 rather than the 200 assumed in this analysis, the upper limit on $f_{\mathrm{DM}}$ would decrease by only $\sim 6\%$.

Additionally, we note that the above estimate should be considered conservative, as we do not account for the potential addition of pulsars to the dataset beyond those that DSA-2000 discovers early in its mission. Moreover, exploiting the spatial cross-correlation of timing deviations across different pulsars to better distinguish the signal from background contributions could further improve this outlook. We leave this promising avenue for future work, which may enable sensitivity to $f_{\mathrm{DM}}\lesssim 1$, even with the observation time $T$ considered in this work.

\subsection{Fifth force}
\label{subsec:pta_fifth}
In addition to gravitational interactions, DM could also interact with the SM via an effective long-range Yukawa potential
\begin{equation}\label{eqn:yukawa_potential}
    \Phi_{\mathrm{fifth}}(r) = -\tilde{\alpha} \frac{GM}{r}e^{-r/\lambda} \, ,
\end{equation}
where $\tilde{\alpha}$ is the effective strength of interaction, and $\lambda\equiv 1/m_{\phi}$ is the force range, inversely proportional to the mediator mass, $m_{\phi}$, which could be either a light scalar or a vector field. The Doppler effect due to this Yukawa force is easily derived by replacing the potential in Eq.~\eqref{eqn:Doppler_effect_formula} by $\Phi_{\mathrm{fifth}}$ in Eq.~\eqref{eqn:yukawa_potential}. In the limit where $\lambda>b$ and $\tilde{\alpha}\gg 1$, the effect of fifth force effectively boosts the potential by a factor of $\tilde{\alpha}$, and hence the SNR in Eq.~\eqref{eqn:max_SNR_percentile} should be modified as (assuming $f_{\mathrm{DM}}=1$, \textit{i.e.} the DM substructure saturates the local DM density and all of them are charged under the fifth force)
\begin{equation}\label{eqn:fifth_SNR_PTA_estimate}
    \mathrm{SNR_{\max,p}}  \propto  \tilde{\alpha} \sum_{i=1}^{N_P}T_i^3 \, .
\end{equation}
Constraints on $\tilde{\alpha}$ for PTAs have been derived in Ref.~\cite{Gresham:2022biw} using NANOGrav's 11-yr dataset, as well as in Ref.~\cite{NANOGrav:2023hvm} by NANOGrav using their 15-yr dataset. Similar to the gravitational interaction case, we derive the upper limits on $\tilde{\alpha}$ by translating the result in Ref.~\cite{NANOGrav:2023hvm} using Eq.~\eqref{eqn:fifth_SNR_PTA_estimate}, and the stacked parameters of DSA-2000 and NANOGrav 
\begin{equation}\label{eqn:stacking_2}
    \frac{\tilde{\alpha}}{\tilde{\alpha}^{(\mathrm{NG-15})}} = \frac{\sum_j^{N_P^{\mathrm{NG-15}}} (T^{(\mathrm{NG-15})}_j)^{\textcolor{blue}{3}}}{\sum_i^{N_P}T_i^{\textcolor{blue}{3}}}\, ,
\end{equation}
where $\tilde{\alpha}^{(\mathrm{NG-15})}$ is the upper limit on the effective fifth force strength derived by NANOGrav in their 15-yr dataset assuming $f_{\mathrm{DM}}=1$~\cite{NANOGrav:2023hvm}. The result is shown in the second panel of Fig.~\ref{fig:PTA_reach}. We find that with the combined data of NANOGrav and DSA-2000, an order of magnitude improvement on the constraints on $\tilde{\alpha}$ is expected by 2033, and \textcolor{blue}{two orders of magnitude} improvement by 2053.

\section{Neutrino mass measurements with fast radio bursts}
\label{sec:neutrinos}

While the previous section examined how pulsar timing can probe BSM physics through gravitational effects, we now turn to another indirect probe: cosmological inference using FRBs. This approach utilizes the DSA-2000's capability to detect large numbers of FRBs across cosmic distances to study how neutrino masses affect the large-scale structure (LSS) of the universe. Unlike the electromagnetic signatures explored in Sections~\ref{sec:axions}-\ref{sec:darkphoton}, or the gravitational perturbations explored in Section~\ref{sec:pta}, in this Section we will exploit statistical correlations between dispersion measures and weak lensing observations to break degeneracies in cosmological parameter estimation. The strength of this approach lies in its ability to constrain neutrino properties through their impact on cosmic structure formation, providing an independent and complementary path to neutrino mass measurements.

Cosmological constraints on the neutrino masses are derived from their influence on the spatial distribution of structures in the Universe, and its background evolution (see, e.g.,~\cite{lesgourgues_massive_2006, abazajian_neutrinos_2015,gerbino_status_2017} for review). Due to their large abundance, neutrinos contribute to the overall matter density; however, their velocity dispersion is very small, enabling them to escape potential wells, washing out structures below a certain scale, called the free-streaming scale (see, e.g., \cite{marulli_effects_2011}). The sum of the neutrino masses, $\sum m_\nu$, has been constrained by the Planck mission \cite{planck_parameters_2018} and in combination with galaxy surveys such as BOSS \cite{ivanov_cosmological_2020} and DESI \cite{2024arXiv240403002D}. 
These cosmological measurements are approaching sensitivities that probe the minimal neutrino mass hierarchies, which, based on neutrino oscillation measurements \cite{deSalas:2020pgw, Esteban:2020cvm}, predict $\sum m_\nu = 0.058 \, \text{eV}$ in the normal hierarchy, and $\sum m_\nu = 0.11 \, \text{eV}$ in the inverted hierarchy. In particular, measurements of the Dark Energy Spectroscopic Instrument (DESI) inferred a $95\%$ upper bound of neutrino masses of $\sum m_\nu < 0.072 \, \text{eV}$ for a $\sum m_\nu > 0 \, \text{eV}$ prior when combined with CMB data, excluding the inverted mass hierarchy, and showing posteriors that are peaked around $\sum m_\nu = 0 \, \text{eV}$ \cite{2024arXiv240403002D}.

It has since been argued that the DESI measurement may even point to a cosmological preference for negative  neutrino masses \cite{Craig:2024tky,Green:2024xbb}, driven by the combination of a low inferred value of the fractional matter density from DESI BAO 
\cite{Loverde:2024nfi}, and excess measured lensing power of CMB photons, opposite to what one
would expect as signatures of neutrino mass \cite{Craig:2024tky, Lynch:2025ine}.

In light of these developments, upcoming
surveys such as Euclid or the Legacy Survey of Space and Time (LSST) will provide valuable independent insights into the neutrino mass puzzle \cite[see e.g.][]{euclid_sensitivity_2024}. 
These upcoming late universe probes
of weak gravitational lensing are sensitive to neutrino masses via their impact on both background evolution, and spatial distribution of structure. However, to fully realise their potential, these surveys rely on the utilisation of highly non-linear scales. It is precisely at these scales that visible matter, i.e. baryons, and various astrophysical processes will significantly affect the distribution of structures within the Universe. Phenomena such as supernova feedback, cooling processes, stellar winds, and feedback from active galactic nuclei can displace baryonic matter, which in turn influences the surrounding DM. This interaction alters the statistical properties of the overall matter distribution. These collective influences are referred to as baryonic feedback (BF) \cite[see e.g.][]{chisari_modelling_2019,vogelsberger_cosmological_2020,schneider_baryonic_2020}. Understanding and accounting for these complex interactions is crucial for accurately interpreting data from cosmological surveys, as they obscure and mimic the weak gravitational lensing signal.

 Here, we show how the statistical properties of the dispersion measure of the large number of FRBs projected to be discovered by DSA-2000 will improve the uncertainty in the neutrino mass measurement from weak gravitational lensing by a factor of three. This improvement is achieved by breaking degeneracies between baryonic feedback and the impact of neutrino masses, and allows for a competitive neutrino mass measurement relying solely on late universe measurements with independent systematics uncertainties.

\subsubsection*{Cosmic shear}
Cosmic shear is the weak gravitational lensing effect induced by the LSS on background galaxies \cite[see][for reviews]{bartelmann_weak_2001,kilbinger_cosmology_2015}. The primary observable is the shear, $\gamma$, which can be expressed as the second derivative of the scalar lensing potential, itself the line-of-sight projection of the gravitational potential\footnote{More precisely, lensing responds to metric perturbations in both the time-time and space-space components of the metric, i.e. the Bardeen potentials. In General Relativity and in absence of anisotropic stress, however, they are equal.}. The statistical properties of the shear field and the convergence, $\kappa$, on the two-point level, are the same.
By relating the potential to the density contrast via Poisson's equation, the convergence at redshift $z$ and angular position $\hat{\boldsymbol{n}}$ can be written as (assuming a spatially flat Universe):
\begin{equation}
\label{eq:kappa}
    \kappa(z,\hat{\boldsymbol{n}}) = \frac{3\Omega_{m}}{2\chi_H^2}\int_0^{\chi(z)}{d}\chi^\prime \frac{\chi^\prime(\chi(z)-\chi^\prime)}{\chi(z)} \frac{\delta(\chi^\prime \hat{\boldsymbol{n}},\chi^\prime)}{a(\chi^\prime)}\,,
\end{equation}
here $\chi$ is the comoving distance:
\begin{equation}
    \chi(z) = \frac{c}{H_0}\int_0^z \frac{{d}z'}{E(z')}\,,
\end{equation}
$E(z)$ is the dimensionless Hubble parameter and $\chi_{H}\coloneqq \chi(\infty)$ the distance to the Hubble horizon. $\Omega_\mathrm{m}$ is the matter density parameter today and $\delta$ is the density contrast of the total matter field. Note that the mean of the convergence field is zero, i.e. $\langle \kappa \rangle = 0$ and is not an observable, as a homogeneous density field will not deflect light rays.

By averaging over an ensemble of sources distributed over redshift, one obtains the effective convergence by suitably rearranging integration limits:
\begin{equation}
\label{eq:kappa_tomo}
    \kappa_i(\hat{\boldsymbol{n}}) = \int_0^\infty {d}z \,n^{s}_i(z) \kappa(z,\hat{\boldsymbol{n}}) = \int_0^{\chi_{H}}{{d}\chi} \,W_{\kappa_i}(\chi) \delta(\chi \hat{\boldsymbol{n}},\chi)\,, 
\end{equation}
where $n^{s}_i(z)$ is the normalised redshift distribution of the $i$-th source ensemble and $W_{\kappa_i}(\chi)$ is the lensing weight function:
\begin{equation}
        W_{\kappa_i}(\chi) = \frac{3\Omega_{m}}{2\chi^2_{H}}\frac{\chi}{a(\chi)}\int_\chi^{\chi_{H}}{ d}\chi'n^{s}_i(\chi')\frac{\chi-\chi'}{\chi'}\,.
\end{equation}
The splitting of the entire source sample into subsamples is called tomography \cite{hu_power_1999} and is a way to restore information about the redshift evolution of structures. Typically, the information almost saturates after splitting the sample into six tomographic bins \cite{sipp_marvin_2021}.

\subsubsection*{Dispersion measure}
The total dispersion measure, $\mathcal{D}_\mathrm{tot}$ is estimated from the temporal delay of an FRB's pulse as a function of frequency $\nu$:
\begin{equation}
\label{eq:time_delay}
    \Delta t \propto \mathcal{D}_\mathrm{tot}(\hat{\boldsymbol
    {n}}, z) \, \nu^{-2} \, .
\end{equation}
 The dispersion itself is caused by interaction with the free electrons along the line of sight. These electrons are either associated with the host galaxy and halo, with the Milky Way (MW), or with the LSS:
\begin{equation}
\label{eq:dm_contributions}
    \mathcal{D}_\mathrm{tot}(\hat{\boldsymbol
    {x}}, z) = \mathcal{D}_\mathrm{host}(z) + \mathcal{D}_\mathrm{MW}(\hat{\boldsymbol
    {x}}) + \mathcal{D}_\mathrm{LSS}(z,\hat{\boldsymbol
    {x}}) \, .
\end{equation}
The LSS contribution, in contrast to cosmic shear, obtains contributions from perturbations and the cosmological background (compare to Equation \ref{eq:kappa}):
\begin{equation}
\label{eq:DM_LSS}
    \mathcal{D}_\mathrm{LSS}(\hat{\boldsymbol{n}},z) = \frac{3 \Omega_{b} H_0}{8 \pi G m_{p}} \chi_{e} \,  \int_0^z {d} z'  \, \frac{1+z'}{E(z')}  f_\mathrm{IGM}(z^\prime)\big(1+\delta_{e}(\hat{\boldsymbol{n}},z')\big)\, ,
\end{equation}
with the proton mass $m_{p}$, the Hubble constant $H_0$, the baryon density parameter today $\Omega_{b}$, the electron fraction $\chi_{e}$ and the fraction of electrons in the inter galactic medium $f_\mathrm{IGM}$
which we model following \cite{macquart_census_2022} (see also \cite{2024arXiv240916952C} for an updated measurement). Lastly, $\delta_\mathrm{e}$ is the electron density contrast.

The mean of \Cref{eq:DM_LSS}, $\langle\mathcal{D}_\mathrm{LSS}\rangle (z)$ is an actual observable and is often referred to as the dispersion measure-redshift relation or the Marcquart relation. Given observations of $N_\mathrm{FRBs}$, with redshifts $z_\alpha$, positions $\hat{\boldsymbol{n}}_\alpha$, and total dispersion measures $\mathcal{D}_{\mathrm{tot}, \alpha}$,
it contains information about the mean baryon density and the expansion history of the Universe \cite[see e.g.][]{zhou_fast_2014,walters_future_2018,macquart_census_2022,hagstotz_new_2022,james_measurement_2022,wu_eight_2022}. Measuring $\langle\mathcal{D}_\mathrm{LSS}\rangle (z)$ requires the specification of the MW component, for which there are different models available \cite{cordes_new_2001,yao_new_2017,yamasaki_galactic_2020}. Furthermore, the host component must be marginalised over for unbiased results \cite{hagstotz_new_2022}. The covariance is given by
\begin{equation}
    \boldsymbol{C} = \boldsymbol{C}_\mathrm{LSS} + \boldsymbol{C}_\mathrm{host} + \boldsymbol{C}_\mathrm{MW}\,,
\end{equation}
assuming that the contributions in \Cref{eq:dm_contributions} are uncorrelated. 
For the MW contribution, we simply assume a constant uncertainty, independent of the FRB with $\sigma_{\mathrm{MW},\alpha} = 30\;\mathrm{pc}\,\mathrm{cm}^{-3}$. The host contribution's uncertainty is assumed to be $\sigma_{\mathrm{host},\alpha} = 100/(1+z_\alpha)\;\mathrm{pc}\,\mathrm{cm}^{-3}$ (as also found in \cite{2024arXiv240916952C})
, also uncorrelated between FRBs. Lastly, the covariance induced by the LSS is described in \cite{reischke_cosmological_2023}\footnote{An implementation of the covariance is provided on \href{https://github.com/rreischke/frb_covariance}{https://github.com/rreischke/frb\_covariance}.}.

Following the procedure of cosmic shear, we can also average over ensembles of FRBs distributed as $n^{\mathrm{FRB}}_i(z)$, so that the averaged, projected dispersion measure is given by \cite{reischke_probing_2021,reischke_consistent_2022}:
\begin{equation}
\label{eq:disper_tomo}
    \mathcal{D}_{\mathrm{LSS},i}(\hat{\boldsymbol{n}}) =  \int_0^{\chi_{H}}\!\!{d}\chi\; W_{\mathcal{D}_i}(\chi)\delta_{e}
    \big[\hat{\boldsymbol{x}},z(\chi)\big]\,,
\end{equation}
where the dispersion measure weight is given by:
\begin{equation}
\label{eq:averaged_dispersion_measure_fluctuation_weighting_function}
    W_\mathcal{D}(\chi) = \frac{3 H_0^2 \Omega_{{b}}}{8\pi G m_{p}}\chi_\mathrm{e}[1+z(\chi)] f_\mathrm{IGM}(z)\int_\chi^{\chi_{H}}{d}\chi'\, n^\mathrm{FRB}_i(\chi') \, .
\end{equation}

\subsubsection*{Large-scale structure probes}
The most basic summary statistic of LSS observables is the angular two-point function or its harmonic space equivalent, the angular power spectrum, $C(\ell)$. For any two projected tracers $A$ and $B$ of the density field of the form of \Cref{eq:kappa_tomo} and \Cref{eq:disper_tomo} the angular power spectrum can be expressed it as:
\begin{equation}
   C_{AB}(\ell)  = \int_0^{\chi_{H}}\frac{{d}\chi}{\chi^2}W_A(\chi) W_B(\chi) P_{AB}\left(\frac{\ell + 1/2}{\chi},z(\chi)\right)\,.
\end{equation}
 assuming the Limber approximation \cite{2008PhRvD..78l3506L}.
Here $W_{A}$ are probe-specific weights and $P_{AB}$ is the three-dimensional power spectrum of the two probes under consideration. In the case considered here, we need the electron, matter and electron $\times$ matter power spectra. For this, we use the \texttt{pyhmcode} \cite{mead_hydrodynamical_2020,troster_joint_2022} which was calibrated to the BAHAMAS simulations and emulated in \cite{neumann_fast_2024}. The angular power spectrum gives the variance of the spherical harmonic modes of the project field, $a_{\ell m}$, as a function of inverse angular scale $\ell$. Due to the symmetries of the cosmological model, statistical properties of different $\ell$ do not couple in this estimator and the fluctuations are independent of $m$. 
 Any observed spectrum, $\hat{C}_{AB}(\ell)$ will contain a white noise component due to the discreteness of the traced field:
\begin{equation}
    \hat{C}_{AB}(\ell) = {C}_{AB}(\ell) + N_{A}\delta^{K}_{AB}\,,
\end{equation}
with the Kronecker delta $\delta^{K}_{AB}$ and the shot noise level $N_{A}$. The former ensures that different tracers are uncorrelated if they constitute different objects, since the noise level only arises from the Poisson sampling of the traced field. The noise levels for cosmic shear and FRBs are:
\arraycolsep=1.3pt\def\arraystretch{1.5}
\begin{equation}
\label{eq:noiselevels}
    N_{A} = 
    \left\{\begin{array}{cl}
        \frac{\sigma^2_{\epsilon,i}}{2\bar{n}^{s}_i} & \mathrm{if\;}  A = \kappa_i\\
        \frac{\sigma^2_\mathrm{host}}{\bar{n}^\mathrm{FRB}_i(1+\bar{z}_i)} & \mathrm{if\;}  A = \mathcal{D}_i\,,
    \end{array}\right.
\end{equation}
where $\sigma^2_{\epsilon,i}$ is the total ellipticity dispersion of the source sample in the $i$-th tomographic bin with average number density $\bar{n}^{s}_i$. Likewise, $\bar{n}^\mathrm{FRB}_i$ is the average number density of FRBs in the tomographic bin $i$ in the FRB sample with average redshift $\bar{z}_i$. \Cref{eq:noiselevels} shows the great advantage of FRBs compared to cosmic shear: the intrinsic ellipticity of galaxies is much larger than the cosmological signal, requiring many galaxies to statistically observe a significant signal. For FRBs, the intrinsic noise is of the same order as the cosmological signal.

Measuring the angular power spectra is limited by cosmic variance and pure shot noise. Cosmic variance is simply the sample variance of cosmology and originates from the fact that we observe only one realisation of the Universe. For example, the estimation of the angular power spectrum at $\ell = 1$ will only be possible using three independent 
modes: $m=-1,0,1$, limiting further shrinking of the error. With this in mind, the Gaussian covariance between $\hat{C}_{AB}(\ell)$ and $\hat{C}_{CD}(\ell^\prime)$ can be calculated by means of Wick's theorem:
\begin{equation}
\label{eq:Gaussian_cov}
    \mathrm{Cov}[C_{AB}(\ell), C_{CD}(\ell^\prime)] = \frac{\delta^{K}_{\ell\ell^\prime}}{f_\mathrm{sky}N_\ell}\left(\hat C_{AC}(\ell)\hat C_{BD} (\ell^\prime) + \hat C_{AD}(\ell)\hat C_{BC} (\ell^\prime)\right)\;,
\end{equation}
where $N_\ell = (2\ell + 1)\Delta \ell$ is the number of modes available in each multipole band with width $\Delta\ell$ and $f_\mathrm{sky}$ is the sky fraction. For the comparison presented here, we will ignore the non-Gaussian and super-sample covariance terms. 

\begin{figure}
    \centering
    \includegraphics[width=0.8\textwidth]{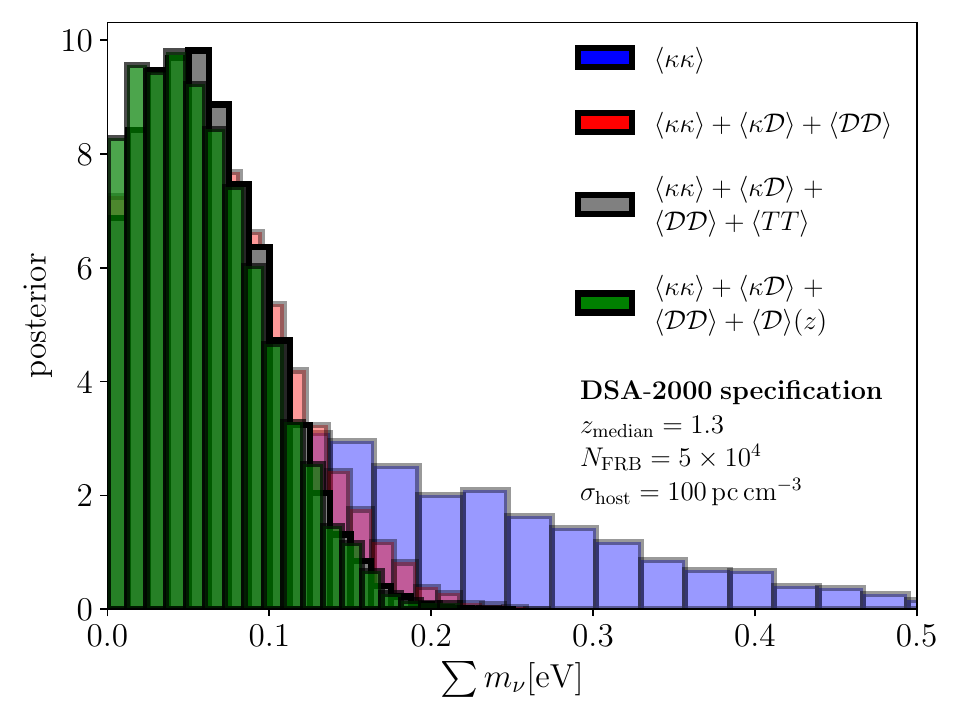}
    \caption{Projected improvement for a joint measurement between Euclid-WL (i.e. cosmic shear $\kappa$) and DSA-2000 (i.e. dispersion measure $\mathcal{D}$). The blue curve ($\sum m_\nu < 0.21\,\mathrm{eV}$) shows the Euclid-WL constraints for the optimistic scenario given in \cite{blanchard_euclid_2020}. In case of the three other histograms shown, the labelled probe is added to Euclid-WL. The red histogram ($\sum m_\nu < 0.09\,\mathrm{eV}$) adds all possible two-point correlations obtained from $5\times 10^4$ FRBs, with a median redshift of 1.3. By adding temperature anisotropies ($TT$) from the CMB as measured by Planck \cite{Planck:2018nkj}, we arrive at the grey histogram ($\sum m_\nu < 0.08\,\mathrm{eV}$). Lastly, the dispersion measure-redshift relation is added to the measurements as well, yielding the green bars ($\sum m_\nu < 0.10\,\mathrm{eV}$). All upper limits given are the 68$\,\%$ upper limits.}
    \label{fig:neutrino_WL_DSA2000}
\end{figure}

\subsection{Cosmological inference of the sum of the neutrino masses}
We create mock noisy data vectors for the fiducial cosmology specified in \Cref{tab:parameters_and_priors}. 
We then sample the posterior distribution over the parameter and priors outlined in \Cref{tab:parameters_and_priors} using \texttt{emcee} \cite{goodman_2010,foreman_emcee_2013}. Our interest here is the sum of the neutrino masses, we therefore only show the marginalised constraints on $\sum m_\nu$.  
For the analysis here, we distinguish four scenarios and display the results in \Cref{fig:neutrino_WL_DSA2000}:
\begin{itemize}
    \item[$(i)$] Cosmic shear only ($\langle\kappa\kappa\rangle$, blue) assuming a Stage IV cosmic shear survey with optimistic settings as described in \cite{blanchard_euclid_2020}.

    \item[$(ii)$] Adding the dispersion measure $\mathcal{D}$ into the cosmic shear survey, considering all three possible correlations: $\langle\kappa\kappa\rangle + \langle\kappa\mathcal{D}\rangle + \langle\mathcal{D}\mathcal{D}\rangle$ (shown in red).

    \item[$(iii)$] The same as $(ii)$ adding as well the primary temperature anisotropies $T$ from Planck: $\langle\kappa\kappa\rangle + \langle\kappa\mathcal{D}\rangle + \langle\mathcal{D}\mathcal{D}\rangle + \langle TT\rangle$ (shown in grey). 
 \item[$(iv)$] The same as $(ii)$ but also adding the dispersion measure-redshift relation as a probe: $\langle\kappa\kappa\rangle + \langle\kappa\mathcal{D}\rangle + \langle\mathcal{D}\mathcal{D}\rangle + \langle \mathcal{D}\rangle(z)$ (shown in green). The covariance between the dispersion measure correlations and the dispersion measure-redshift relation is ignored in this analysis. 
\end{itemize}
Our findings indicate that FRBs can effectively provide the same enhancement in constraining power as adding CMB measurements to a Stage IV cosmic shear survey. This enhancement is primarily driven by the breaking of degeneracies involving baryonic feedback, as demonstrated in \cite{reischke_calibrating_2023}, alongside additional constraints on the Hubble constant and baryon density. Alternative methods to measure BF are the Sunyaev-Zel'dovich (SZ) effects \citep{sunyaev_observations_1972,sunyaev_microwave_1980}. In \citep{troster_joint_2022} the thermal SZ effect  was used to calibrate the feedback strength and breaking degeneracies. Alternatively, the kinetic SZ effect was employed in \citep{Hand:2012ui,Planck:2015ywj,DES:2016umt,PhysRevD.93.082002} and has also the potential to constrain feedback models \citep{PhysRevD.103.063513, 2021PhRvD.103f3514A}.

The $68\,\%$ upper limit on neutrino mass is improved by approximately a factor of three when incorporating 50,000 FRBs into a cosmic shear analysis. This competitive constraint, entirely independent of the CMB, introduces a high degree of complementarity to the measurements, which is essential for consistent constraints in an era of precision cosmology, especially in light of increasing tensions \cite[see][for a review]{abdalla_cosmology_2022}.

\begin{table}[t]
\centering
\renewcommand{\arraystretch}{1.15}
\begin{tabular}{c|c|c|c}
\hline
\hline
\textbf{Parameter} & \textbf{Description} & \textbf{Fiducial value} & \textbf{Prior range} \\
\hline
\hline
$h$ & reduced Hubble constant & $0.674$ & $[0,\infty)$ \\ 
\hline
$\Omega_{b}$ & baryon density parameters & $0.045$ & $[0,\Omega_\mathrm{m})$ \\
\hline
$\sigma_8$ & fluctuation amplitude at $8\,\mathrm{Mpc}\,h^{-1}$ & $0.83$ & $[0,\infty)$ \\
\hline
$\sum m_\nu\, [\mathrm{eV}]$ & sum of neutrino mass & $0.05$ & $[0,\infty)$ \\
\hline
$n_{s}$ & spectral index of primordial fluctuations & $0.963$ & $[-4,4]$ \\
\hline
$w_0$ & constant dark energy equation of state & $-1$ & $[-1,\infty)$ \\
\hline
$w_{a}$ & varying dark energy equation of state & $0$ & $[-10,10]$ \\
\hline
$\log_{10}T_\mathrm{AGN}$ & baryonic feedback strength & $7.8$ & $[6.8,8.8]$ \\
\hline
$\Omega_{m}$ & matter density parameter & $0.31$ & $[0,1]$ \\
\hline
$\langle \mathcal{D}_\mathrm{host}\rangle\,[\mathrm{pc}\, \mathrm{cm}^{-3}]$ & mean host contribution & $100$ & $[0,\infty)$ \\
\hline
\hline
\end{tabular}
\caption{Parameter varied to infer the neutrino mass scale. For the CMB, we additionally marginalise over a few other parameters, such as the optical depth.}
\label{tab:parameters_and_priors}
\end{table}

There are several limitations to consider in this analysis, which we will outline below. We assumed a simplistic one-parameter model for the effect of baryonic feedback. More sophisticated models would include a greater number of parameters (typically around six, as seen in \cite{schneider_new_2015,arico_bacco_2021,giri_emulation_2021}). However, we can anticipate that the degradation of constraints provided by cosmic shear becomes even more pronounced for more complex models of baryonic feedback, making their calibration even more crucial. Moreover, the analysis overlooks systematic effects, but since these are not included for any of the probes, the comparison of the final constraining power remains consistent. This also applies to the general simplicity of the inference using a simple Gaussian likelihood and covariance matrix. For the latter, one might expect that the benefit from FRBs is greater, as the non-Gaussian covariance should be more significant for cosmic shear due to its larger total signal from Stage IV surveys compared to DSA-2000.

\section{Discussion and conclusions}
\label{sec:conc}

The DSA-2000's capabilities offer distinct opportunities for probing BSM physics during its planned operation beginning in 2028.
Here, we have quantified the discovery potential for select hypotheses, exploiting electromagnetic signatures as well as gravitational imprints on measured radio observables from new physics.

\begin{table}[h]
\centering
\begin{tabular}{p{2.2cm}|p{1.5cm}|p{1.5cm}|p{2cm}|p{2.5cm}|p{2.8cm}}  
\hline
\hline
\textbf{Science case} & \textbf{Data } & \textbf{Time} & \textbf{Resolution} & \textbf{Sensitivity ($g_{a\gamma\gamma}/\epsilon$)} & \textbf{Mass ($\mu$eV)} \\
\hline
\hline
Axion DM & high-res beam & $10$ h & $10$ kHz & $2 \times 10^{-13}$ GeV$^{-1}$ & $2.9 - 8.3$ \\
\hline
Axion clouds & imaging/ high-res beam & $10$ h/ $4$ h & $50$ MHz & $4 \times 10^{-15}$ GeV$^{-1}$ & $2.9 - 8.3$ \\
\hline
Dark photon superradiance & imaging & $15$ min & $1.3$ GHz & $2 \times 10^{-8} - 10^{-7}$ & $6 \times 10^{-8} - 5 \times 10^{-7}$ \\
\hline
\hline
\end{tabular}
\caption{ Table summarizing requirements and resulting sensitivity for three of the considered science cases with electromagnetic signatures. The imaging data product is the standard continuum all-sky survey image cube. High-resolution beamforming data refers to placing a single beam on the position of a known source, preserving data down to 100 $\mu$s sampling time and 10 kHz.}
\label{tab:your_label}
\end{table}

One significant finding concerns axion cloud formation around NSs, where we have developed an analytical treatment of the expected radio signal as a function of period and magnetic field, previously examined only through simulations. Our analytical framework conservatively captures key features of the full simulation while providing physical insight into the underlying processes. Using targeted observations not yet included in the planned observing strategy, DSA-2000 could potentially reach sensitivity to the theoretically significant QCD axion target band in the mass range 2.9$-$8.3 $\mu$eV. This represents a promising avenue for detecting axions, with potential implications for axion direct detection. In Table~\ref{tab:your_label} we summarize the achievable sensitivities with the DSA 2000.

The pulsar timing capabilities of DSA-2000 offer a powerful probe of fundamental physics through gravitational effects. Our analysis shows that the planned discovery of $\sim$130 new millisecond pulsars will improve constraints on both DM substructure and fifth forces by approximately an order of magnitude compared to current limits. 
While the initial five-year dataset leaves us short of probing the $f_{\mathrm{DM}} = 1$ threshold, extended timing over a 25-year period would dramatically enhance sensitivity to levels corresponding to probing the local DM density. This long-term potential of pulsar timing arrays represents a promising approach to constraining compact DM models. Additionally, exploiting the timing correlation between pulsars—which has not been considered in this work—could potentially further increase the sensitivity to $f_{\mathrm{DM}}$, possibly bringing its projected upper limit down to unity. We leave this analysis to future work.

Additionally, DSA-2000's detection of FRBs will contribute to cosmological neutrino mass measurements by breaking degeneracies with baryonic feedback effects in weak lensing observations, providing a competitive exclusively late universe measurement.

There are other compelling BSM radio signals that we have not pursued in this work, some of which we briefly highlight below. 

\begin{itemize}

\item Axion miniclusters have observable radio signals. However, the prediction for axion miniclusters constituting all of the DM lies above the DSA-2000 frequency band \cite{Gorghetto:2018myk, Buschmann:2019icd, Gorghetto:2020qws, Edwards:2020afl, Gorghetto:2024vnp}. 

\item  Stimulated axion decay \cite{Buen-Abad:2021qvj} triggered by sources in the sky such as remnant supernova, lead to radio emission. The signal is most competitive for frequencies below the DSA-2000 band. 

\item The decay or scattering of weakly interacting massive particles into leptons induces a radio signal via synchrotron radiation, if the leptons are accelerated by astrophysical magnetic fields   \cite{Colafrancesco:2005ji, 2013ApJ...768..106S, Leite:2016lsv, Regis:2021glv, Bhattacharjee:2020phk}. This promising signal is sensitive to the astrophysical environment. 

\item Collisions between DM in the form of ultraheavy composite bound states, so-called dark blobs, could induce transient radio signals \cite{Kaplan:2024dsn}
which DSA-2000 is optimized for, a promising target to pursue in future work.

\item The large number of FRBs projected to be detected by DSA-2000 lead to a small chance of  measuring a repeating lensed FRB. Such a lucky measurement would provide a very sensitive probe of DM substructure \cite{Xiao:2024qay}. Unfortunately, due to the scanning strategy of DSA-2000, and the theoretical gaps in being able to predict in which interval an FRB may repeat, make this an extremely challenging measurement. 
\end{itemize}

Several promising directions have emerged from our analysis. Extended pulsar timing observations beyond the initial five-year program would substantially enhance sensitivity to DM substructure, potentially reaching theoretically significant thresholds. For axion cloud searches the newly discovered population of NS will play an important role, and could identify an ideal target for dedicated follow-up observations that could reach the QCD band. 
In conclusion, the DSA-2000 represents not just an advancement in radio astronomy capabilities, but a versatile tool for probing fundamental physics across multiple fronts.
In future work, we will be searching for the BSM signals discussed in this work with data taken by the DSA-2000.

\acknowledgments
We are grateful to Masha Baryakhtar, Abhiram Cherukupalli, Joshua Foster, David E. Kaplan, Andrea Mitridate, Vikram Ravi, Benjamin Safdi, Kai Schmitz, Myles Sherman, Erwin Tanin and Tanner Trickle for stimulating discussion. We are especially grateful to Joseph Lazio, Cristina Mondino, Stephen Taylor and Samuel Witte for providing comments on the manuscript. K.B. and Y.D. thank the U.S. Department of Energy, Office of Science, Office of High Energy Physics, under Award Number DE-SC0011632 and the Walter Burke Institute for Theoretical Physics. V.L. is supported by NSF Physics Frontier Center Award \#2020275 and the Heising-Simons Foundation. A.P. acknowledges support from the Princeton Center for Theoretical Science. K.Z. is supported by a Simons Investigator Award, the U.S. Department of Energy, Office of High Energy Physics under Award Number DE-SC0011632, and by the Walter Burke Institute for Theoretical Physics. 

\appendix
\section{Appendix}

We use a combination of natural units, $c = \hbar = 1$, and radio astronomy units for the observed radio flux density $F_{\text{obs}}, [\text{Jy} = 10^{-26}\, \text{Watts}/{\text{m}^2 \text{Hz}} = 10^{-23}\, \text{erg}/\text{cm}^{2}\text{s}\,\text{Hz}]$.
The observational time required to reach a sensitivity limit $\sigma [\mu\text{Jy}]$ for a radio point source is given by 
\begin{equation}\label{eqn:SEFD}
\Delta t_{\text{obs}} = \left(\frac{2 k T}{A_{\text{eff}}} \right)^2 \frac{1}{\sigma^2 \mathcal{B} n_p} 
=(\text{SEFD})^2 \frac{1}{\sigma^2 \mathcal{B} n_p} \,,
\end{equation}
where $T$ is the system temperature, $k$ is the Boltzmann constant, $\mathcal{B} \equiv \Delta f$ is the signal bandwidth, $n_p =2$ is the number of polarizations, and $A_{\text{eff}}$ is the effective dish area, taking into account the collecting area of a single element and the number of elements. 
The system equivalent flux density is 
\begin{equation}\label{eqn:SEFD_DSA2000}
\text{SEFD} [\text{Jy}] = \left(\frac{2 k T}{A_{\text{eff}}} \right) \, ,
\end{equation}
where for DSA-2000 $A_{\text{eff}} = 10240 \, \text{m} $, and $T = 25 \, \text{K}$.

\bibliographystyle{JHEP}
\bibliography{bib,nsdm1}

\end{document}